\documentclass[conference]{IEEEtran}
\usepackage{cite}
\usepackage[pdftex]{graphicx}
\usepackage[cmex10]{amsmath}
\usepackage{array}
\usepackage[caption=false,font=footnotesize]{subfig}
\usepackage{url}
\usepackage{booktabs,multirow,multicol,tabularx,pifont,wasysym,hyperref,xcolor}
\usepackage[noend]{algpseudocode}
\usepackage{capt-of}
\usepackage{ifthen}

\newcommand{\fu}[1]{\dofu#1}
\newcommand{\dofu}[1]{\underline{#1}\nobreak\hspace{0pt}}

\renewcommand{\paragraph}[1]{
\noindent
{\bf #1.}\hspace{.5em}
}

\begin{document}

\title{Passive Inference Attacks on Split Learning via~Adversarial~Regularization}

\author{\IEEEauthorblockN{Xiaochen Zhu\IEEEauthorrefmark{1}\IEEEauthorrefmark{2},
        Xinjian Luo\IEEEauthorrefmark{1}\IEEEauthorrefmark{3},
        Yuncheng Wu\IEEEauthorrefmark{4}\textsuperscript{$\diamond$}\textsuperscript{1},
        Yangfan Jiang\IEEEauthorrefmark{1},
        Xiaokui Xiao\IEEEauthorrefmark{1}, and
        Beng Chin Ooi\IEEEauthorrefmark{1}}
    \IEEEauthorblockA{\IEEEauthorrefmark{1}National University of Singapore, \IEEEauthorrefmark{2}Massachusetts Institute of Technology,\\\IEEEauthorrefmark{3}Mohamed bin Zayed University of Artificial Intelligence, \IEEEauthorrefmark{4}Renmin University of China\\xczhu@mit.edu, xinjian.luo@mbzuai.ac.ae, wuyuncheng@ruc.edu.cn,\\yangfan.jiang@comp.nus.edu.sg, xkxiao@nus.edu.sg, ooibc@comp.nus.edu.sg
    }
}

\maketitle
\begingroup\renewcommand\thefootnote{}
\footnotetext{This is the full and corrected version of our paper that appeared at NDSS 2025 \cite{zhu2024passive}.\\}
\endgroup
\begingroup\renewcommand\thefootnote{$\diamond$}
\footnotetext{This work was done at National University of Singapore.}
\endgroup
\begingroup\renewcommand\thefootnote{1}
\footnotetext{Corresponding author.}
\endgroup

\begin{abstract}
Split Learning (SL) has emerged as a practical and efficient alternative to traditional federated learning. While previous attempts to attack SL have often relied on overly strong assumptions or targeted easily exploitable models, we seek to develop more capable attacks. We introduce SDAR, a novel attack framework against SL with an honest-but-curious server. SDAR leverages auxiliary data and adversarial regularization to learn a decodable simulator of the client's private model, which can effectively infer the client's private features under the vanilla SL, and both features and labels under the U-shaped SL. We perform extensive experiments in both configurations to validate the effectiveness of our proposed attacks. Notably, in challenging scenarios where existing passive attacks struggle to reconstruct the client's private data effectively, SDAR consistently achieves significantly superior attack performance, even comparable to active attacks. On CIFAR-10, at the deep split level of 7, SDAR achieves private feature reconstruction with less than 0.025 mean squared error in both the vanilla and the U-shaped SL, and attains a label inference accuracy of over 98\% in the U-shaped setting, while existing attacks fail to produce non-trivial results.
\end{abstract}

\section{Introduction}\label{sec:intro}

To bridge the isolated data repositories across different data owners, federated learning (FL) \cite{mcmahan2017communication,kairouz2021advances,wu2020privacy,wu2023falcon,bonawitz2017practical} has been proposed as a solution to privacy-preserving collaborative learning.
However, participants engaged in FL often suffer from low communication efficiency and heavy computational overhead. This is often imposed by the iterative process of local model training and the frequent exchange of model parameters, especially for deep neural network (NN) models.
Naturally, as a simple adaption with enhanced communication and computation efficiency of FL, split learning (SL) \cite{vepakomma2022split,poirot2019split,gupta2018distributed,vepakomma2018split,thapa2021advancements,yin2021comprehensive,singh2019detailed} has drawn increasing attention in various applications, such as healthcare \cite{ha2022feasibility,vepakomma2018split,poirot2019split} and open source packages \cite{pysyft,hall2020split}.
The general idea of SL is to \emph{split} an NN model into smaller partial models, with simpler models being allocated to clients, and more intricate ones hosted on a computationally capable server. 
During training, clients first send the intermediate representations produced by their partial models to the server. Subsequently, the server performs a forward pass on its own partial model and back-propagates the gradients back to the clients for model updates.
In this way, SL enables privacy-preserving collaborative learning by sharing only the intermediate representations without revealing the original private data from the clients.
Compared to FL, SL adopts a more computationally efficient approach, yielding improved communication efficiency and scalability without compromising model utility~\cite{yin2021comprehensive,singh2019detailed,gao2020end}.
Nonetheless, privacy concerns still exist within the SL framework. Given that clients share model intermediate representations with the server, one may naturally wonder \emph{if it is possible for the server to infer private data of clients from the shared \mbox{intermediate} representations}.

\paragraph{Related work} To address this question, several inference attacks~\cite{pasquini2021unleashing,erdougan2022unsplit,gao2023pcat,yin2023ginver,qiu2023exact,he2019model} have been devised to investigate the privacy risks of SL.
However, \emph{these attacks typically consider overly strong threat models and target non-standard settings favorable for the adversary.}
For example, Pasquini et al.~\cite{pasquini2021unleashing} assume a malicious server that \emph{actively} tampers the back-propagated gradients. This deviates from the original training protocol of SL and can be easily detected as an agreement violation \cite{fu2023focusing, erdogan2022splitguard}.
In addition, Erdo{\u{g}}an et al.~\cite{erdougan2022unsplit}, Qiu et al.~\cite{qiu2023exact} and Gao and Zhang~\cite{gao2023pcat} propose passive inference attacks, in which the server honestly follows the SL training protocol but attempts to infer clients' private data by analyzing the shared intermediate outputs.
Although these attacks are more stealthy, they often target non-standard and easily exploitable models, as did in \cite{pasquini2021unleashing}.
Specifically, the targeted models on the client side are typically non-standard models that are overly wide for their input dimensions. This results in higher dimensional intermediate representations that encode more information about the private input, thus making the attacks more effective.
Additionally, existing attacks are often evaluated on split configurations where the clients' models are shallow, which are commonly believed to be more vulnerable to attacks \cite{vepakomma2020nopeek,pasquini2021unleashing,gao2023pcat}. %
Correspondingly, as we later show in the paper, these attacks can be mitigated by considering models with adequate width and allocating more layers to the clients' partial models.
Furthermore, the visual features of reconstructed images produced by existing passive attacks are hardly comparable to that of the active attack proposed by Pasquini et al.~\cite{pasquini2021unleashing}.
Therefore, it is still underexplored whether passive inference attacks can achieve reasonable effectiveness comparable to their active counterparts with less exploitable model structures. %

In this paper, we address this gap by developing passive feature and label inference attacks on more challenging SL settings where existing attacks~\cite{pasquini2021unleashing, erdougan2022unsplit, gao2023pcat} fail to work, i.e., we target standard models with adequate width at deeper split levels.
By considering passive attacks, we can ensure the stealthiness of our attacks, making them hardly susceptible to detection by active defense mechanisms~\cite{fu2023focusing,erdogan2022splitguard}.
By considering the more challenging settings, we can reveal the privacy vulnerabilities of SL that may exist in real-world applications but have not been explored. %
Nonetheless, before devising the proposed attacks, several challenges must be addressed.

\paragraph{Challenges} First, the main challenge in server-side attacks on SL is the absence of access (including the black-box access) to the client's partial model. Namely, the server cannot feed specific input to the client's model and observe the corresponding output. Such lack of access renders the traditional approach of model inversion~\cite{fredrikson2015model,zhao2021exploiting,fredrikson2014privacy,yang2019neural,truong2021data} inapplicable.
Second, in a passive setting where the server cannot manipulate the client's model, the only information available to the server is the shared intermediate representations during training.
As the complexity of the client's partial model increases, e.g., with deeper split levels, the information encoded in these representations diminishes. This would make it more challenging to decode the private data solely from the received representations.  
Third, during SL training, the client's model undergoes continuous updates in each iteration. This dynamic nature results in entirely different representations over iterations, further complicating inference attacks, compared to attacks on finalized models~\cite{he2019model,yin2023ginver}.

Aside from the typical SL setting, we also consider a more intricate SL configuration, known as U-shaped SL. In this setting, the last few layers of the model are also kept privately by the clients, and the server's responsibility is limited to training the intermediate layers of a neural network. 
This introduces two new challenges. First, the absence of access to training examples' labels poses the need for the server to simultaneously recover private features and labels to achieve a reasonable attack. Second, established practices and theoretical insights of transfer learning indicate that the last few layers of a network are crucial to learn domain-specific representations~\cite{yosinski2014transferable}. The lack of these layers makes it harder for the server to recover private training samples.

\paragraph{Contributions} To address these challenges, we propose a novel class of passive attacks on SL, namely, \fu{S}imulator \fu{D}ecoding with \fu{A}dversarial \fu{R}egularization (SDAR). The foundation of SDAR is to let the server train a simulator that emulates the client's private model while simultaneously training a decoder tailored to the trained simulator, on auxiliary data disjoint from the private target data.
Ideally, a well-trained decoder should also be capable of decoding the client's private model. However, due to the intrinsic differences in distribution between the client's private data and the server's auxiliary data, the simulator trained by the server tends to overfit the auxiliary data. Consequently, the simulator cannot faithfully replicate the behaviors of the client's model. This renders the decoder incapable of reconstructing the client's data albeit proficient at reconstructing the auxiliary data.

To solve this issue, we draw inspiration from generative adversarial networks (GANs) \cite{goodfellow2020generative,goodfellow2014generative} and innovatively propose to regularize the simulator and decoder through an adversarial loss, encouraging them to learn more generalized representations that are transferable to the client's private data. 
For vanilla SL where the server has access to the labels of training records, SDAR utilizes the label information in the style of conditional GAN \cite{mirza2014conditional}. 
In the more challenging U-shaped SL setting, where the client privately retains the last few layers of the model along with the record labels, SDAR trains an additional simulator to mimic the behavior of these last few layers. 
To mitigate overfitting of this supplementary simulator, we introduce label random flipping, which encourages this simulator to learn the general representations rather than mere memorization of the server's data. 
Furthermore, we utilize the predictions of this simulator on private samples to achieve label inference attacks in the context of the U-shaped setting.

We conduct extensive experiments on various real-world datasets and widely-used model architectures in both SL configurations to demonstrate the effectiveness of the proposed attacks. The results show that under challenging settings where existing passive attacks fail to effectively reconstruct clients' private data, SDAR achieves consistent and distinctive attack performance, and is even able to almost match the performance of active hijacking attacks. This is the first time that passive attacks have been shown to be comparably effective as their active counterparts.
Also, SDAR remains effective when the server has limited access to auxiliary dataset or possesses no knowledge of the client's model architecture.
Additionally, our results reveal that inference attacks become more challenging with the increase of split level or the decrease of the client's model width. While the former effect has been investigated in the literature \cite{pasquini2021unleashing,erdougan2022unsplit,gao2023pcat,qiu2023exact}, we are the first to demonstrate that wider models, due to their higher dimensional intermediate representations, are also more vulnerable to inference attacks. This is a critical insight for practitioners to consider when deploying SL in real-world applications.
Finally, we evaluate potential countermeasures and show the robustness of our attacks against such defenses.

\section{Preliminaries}\label{sec:pre}

\begin{figure}
    \centering
    \subfloat[Vanilla SL]{%
        \centering
        \includegraphics[height=55mm]{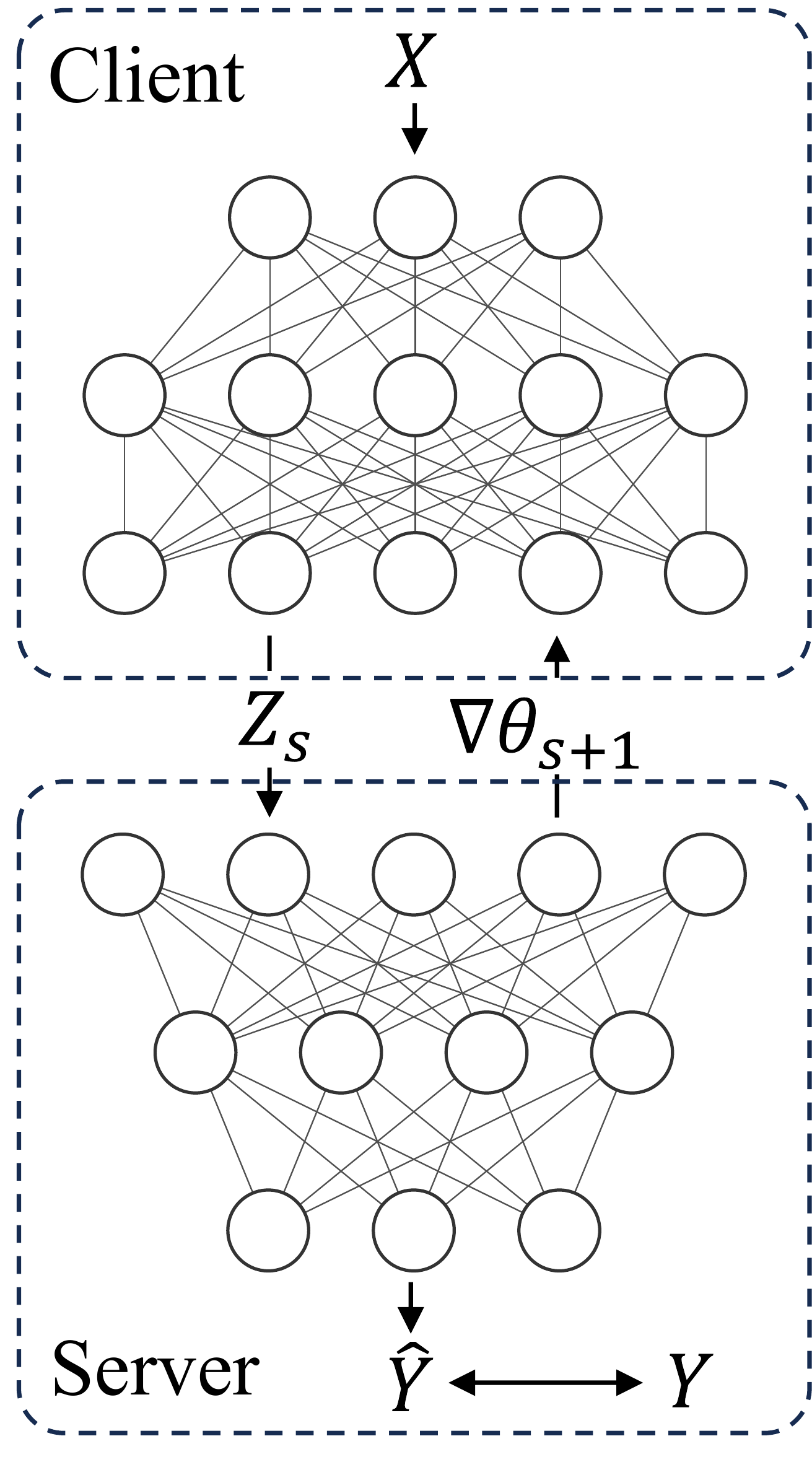}
        \label{fig:vanilla_sl}
    }
    \hfil
    \subfloat[U-shaped SL]{%
        \centering
        \includegraphics[height=55mm]{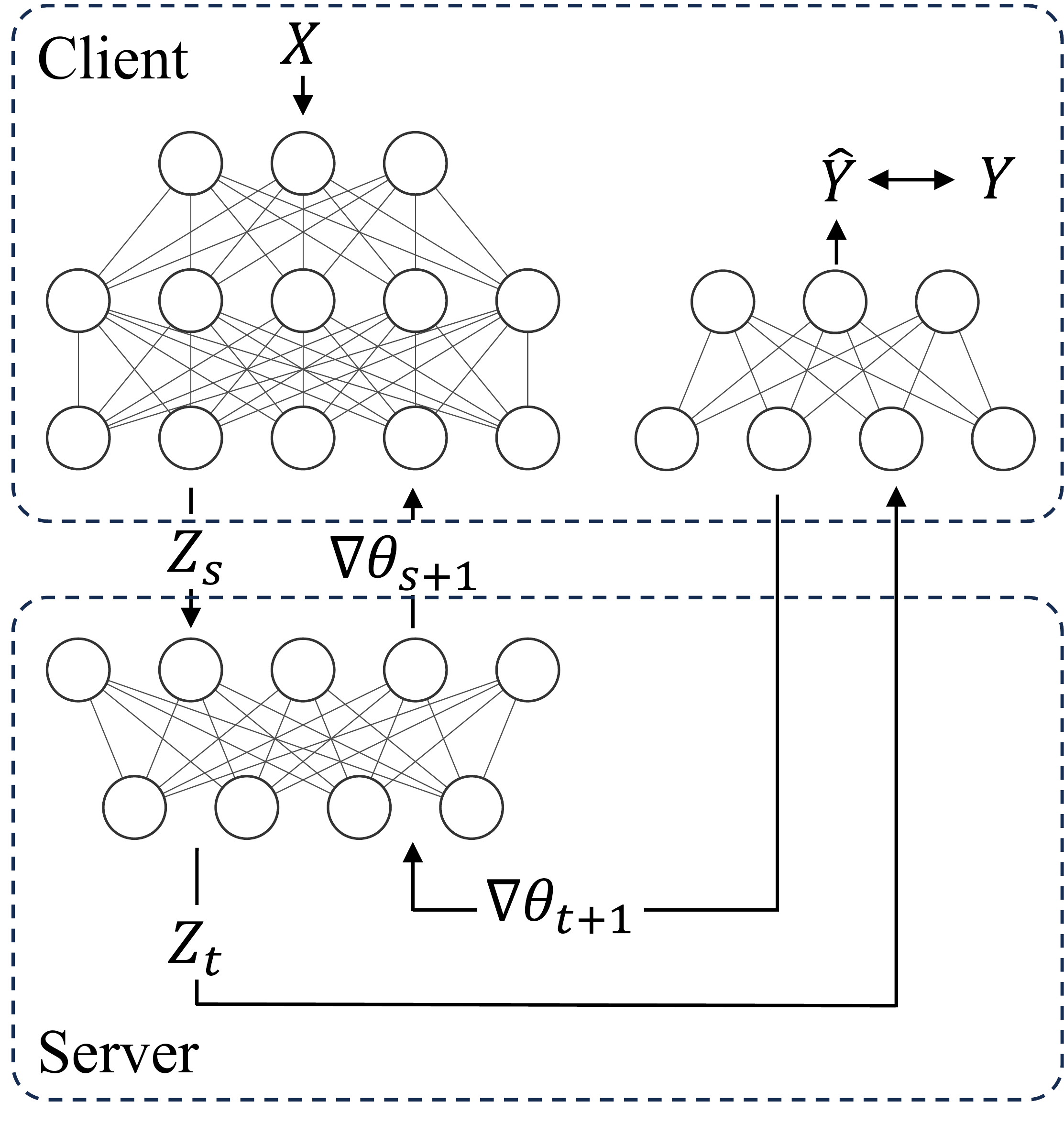}
        \label{fig:u_shape_sl}
    }
    \caption{Vanilla and U-shaped configurations of SL.}
    \label{fig:sl}
\end{figure}

\paragraph{Split learning} We consider training an NN model $H$ on dataset $D=\{(x_t,y_t):t=1,\ldots,N\}$. $H$ consists of $n$ layers $H=L_n\circ L_{n-1}\ldots \circ L_1$. The key idea of SL \cite{poirot2019split,gupta2018distributed,vepakomma2018split} is to \textit{split} the execution of $H$ by a \textit{split layer} (namely, $L_s$) and assign the first half $f=L_s\circ\ldots\circ L_1$ to the client and the second half $g=L_n\circ\ldots\circ L_{s+1}$ to the server. Then, $H=g\circ f$. 
In the vanilla SL setting, the server also holds the labels of training examples. For a batch of examples $X$, we define the representations returned by layer $L_i$ as $Z_i$. Then, on the forward pass, the client sends intermediate representations $Z_s=f(X)$ (also known as smashed data) to the server such that the latter can complete the forward pass via $Z_n=g(Z_s)$. 
Since the server has access to the record labels $Y$, it can evaluate the loss $\ell(Z_n, Y)$ where $\ell(\cdot,Y)$ is the loss function given ground truth labels $Y$. In backpropogation, for parameters $\theta_i$ of $L_i$, chain rule gives \begin{equation}\label{eq:sl}
    \begin{aligned}
    \nabla\theta_i&=\frac{\partial\ell(Z_n,Y)}{\partial Z_n} \frac{\partial L_n(Z_{n-1})}{\partial\theta_{n}}\cdots \frac{\partial L_{i}(Z_{i-1})}{\partial\theta_i}\\
    &=\nabla\theta_{i+1}\frac{\partial L_{i}(Z_{i-1})}{\partial\theta_i}.
    \end{aligned}
\end{equation}
By \eqref{eq:sl}, one only needs the gradients of the next layer and the representations returned by the previous layer to differentiate layer $L_i$. Thus, the server can update parameters $\theta_g$ of its model $g$ after receiving $Z_{s}$ from the client, while the client can update $f$ after receiving $\nabla\theta_{s+1}$ from the server (see Fig.~\ref{fig:vanilla_sl} for an example). 

\paragraph{U-shaped SL} If the client's record labels are considered private and not shared with the server, SL can be configured as a U-shaped structure~\cite{gupta2018distributed}, as illustrated in Fig.~\ref{fig:u_shape_sl}.
Under U-shaped SL, the NN model is partitioned into three parts, i.e., $H=h\circ g\circ f$, where $f=L_s\circ\ldots\circ L_1$, $g=L_t\circ\ldots\circ L_{s+1}$, and $h=L_n\circ\ldots\circ L_{t+1}$. The client owns $f$ and $h$ while the server only hosts $g$. 
During the training on samples $X$, the client sends $Z_s$ to the server, which computes $Z_t$ and sends it back to the client. The client then computes the final prediction $Z_n=h(Z_t)$. In the backpropogation phase, the client first updates its model $h$ and sends $\nabla\theta_{t+1}$ to the server. Then, the server updates $g$ and sends $\nabla\theta_{s+1}$ to the client for updating the partial model $f$ on the client side.

\paragraph{SL with multiple clients} Due to its simplicity, the SL framework can be easily scaled to support multiple clients on either horizontally-partitioned data~\cite{vepakomma2018split} or vertically-partitioned data~\cite{ceballos2020split}.  
Note that in the multi-client SL scenarios, each client still needs to communicate with the server via the above-described protocol, i.e., sharing intermediate representations with the server.

\section{Problem Statement}\label{sec:prob}
 
\paragraph{System model} For ease of presentation, we introduce the proposed attacks in the SL setting where a client and a server collaboratively train a deep model $H$ in either the vanilla or the U-shaped configuration. 
Our attacks can be seamlessly extended to the multi-client SL scenario for inferring the private data from different clients without further modifications, given that the shared information between clients and the server in the multi-client scenario is the same as that in the single-client scenario.
Note that image datasets are typically used to train deep models under various SL settings~\cite{gupta2018distributed}.

\paragraph{Threat model} We investigate the privacy risks of SL under the honest-but-curious threat model~\cite{luo2021feature}. Namely, we consider a server that honestly adheres to the SL protocol and does not tamper with the training process. Meanwhile, the server tries to infer the client's private information from the received messages.
In addition, we assume that the server has background knowledge of an auxiliary public dataset $D'=\{(x'_t,y'_t):t=1,\ldots,m\}$ such that $D\cap D'=\emptyset$, where $D$ denotes the client's private dataset and shares a similar distribution with $D'$. 
This is a common and reasonable assumption in related studies~\cite{pasquini2021unleashing,gao2023pcat} because the task types and data domain should be negotiated between the client and the server before initiating an SL training. This is a necessary step for the participating parties to determine the appropriate network structure and split level before training, for eliminating possible overfitting or underfitting issues~\cite{pasquini2021unleashing,gao2023pcat}.
Consequently, the server can collect a public dataset from the same data domain for attack implementation~\cite{pasquini2021unleashing,gao2023pcat}. 
Note that we assume no access to any example in the client's private dataset $D$ for the server, which is more stringent than \cite{gao2023pcat} where $D'$ can be a subset of $D$. 
In addition, the server and the client should agree on the model architecture of $H$ and the split configuration beforehand for information exchange and model convergence~\cite{erdougan2022unsplit}.
That is, we start by assuming that the server knows the architecture of $f$. We later relax this, assuming that the server does not have any knowledge of the architecture of $f$, but only the input and output dimensions of the model.
During model training, the server has neither black-box nor white-box access to the client's models, and can only receive intermediate representations from the client based on the SL protocol.

In \emph{vanilla SL}, the client hosts model $f$ while the server hosts model $g$ and has access to the record labels. For a batch of training records $(X,Y)$, the server aims to reconstruct the private features $X$ given the received intermediate representations $Z_s$, i.e., $\hat X=\mathcal A(Y,Z_s,\theta_g,D')$, where $\hat X$ is the inferred features, $\theta_g$ represents the parameters of $g$, and $\mathcal A$ denotes the attack algorithm.
In \emph{U-shaped SL}, the client hosts partial models $f$ and $h$ while the server hosts model $g$. For a batch of training records $(X,Y)$, the server aims to reconstruct both the private features $X$ and the labels $Y$ given the received $Z_s$ and the gradient vector $\nabla \theta_{t+1}$, i.e., $\hat X,\hat Y=\mathcal A(Z_s,\theta_g,\nabla\theta_{t+1},D')$.

In Section~\ref{sec:related}, we discuss the threat models of existing attacks in greater details and compare them with ours. We also provide a taxonomy of inference attacks on SL in Table~\ref{tab:threat}.

\section{Passive attacks against SL}\label{sec:approach}

In this section, we introduce SDAR, a novel framework for an honest-but-curious server to infer the client's private data in SL. 
The core idea of SDAR is to train a simulator model that learns similar representations as the client model in a way that their outputs are indistinguishable from each other. This is achieved by introducing an adversarial discriminator that regularizes the training of the simulator model.
After that, a corresponding decoder is trained on $D'$ for decoding the intermediate representations output by the simulator. 
By regularizing this decoder via an additional adversarial discriminator, this decoder can be generalized to effectively decode the client's intermediate representations despite $D\cap D'=\emptyset$. 
SDAR is capable of inferring private features in the vanilla SL and both private features and labels in the U-shaped SL.

\begin{figure}[t]%
  \subfloat[Private $X$ (upper) vs $\hat X$ (lower) reconstructed by the na\"ive SDA]{%
      \begin{minipage}{249.54981pt}
      \includegraphics[width=\linewidth]{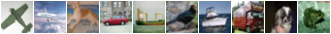}\\
      \includegraphics[width=\linewidth]{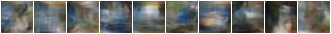}
      \end{minipage}
    \label{fig:naive_sda_recon}
  }\\
  \subfloat[Training loss of $g\circ f$ vs $g\circ \tilde f$]{%
    \includegraphics{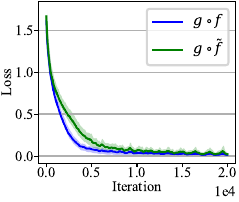}
    \label{fig:naive_sda_loss}
  }%
  \hfil
  \subfloat[Decoding MSE on $X'$ vs $X$]{%
    \includegraphics{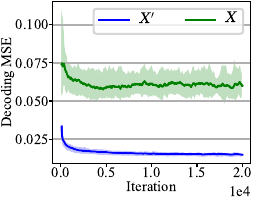}
    \label{fig:naive_sda_mse}
  }
  \caption{Failure of na\"ive SDA on CIFAR-10 with ResNet-20 at split level 7. Solid lines are mean values over 5 runs, and values between min/max boundaries are shaded. All later figures follow the same convention.}
  \label{fig:naive_sda}
\end{figure}

\subsection{Feature inference attacks in vanilla SL}\label{sec:vanilla}
We start with a simple attack prototype, denoted as na\"ive simulator decoding attack (na\"ive SDA). 
As the server has access to an auxiliary dataset $D'$, it can first initialize a simulator $\tilde f$ on its own and then train the model $g\circ \tilde f$ on $D'$. Depending on particular settings, this simulator $\tilde f$ may or may not share the same architecture as $f$. For now, we assume they share the same architecture but $\tilde f$ is initialized independently and randomly as the server has no access to the weights of $f$.
Specifically, for each batch of training examples $(X,Y)$, after the parameters of $f,g$ are updated via the training loss $\ell(g(f(X)),Y)$, the server samples another batch of examples $(X',Y')$ from $D'$ and trains $\tilde f$ to minimize loss $\ell(g(\tilde f(X')),Y')$ with the model $g$ frozen. 
The reason to fix $g$ in this step is mainly twofold. First, as an honest-but-curious party, the server should ensure that the parameters of $g$ are only updated based on $D$ as specified in the SL protocol. 
Second, fixing $g$ while training $\tilde f$ forces $\tilde f$ to learn representations that are compatible with $g$. As $g$ is trained collaboratively with the client's model $f$ on $D$, it is expected to memorize information of $D$~\cite{shokri2017membership} which may be implicitly leaked to $\tilde f$ while training $g\circ\tilde f$.
In this way, the encoder $\tilde f$ trained by the server is expected to mimic the behaviors of $f$. 
In the meantime, the server trains a decoder $\tilde f^{-1}$ with the transposed architecture of $\tilde f$ to decode $Z_s'=\tilde f(X')$, i.e., the decoder parameters $\theta_{\tilde f^{-1}}$ are updated via the loss $\ell_{\text{MSE}}(X',\tilde f^{-1}(Z_s'))$. 
One may expect the decoder $\tilde f^{-1}$ can not only decode the output of the simulator $\tilde f$ but also effectively decode the output $Z_s$ of $f$, i.e., $\tilde f^{-1}(Z_s)\approx X$ for private features $X$. 

Unfortunately, the na\"ive SDA fails to achieve this objective as illustrated in Fig.~\ref{fig:naive_sda_recon}. 
The main reason behind such failure is that although the training loss of $g\circ \tilde f$ on $X'$ converges to a similar minimum as that of $g\circ f$ on $X$ (see Fig.~\ref{fig:naive_sda_loss}), $\tilde f$ fails to learn the same representations as $f$. This generalization gap is rooted in the distributional discrepancy between the disjoint datasets $D$ and $D'$, and the fact that $\tilde f$ is trained solely on $D'$ makes it prone to overfitting to $D'$. As a result, even if the decoder $\tilde f^{-1}$ can precisely decode the simulator $\tilde f$ with very low reconstruction error, its reconstruction error on $Z_s$, the output of the client's private model $f$ on unseen features $X$, does not converge (see Fig.~\ref{fig:naive_sda_mse}).

In response, we propose to enhance na\"ive SDA with adversarial regularization, to encourage the simulator and the decoder to learn more general representations transferable to the client's private data $X$.

\begin{figure}[t]
  \centering
  \includegraphics[width=0.9\linewidth]{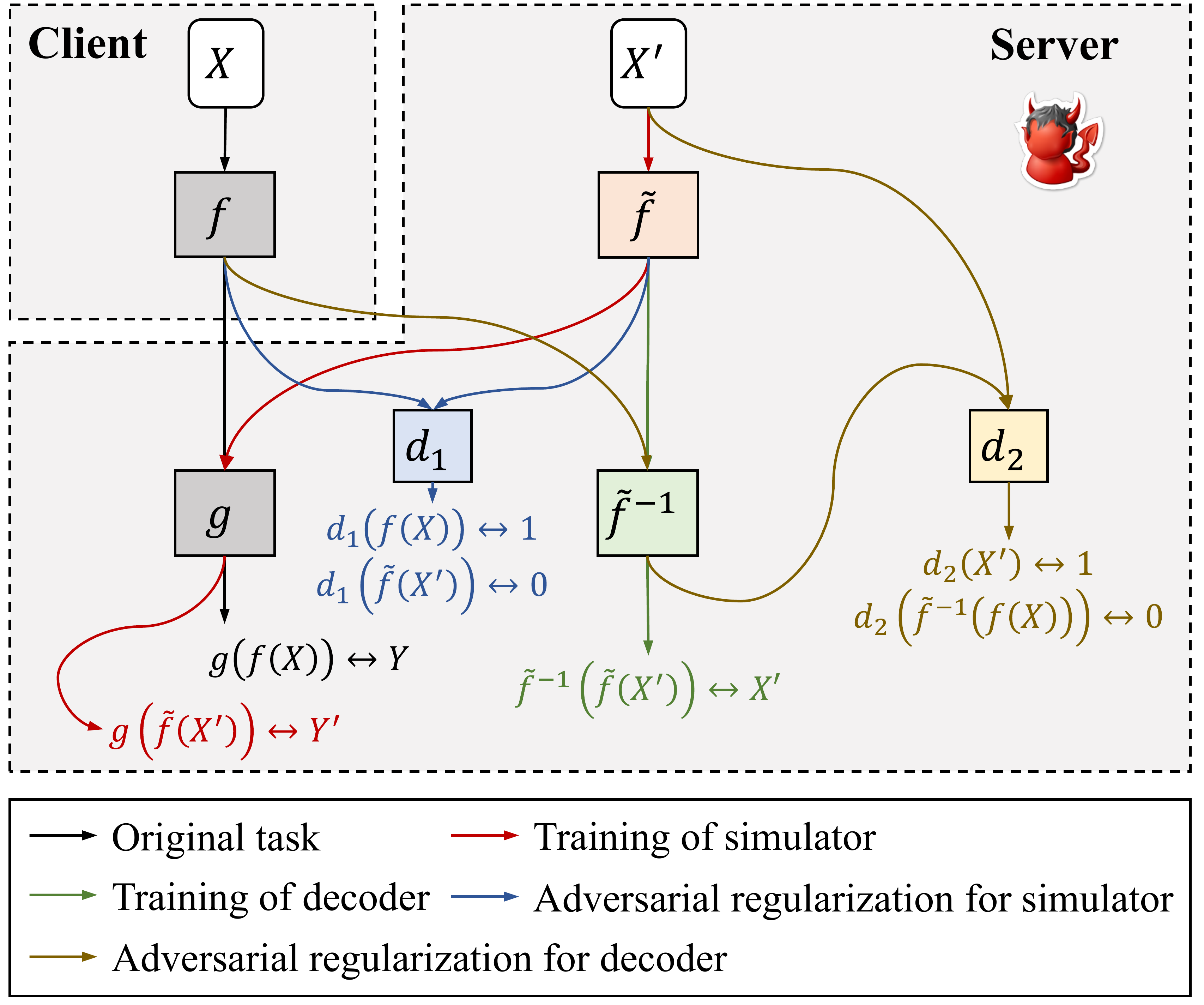}
  \caption{Overview of SDAR in the vanilla SL setting.}
  \label{fig:diagram}
\end{figure}

\paragraph{Adversarial regularization for $\tilde f$} To regularize the simulator $\tilde f$ such that it behaves more similarly to the client's private model $f$, we introduce a server-side discriminator $d_1$ that is trained to distinguish $Z_s'=\tilde f(X')$ (the fake data) against $Z_s=f(X)$ (the real data). In each iteration, $d_1$ is updated to minimize the loss \begin{equation}\label{eq:d1_loss}\mathcal L_{d_1}=\ell_{\text{BCE}}(d_1(Z_s'),0)+\ell_{\text{BCE}}(d_1(Z_s),1)\end{equation} to correctly classify $Z_s'$ against $Z_s$, 
where $\ell_{BCE}$ is the binary cross-entropy loss. 
In the meantime, the simulator is updated in an adversarial way that maximizes the likelihood of being misclassified by the discriminator, i.e., to minimize the loss $\ell_{\text{BCE}}(d_1(\tilde f(X')),1)$. 
We introduce this adversarial loss as a regularization term with penalty parameter $\lambda_1$, i.e., the simulator $\tilde f$ is updated to minimize the loss \begin{equation}\label{eq:e_loss}\mathcal{L}_{\tilde f}=\ell(g(\tilde f(X')), Y')+\lambda_1\cdot\ell_{\text{BCE}}(d_1(\tilde f(X')),1).\end{equation} 
Thus, with the training loss on $D'$ as the main objective and the adversarial loss as the regularization term, the simulator is expected to output representations indistinguishable from the output of $f$, thus better simulating $f$'s behaviors on the private data $X$.

\paragraph{Adversarial regularization for $\tilde f^{-1}$} As shown in Fig.~\ref{fig:naive_sda_mse} and \ref{fig:naive_sda_recon}, the decoder $\tilde f^{-1}$ can accurately decode the output of $\tilde f$ since $\tilde f^{-1}$ is trained in a supervised manner on $Z'_s$. 
However, this reconstruction capability can not be generalized to decode $Z_s$. Specifically, we observe from Fig~\ref{fig:naive_sda_recon} that the reconstructed samples $\hat X$ exhibit obfuscated visual features and can be easily distinguished from real images by a well-trained discriminator. 
Motivated by this observation, we introduce another discriminator $d_2$ to distinguish $\tilde f^{-1}(Z_s)$ (the fake data) against $X'$ (the real data) owned by the server. 
Similar to Eq.~\eqref{eq:d1_loss}, $d_2$ is trained by minimizing 
\begin{equation}\label{eq:d2_loss}\mathcal L_{d_2}=\ell_{\text{BCE}}(d_2(\tilde f^{-1}(Z_s)),0)+\ell_{\text{BCE}}(d_2(X'),1).\end{equation}
Meanwhile, the decoder $\tilde f^{-1}$ needs to be trained in an adversarial manner to maximize the likelihood of being misclassified by the discriminator. 
Similar to Eq.~\eqref{eq:e_loss}, we introduce a regularization term with penalty parameter $\lambda_2$, and the decoder is trained to minimize \begin{equation}\label{eq:decoder_loss}\mathcal L_{\tilde f^{-1}}=\ell_{\text{MSE}}(X',\tilde f^{-1}(Z_s'))+\lambda_2\cdot\ell_{\text{BCE}}(d_2(\tilde f^{-1}(Z_s)),1).\end{equation}
In this way, the decoder is trained to reconstruct private images from $Z_s$ with plausible visual features as real images.

\paragraph{Use of labels} In the vanilla SL, the server holds labels of all private training images, indicating that the server has access to the labels of both the private data $X$ and the auxiliary data $X'$. 
This enables the server to adopt the decoder $\tilde f^{-1}$ and the discriminators $d_1,d_2$ in a conditional manner similar to the conditional GANs \cite{mirza2014conditional}. 
Take the decoder $\tilde f^{-1}$ as an example. Instead of only taking the intermediate representations $Z_s, Z_s'$ as the input, a conditional $\tilde f^{-1}$ can additionally take the corresponding labels $Y,Y'$ as the input.
These labels can be first transformed into a high-dimensional embedding and then concatenated with the intermediate representations. 
A similar modification can be applied to the discriminators.
This adaptation enables the decoder to more effectively decode the intermediate representations, and enhances the discriminators' capability to distinguish between real and synthetic examples.

With the integration of the aforementioned enhancements, we denote the refined simulator decoding attack as Simulator Decoding with Adversarial Regularization (SDAR). We describe the complete attack in Fig.~\ref{fig:diagram} and provide pseudocode of our attack in Fig.~\ref{alg:sdar} of Appendix~\ref{app:pseudocode}.

\subsection{Feature \& label inference attacks in U-shaped SL}\label{sec:sdar_u}

In the U-shaped SL, the model $H$ is split into three parts $H=h\circ g\circ f$ such that only the intermediate partial model $g$ is kept by the server, and the other two are trained by the client.
An intuitive approach to adapt SDAR to the U-shaped SL is that the server trains an additional simulator $\tilde h$ to mimic the behaviors of $h$, which can be updated by minimizing the training loss of $\tilde h\circ g\circ \tilde f$ on $D'$ when freezing $g$.
If the simulator $\tilde h$ is well-trained on $D'$ such that it is able to generalize well to classify unseen private examples in $D$, the server can effectively reconstruct private labels $\hat Y=\tilde h(g(Z_s))$ in addition to private features.
However, the last few layers (i.e., $h$) of the model has a strong expressive capacity of the private information suggested by the practice and theory of transfer learning \cite{yosinski2014transferable}. 
Consequently, $\tilde h$ can easily overfit to $D'$ such that the training loss on $D'$ converges to a local minimum yet $\tilde h$ and $\tilde f$ fail to learn the similar behaviors of $h$ and $f$. 
In this case, the decoder $\tilde f^{-1}$ will not be able to reconstruct the private input $X$ from $Z_s$ effectively when trained along with the overfitted $\tilde f$. Thus, it is crucial to ensure that the simulator $\tilde h$ learns the general and transferable data representations, instead of overfitting to $D'$. 

To this end, we introduce \textbf{random label flipping} in the training of $\tilde h\circ g\circ \tilde f$. 
Specifically, for an incoming batch of auxiliary examples $(X',Y')$ sampled from $D'$, we first independently flip each label $Y'_i$ to a random label $Y_i'$ with a probability of $p$. Namely, for each $Y_i'\in Y'$, we have $$Y_i'=\begin{cases}Y_i' & \text{w.p. } 1-p\\ \text{Uniform}(\mathcal Y) & \text{w.p. } p\end{cases},$$ where $\mathcal Y$ is the set of all possible labels. This way, the server generates a new noisy label set $\tilde Y'$ and trains $\tilde h\circ g\circ \tilde f$ based on the new training batch $(X', \tilde Y')$. 
These randomly flipped labels can help regularize the training of $\tilde h$, encourage $\tilde h$ not to overly fit the auxiliary data, but to produce more general representations that are compatible with the output of $g$ and meanwhile transferable to the private data $D$.
Other components of SDAR except for the use of labels in Section~\ref{sec:vanilla} are directly applicable to the U-shaped SL setting, including the adversarial regularization on $\tilde f$ and $\tilde f^{-1}$.
We describe the complete attack in Fig.~\ref{alg:sdar_u} of Appendix~\ref{app:pseudocode}.

\subsection{Computational complexity}\label{sec:complexity}

Let $b$ be the batch size, i.e., the number of examples used for SL training in each iteration. For each iteration, if there's no attack, the SL protocol requires the client and the server to update the model $g\circ f$ via a forward pass and backpropagation. Let $F_f$ and $F_g$ denote the number of floating-point operations (FLOPs) required by a single forward pass on models $f$ and $g$, respectively. The total computation cost of the SL protocol is $\mathcal{O}(b\cdot (F_f+F_g))$ FLOPs, as the forward pass costs $b(F_f+F_g)$ FLOPs and the backpropagation costs approximately twice as many FLOPs \cite{goodfellow2016deep}.
If the server is to reconstruct the private $b$ training images via na\"ive SDA, it will then also train $g\circ \tilde{f}$ and $\tilde{f}^{-1}$ using a batch of $b$ auxiliary images, before decoding $Z_s$ via $\tilde f^{-1}$. This introduces additional $\mathcal{O}(b(F_g+F_{\tilde f}+F_{\tilde{f}^{-1}}))$ FLOPs. Hence, the SL training and server's inference attack via na\"ive SDA require $\mathcal{O}(b(F_f+F_g+F_{\tilde{f}}+F_{\tilde f^{-1}}))$ FLOPs per iteration. Furthermore, if the server adopts full SDAR and introduce discriminators $d_1,d_2$, additional cost of $\mathcal{O}(bF_{d_1}+bF_{d_2})$ FLOPs will be introduced. Therefore, in total, SL training and SDAR will require $\mathcal{O}(b(F_f+F_g+F_{\tilde{f}}+F_{\tilde f^{-1}}+F_{d_1}+F_{d_2}))$ FLOPs of computation, which is asymptotically as efficient as SL training itself if model complexity is considered as a constant. We later compare the computational complexity of SDAR with other state-of-the-art attacks in Section~\ref{sec:exp}.

\subsection{Technical novelty}\label{sec:novelty}

First, SDAR identifies an underexplored vulnerability of SL, i.e.,  the risks of information leakage from the parameters of the server's model $g$. 
Since $g$ is trained together with $f$ on the private examples $(X,Y)$, it unintentionally memorizes information about the private data \cite{shokri2017membership}, and a carefully designed simulator trained with $g$ and $(X',Y')$ can accurately simulate the behaviors of $f$. 
On the contrary, most existing attacks \cite{pasquini2021unleashing,erdougan2022unsplit} focus on only exploiting the privacy risks of the client's intermediate representations $Z_s$, which is often insufficient to reconstruct private data under practical settings.

Second, SDAR utilizes a novel adversarial regularization method to considerably improve the generalization performance of the simulator and decoder on the client's private data, which can achieve consistent and robust attack performance on less vulnerable models where existing attacks~\cite{pasquini2021unleashing,erdougan2022unsplit,gao2023pcat} cannot work.
To the best of our knowledge, this is the first time that adversarial regularization is utilized in the design of inference attacks on FL or SL.

One parallel study, PCAT~\cite{gao2023pcat}, adopts a similar attack framework coinciding with our na\"ive SDA discussed in Section~\ref{sec:vanilla}, with some minor improvements.
However, as demonstrated in Fig.~\ref{fig:naive_sda} and our comparative experiments in Section~\ref{sec:exp}, PCAT fails to reconstruct reasonable images in challenging settings. Compared to naïve SDA or PCAT, SDAR's main advantage is our novel adversarial regularization framework, boosting the attack performance by improving the simulator and decoder's generalization.

\section{Experiments}\label{sec:exp}

\subsection{Experimental settings}

\paragraph{Datasets} We experiment with four popular benchmarking datasets: CIFAR-10~\cite{krizhevsky2009learning}, CIFAR-100~\cite{krizhevsky2009learning}, Tiny ImageNet~\cite{tinyimagenet}, and STL-10~\cite{coates2011analysis}.
CIFAR-10 and CIFAR-100 consist of 60,000 $32\times32\times3$ colored images in 10 and 100 classes, respectively. To demonstrate the scalability of our attacks to larger images, we experiment with image datasets with higher resolutions: Tiny ImageNet is a subset of the ImageNet dataset~\cite{deng2009imagenet} consisting of 120,000 colored images of size $64\times64\times3$ of 200 classes, while STL-10 is another subset of the ImageNet dataset with 13,000 colored images of size $96\times96\times3$ in 10 classes.
Due to limited space, we report results on CIFAR-10 and CIFAR-100 in the main text, and defer full results to Appendix~\ref{app:exp_res}. %
We normalize all images to $[0,1]$ beforehand to match the input configuration of the convolutional models and partition each dataset into two disjoint subsets $D,D'$ where $D$ belongs to the client, and $D'$ belongs to the server. 
We start with $|D|=|D'|$ where the server acquires auxiliary data of the same size as the client's private dataset. Then, we experiment with $|D'|\ll|D|$ to discuss the effectiveness of our attacks with much less data.

\paragraph{Models} We experiment SDAR and other baseline attacks on ResNet-20 and PlainNet-20 \cite{he2016deep} as model $H$. Both models have standard architectures specifically designed to classify $32\times32\times3$ datasets. 
We chose these models because they are widely adopted in real-world applications, and are recognized to be hard to invert \cite{behrmann2019invertible}. 
For both architectures, the complete model $H$ consists of 20 layers (19 convolutional layers of 16/32/64 filters and one additional fully-connected output layer) with batch normalization.
Particularly, ResNet-20 is equipped with 9 residual blocks, while PlainNet-20 is a VGG-style CNN model with 9 convolutional building blocks without residual connections.
We experiment with different split levels denoted by $s\in\{4,5,6,7\}$ on both models, characterized by the number of building blocks in the client's model $f$. 
The deepest client's model that we consider has 7 building blocks as adding any more blocks will render the client to host more parameters than the server, invalidating SL's purpose of enabling a powerful server to relieve the client from heavy computation.
In the vanilla SL, all the remaining layers are assigned to the server as shown in Table~\ref{tab:model_vanilla}. 
In the U-shaped SL, the last few layers, namely, the average pooling and the fully connected output layer, are also assigned to the client, while the server only hosts the layers in between. 
It is worth noting that we experiment with ResNet-20 and PlainNet-20 on larger images in Tiny ImageNet and STL-10, instead of wider models with more filters, because the former is more challenging for inference attacks due to the lower dimensionality of the intermediate representations.
By considering the standard convolutional models with 16/32/64 filters and deep split levels up to 7, we target more challenging settings where the client's private model is less exploitable to existing attacks. Our experiments later demonstrate that the wider and shallower models considered by existing attacks~\cite{pasquini2021unleashing,erdougan2022unsplit,gao2023pcat} are more vulnerable to inference attacks.

\begin{table}[t]
    \centering
    \caption{Model Statistics on Various Split Levels of 4--7 in Vanilla Split Learning}
    \label{tab:model_vanilla}
    \begin{tabular}{@{}ccccc@{}}
        \toprule
        \multirow{2}{*}{Level} & \multicolumn{2}{c}{Client's $f$} & \multicolumn{2}{c}{Server's $g$} \\
        \cmidrule(lr){2-3} \cmidrule(lr){4-5} 
        & No. of layers    & No. of parameters & No. of layers & No. of parameters\\ \midrule
        4           & 9            & 29,424      & 11           & 244,618     \\
        5           & 11           & 48,112      & 9            & 225,930     \\
        6           & 13           & 66,800      & 7            & 207,242     \\
        7           & 15           & 124,912     & 5            & 149,130     \\ \bottomrule
    \end{tabular}
\end{table}

For the simulator $\tilde f$ (and $\tilde h$ if in U-shaped SL) of SDAR, we start with the case where the server uses the same model architecture as that of the client's $f$ (and $h$ if in U-shaped SL). We later show that our attacks are still effective even without such information where the simulator has a different architecture. The server uses the decoder $\tilde f^{-1}$ with a structure transposed to $\tilde f$ ending with a sigmoid function to produce tensors within $[0,1]$. 
The discriminators $d_1$ that distinguishes $Z_s'$ from $Z_s$, and $d_2$ that distinguishes $\hat X$ from $X'$, are deep convolutional networks. Architecture details of the attack models are described in Appendix~\ref{app:sdar_impl}.

\paragraph{Baselines} We compare SDAR with existing attacks, including UnSplit \cite{erdougan2022unsplit}, PCAT \cite{gao2023pcat} and FSHA \cite{pasquini2021unleashing}. 
UnSplit \cite{erdougan2022unsplit} is a passive attack which reconstructs private features by minimizing $\ell_{\text{MSE}}(\tilde f(\hat X), Z_s)$ via alternating optimization on $\tilde f$ and $\hat X$. 
PCAT \cite{gao2023pcat} is also a passive attack that adopts a similar attack framework as our na\"ive SDA discussed in Section~\ref{sec:vanilla}.
FSHA \cite{pasquini2021unleashing}, on the other hand, is an active attack framework, where the server actively hijacks the gradients transmitted to the client such that the client's model $f$ is induced to behave similarly to the server's own encoder. To ensure the fairness of comparison, we use the same model architectures and auxiliary datasets for all attackers unless specified otherwise. Implementation details of the baselines are described in Appendix~\ref{app:baseline_impl}.

\begin{figure}[t]
    \centering
    \includegraphics{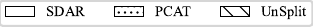}\\
    \subfloat[ResNet-20]{%
        \includegraphics{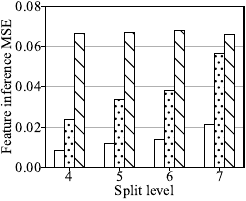}
        \label{fig:bars/vsl_resnet_cifar10_mse}
    }
    \hfil
    \subfloat[PlainNet-20]{%
        \includegraphics{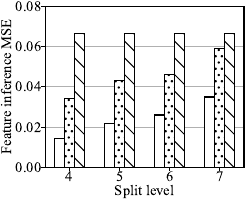}
        \label{fig:bars/vsl_plainnet_cifar10_mse}
    }
    \caption{Mean feature inference MSE on CIFAR-10 with ResNet-20 and PlainNet-20 at the split levels of 4--7 on CIFAR-10 in vanilla SL. Refer to Table~\ref{tab:vanilla_full} in the appendix for full results.}
    \label{fig:vsl_bars_cifar}

    \vskip 1.5\baselineskip

    \subfloat[ResNet-20]{%
        \small%
        \renewcommand{\arraystretch}{0}%
        \begin{tabular}{@{}m{3mm}@{}m{4.2cm}@{\hspace{2pt}}m{4.2cm}@{}}
            & \includegraphics[width=0.99\linewidth]{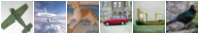} & \includegraphics[width=0.99\linewidth]{img/examples/original/cifar10/cifar10_cropped_6.png}\\\midrule
            4 & \includegraphics[width=0.99\linewidth]{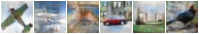} & \includegraphics[width=0.99\linewidth]{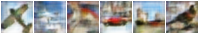}\\
            5 & \includegraphics[width=0.99\linewidth]{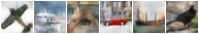} & \includegraphics[width=0.99\linewidth]{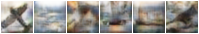}\\
            6 & \includegraphics[width=0.99\linewidth]{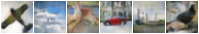} & \includegraphics[width=0.99\linewidth]{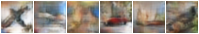}\\
            7 & \includegraphics[width=0.99\linewidth]{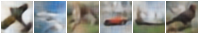} & \includegraphics[width=0.99\linewidth]{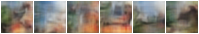}\\\\[6pt]
            &\centering \footnotesize SDAR & \centering \footnotesize PCAT \cite{gao2023pcat}
        \end{tabular}
    }\\
    \subfloat[PlainNet-20]{%
    \renewcommand{\arraystretch}{0}%
        \small%
        \begin{tabular}{@{}m{3mm}@{}m{4.2cm}@{\hspace{2pt}}m{4.2cm}@{}}
            & \includegraphics[width=0.99\linewidth]{img/examples/original/cifar10/cifar10_cropped_6.png} & \includegraphics[width=0.99\linewidth]{img/examples/original/cifar10/cifar10_cropped_6.png}\\\midrule
           4 & \includegraphics[width=0.99\linewidth]{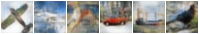} & \includegraphics[width=0.99\linewidth]{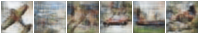}\\
           5 & \includegraphics[width=0.99\linewidth]{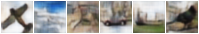} & \includegraphics[width=0.99\linewidth]{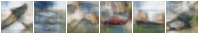}\\
           6 & \includegraphics[width=0.99\linewidth]{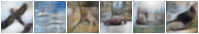} & \includegraphics[width=0.99\linewidth]{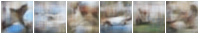}\\
           7 & \includegraphics[width=0.99\linewidth]{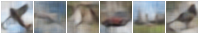} & \includegraphics[width=0.99\linewidth]{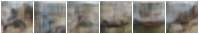}\\\\[6pt]
           &\centering \footnotesize SDAR & \centering \footnotesize PCAT \cite{gao2023pcat}
       \end{tabular}
    }

    \caption{Examples of feature inference attack results with ResNet-20 and PlainNet-20 on CIFAR-10 in vanilla SL. The first row shows the original private images while the following rows show the reconstructed images by SDAR and PCAT at various split levels 4--7. Reconstructions of the best quality among 5 trials are shown. Refer to Fig.~\ref{fig:v_examples_full} in the appendix for full results.}
    \label{fig:vanilla_examples}

\end{figure}

\begin{figure}[t]

    \centering
    \includegraphics{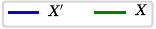}

    \subfloat[SDAR]{%
        \includegraphics{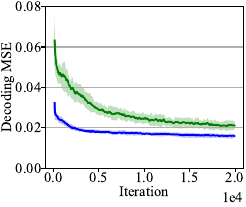}
        \label{fig:decoding_mse/sdar_resnet_cifar10}
    }
    \hfil
    \subfloat[PCAT]{%
        \includegraphics{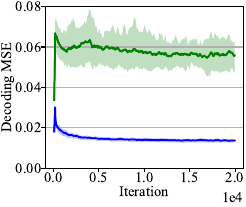}
        \label{fig:decoding_mse/pcat_resnet_cifar10}
    }
    \caption{SDAR and PCAT's decoding MSE on auxiliary data $X'$ versus private data $X$ with ResNet-20 at split level of 7 on CIFAR-10 in vanilla SL.}
    \label{fig:sdar_vs_pcat_decoding_mse}

    \vskip 1.5\baselineskip

    \centering
    \includegraphics{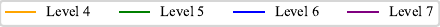}

    \subfloat[ResNet-20]{%
        \includegraphics{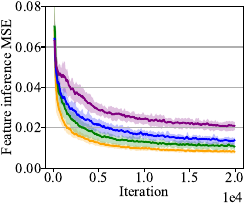}
        \label{fig:curves/vsl_resnet_cifar10_mse}
    }
    \hfil
    \subfloat[PlainNet-20]{%
        \includegraphics{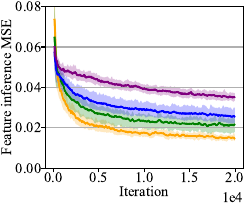}
        \label{fig:curves/vsl_plainnet_cifar10_mse}
    }
    \caption{Feature inference MSE over training iterations with ResNet-20 and PlainNet-20 at the split levels of 4--7 on CIFAR-10 in vanilla SL. Refer to Fig.~\ref{fig:v_sdar} in the appendix for full results.}
    \label{fig:sdar_cifar10_curve}
\end{figure}

\subsection{Experiments on vanilla SL attacks}\label{sec:exp_vanilla}

\paragraph{Attack effectiveness} We report the results of SDAR and other passive feature inference attacks on CIFAR-10 with ResNet-20 and PlainNet-20 at split levels 4--7 in Fig.~\ref{fig:vsl_bars_cifar}, and give concrete examples of private data reconstruction in Fig.~\ref{fig:vanilla_examples}. %
Fig.~\ref{fig:vsl_bars_cifar} shows that both SDAR and PCAT perform better at shallower split levels, and the performance degrades as the split level increases. This is expected as the intermediate representations $Z_s$ become more abstract and less informative about the private features $X$ at deeper layers, hence rendering it harder to reconstruct $X$ from $Z_s$.
We also notice that attacks are slightly more effective with the ResNet-20 than PlainNet-20, which is likely due to the former's residual connections that facilitate the flow of information through the network.

Among the passive feature inference attacks, Fig.~\ref{fig:vsl_bars_cifar} demonstrates that the proposed attack, SDAR, surpasses existing ones \cite{gao2023pcat,erdougan2022unsplit} in terms of reconstruction errors at all configurations by a significant margin.
We note that SDAR achieves nearly perfect reconstruction at shallower split cuts, and is still able to achieve exceptional reconstruction quality even at deep split levels up to 7 (see Fig.~\ref{fig:vanilla_examples}). 
PCAT \cite{gao2023pcat} is only able to recover private features at shallower split levels and of much lower quality than SDAR. In particular, the reconstruction quality of PCAT in the most trivial case (at level 4) is just on par with SDAR at the deepest level of 7. 
Moreover, the effectiveness of PCAT degenerates dramatically at deeper split levels, resulting in even larger performance gaps compared to SDAR. This is demonstrated in Fig.~\ref{fig:vanilla_examples} where PCAT struggles to produce reasonable reconstruction at split levels 6--7, while SDAR is still able to reconstruct images of visibly good quality. In particular, at the deepest and most challenging split level of seven, SDAR reduces the reconstruction MSE by a significant margin of 63\% on CIFAR-10 with ResNet-20.
We attribute the superior performance of SDAR to the introduction of the adversarial regularization. It ensures the simulator $\tilde f$ to generate realistic private features $Z_s'$ that are indistinguishable from the real ones. This leads to its decoder's improved reconstruction quality of unseen private data $X$. As shown in Fig.~\ref{fig:sdar_vs_pcat_decoding_mse}, while both decoders of SDAR and PCAT can successfully reconstruct auxiliary data $X'$ (as they are trained to do so in a supervised manner), only SDAR's decoder is able to generalize well to the unseen client's private data $X$. PCAT's decoder suffers from severe overfitting to the auxiliary dataset, as evidenced by the vast discrepancy between its reconstruction MSE on $X$ and $X'$.
At last, we note that the images reconstructed by UnSplit \cite{erdougan2022unsplit} are indistinguishable at all split levels and are thus omitted here.
This is expected as $\tilde f(\hat X)\approx f(X)$ does not necessarily lead to $\hat X\approx X$.%

To further investigate the dynamics of the proposed SDAR attack, we present its attack MSE over training iterations in Fig.~\ref{fig:sdar_cifar10_curve}. We observe that the attack performance of SDAR converges quickly within the first few epochs, and it quickly surpasses the final attack performance of the baseline attacks. With more training iterations, the server observes more incoming intermediate representations, and the attack performance improves. It is worth noting that the attack performance of SDAR consistently improves over time at all levels without any signs of overfitting, indicating its robustness and effectiveness.

\begin{figure}[t]
    \centering
    \includegraphics{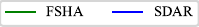}

    \subfloat[Feature reconstruction MSE]{%
        \includegraphics{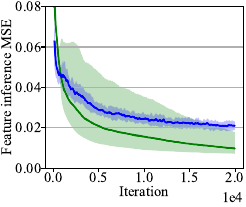}
        \label{fig:fsha_mse_cifar10}
    }
    \hfil
    \subfloat[Training loss of $g\circ f$]{%
        \includegraphics{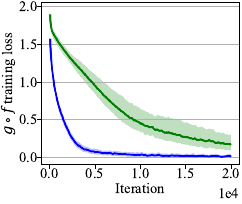}
        \label{fig:fsha_loss_cifar10}
    }
    \caption{Attack MSE and training losses on the original task of SDAR vs FSHA on CIFAR-10 with ResNet-20 at split level 7 in vanilla SL. Refer to Fig.~\ref{fig:fsha_full} in the appendix for full results.}
    \label{fig:fsha_cifar10}

    \vskip 1.5\baselineskip

    \small
    \renewcommand{\arraystretch}{0}
    \begin{tabular}{@{}m{1.3cm}@{}m{7.6cm}@{}}
        Original & \includegraphics[width=0.99\linewidth]{img/examples/original/cifar10/cifar10_cropped_10.png} \\\midrule
        FSHA & \includegraphics[width=0.99\linewidth]{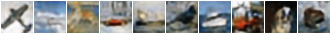}\\\midrule
        SDAR & \includegraphics[width=0.99\linewidth]{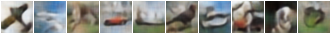}\\
        PCAT & \includegraphics[width=0.99\linewidth]{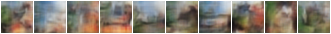}\\
    \end{tabular}
    \normalsize
    
    \caption{Examples of feature inference attack results of FSHA, SDAR, and PCAT on CIFAR-10 with ResNet-20 at split level 7 in vanilla SL.}
    \label{fig:fsha_vs_sdar}
\end{figure}

\paragraph{Comparison to active attacks} We also compare SDAR with the SOTA active feature inference attack, FSHA \cite{pasquini2021unleashing}, and report the attack MSE curve over training iterations at split level 7 on CIFAR-10 in Fig~\ref{fig:fsha_mse_cifar10}. Note that as an active attack where the server actively hijacks the client's gradients to manipulate the training of $f$, FSHA is expected to exhibit substantially better attack performance than passive attacks. 
Yet, we observe that our attack, despite being passive, still can achieve comparable attack performance as FSHA. As shown in Fig.~\ref{fig:fsha_vs_sdar}, the visual quality of the images reconstructed by SDAR is only slightly worse than FSHA, and is significantly better than PCAT.
This indicates the strong capability of our proposed attack, which is unprecedented in existing passive attacks.
The reason that SDAR can achieve comparable performance to FSHA while being passive may be attributed to the adversarial regularization that enables $\tilde f$ to better simulate the client's model $f$ without manipulating the training of $f$.
A particular advantage of SDAR is the preservation of the original training task as the server does not tamper with the SL training protocol. 
This is demonstrated in Fig.~\ref{fig:fsha_loss_cifar10} where SDAR's training loss of $g\circ f$ on $X$ converges much faster than FHSA.
Consequently, SDAR poses a greater risk to real-world applications compared to FSHA,
as the original training task is preserved and the server can perform the attack of robust quality without being detected by the client \cite{erdogan2022splitguard,fu2023focusing}.

\begin{figure}[t]
    \centering
    \includegraphics{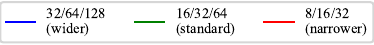}

    \subfloat[ResNet-20]{%
        \includegraphics{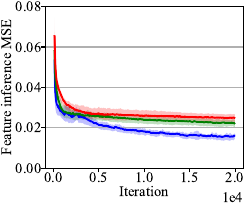}
        \label{fig:width_resnet_cifar100}
    }
    \hfil
    \subfloat[PlainNet-20]{%
        \includegraphics{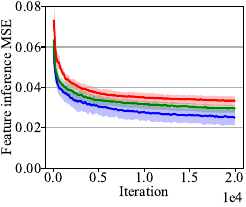}
        \label{fig:width_plainnet_cifar100}
    }
    \caption{Effects of target model widths on SDAR performed on CIFAR-100 with ResNet-20 and PlainNet-20 at split level 7 in vanilla SL. Refer to Fig.~\ref{fig:width_full} in the appendix for full results.}
    \label{fig:width_cifar}

    \vskip 1.5\baselineskip

    \scriptsize
    \centering
    \captionof{table}{Effects of Auxiliary Data Size on SDAR Performed on CIFAR-10 at Split Level 7 in Vanilla SL (Full Results Available in Table \ref{tab:aux_data_size_full} in the Appendix)}
    \label{tab:aux_data_size}
    \begin{tabular}{cccccc}
    \toprule
        $|D'|/|D|$& 1.0 & 0.5 & 0.2 & 0.1 & 0.05 \\\midrule
        Attack MSE & 0.0212 & 0.0225 & 0.0232 & 0.0248 & 0.0297\\
        w/ ResNet-20 & (0.0019) & (0.0014) & (0.0027) & (0.0024) & (0.0011) \\\midrule
        Attack MSE & 0.0350 & 0.0353 & 0.0365 & 0.0363 & 0.0435\\
        w/ PlainNet-20 & (0.0014) & (0.0025) & (0.0011) & (0.0014) & (0.0025) \\\bottomrule
    \end{tabular}
    \normalsize
    
    \vskip 1.5\baselineskip
    
    \small
    \renewcommand{\arraystretch}{0}
    \begin{tabular}{@{}m{1.8cm}@{}m{7.1cm}@{}}
        Original & \includegraphics[width=0.99\linewidth]{img/examples/original/cifar10/cifar10_cropped_10.png} \\\midrule
        $\frac{|D'|}{|D|}=1.0$ & \includegraphics[width=0.99\linewidth]{img/examples/sdar_results/vsl/cifar10/resnet_l7_cropped_10.png}\\
        $\frac{|D'|}{|D|}=0.05$ & \includegraphics[width=0.99\linewidth]{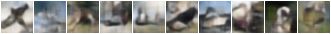}\\
    \end{tabular}
    \normalsize
    \captionof{figure}{Examples of feature inference attack results of SDAR performed on CIFAR-10 with ResNet-20 at split level 7 with auxiliary datasets of different sizes in vanilla SL.}
    \label{fig:aux_cifar10_examples}
\end{figure}

\paragraph{Effects of target model width} We have shown that deeper target model $f$ is less vulnerable to inference attacks, and we discuss the effects of model width (number of filters) on the attack performance of SDAR in Fig.~\ref{fig:width_cifar}. Since ResNet-20 and PlainNet-20 use convolutional layers of 16/32/64 filters, we experiment with wider and narrower models by doubling and halving filters in each layer. Despite the common belief that wider models are more robust to attacks due to their higher model complexity, Fig.~\ref{fig:width_cifar} shows that \emph{attack performance improves with model width} instead. This is because wider models output $Z_s$ of significantly larger dimensions, encoding more information about the private training data. It is worth noting that existing attacks \cite{pasquini2021unleashing,erdougan2022unsplit,gao2023pcat} are often evaluated on \emph{overly wide models at shallower split levels}, which, as discussed, are much easier to attack. For example, Gao and Zhang~\cite{gao2023pcat} evaluate PCAT on a CNN model with up to 512 filters for $32\times32\times3$ images in CIFAR-10, with the deepest split level of merely four convolutional layers allocated to the client. Our experiments with the standard model structures demonstrate that SDAR is able to attack much less vulnerable models with significantly better performance.

\begin{figure}[t]
    \centering
    \captionof{table}{Effects of the Removal of Classes from $D'$ on SDAR Performed on CIFAR-100 with ResNet-20 and PlainNet-20 at Split Level 7 in Vanilla SL}
    \label{tab:aux_data_distr}
    \scriptsize
    \begin{tabular}{ccccc}
    \toprule
            No. of classes removed & 0 & 1 & 5 & 10 \\\midrule
            Attack MSE & 0.0220 & 0.0241 & 0.0231 & 0.0329\\
            w/ ResNet-20& (0.0014) & (0.0015) & (0.0008) & (0.0070)\\\midrule
            Attack MSE & 0.0301 & 0.0305 & 0.0338 & 0.0423\\
            w/ PlainNet-20 & (0.0014) & (0.0029) & (0.0012) & (0.0037)\\\bottomrule
    \end{tabular}
    \normalsize
    
    \vskip 1.5\baselineskip
    
    \small
    \renewcommand{\arraystretch}{0}
    \begin{tabular}{@{}m{2.5cm}@{}m{6.4cm}@{}}
        Original & \includegraphics[width=0.99\linewidth]{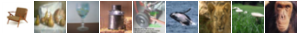} \\\midrule
        w/o 0 class & \includegraphics[width=0.99\linewidth]{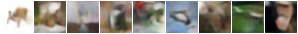}\\
        w/o 10 classes & \includegraphics[width=0.99\linewidth]{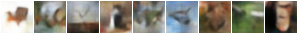}\\
    \end{tabular}
    \normalsize
    \captionof{figure}{Examples of feature inference attack results of SDAR performed on CIFAR-100 with ResNet-20 at split level of 7 with classes removed from the auxiliary datasets in vanilla SL.}
    \label{fig:aux_data_distr}

\end{figure}

\paragraph{Effects of auxiliary data}
Naturally, one is curious to ask if the proposed attacks are still effective \emph{when the server cannot curate a large auxiliary dataset that shares the same distribution as the client's dataset}. We hereby discuss the effects of the size and distribution of $D'$ on SDAR. We report the attack performance of SDAR with different sizes of auxiliary dataset $D'$ in Table~\ref{tab:aux_data_size}. We observe only a limited effect of the shrinkage of the auxiliary dataset on the performance of SDAR, as the attack MSE only slightly increases when $D'$ shrinks.
Fig.~\ref{fig:aux_cifar10_examples} also demonstrates precise reconstructions even with auxiliary data of size 5\% of the client's dataset.
As for the effects of distributional discrepancies between $D'$ and $D$, we experiment with SDAR on CIFAR-100 while removing examples of up to ten classes (out of 100 classes) from the auxiliary dataset in Table~\ref{tab:aux_data_distr}. With the removal of more classes in $D'$, the attack performance degrades, but to a limited extent. Fig.~\ref{fig:aux_data_distr} demonstrates that SDAR can still reconstruct images of visibly good quality even when 10 out of 100 classes are missing in the auxiliary dataset.

\begin{table}[t]
    \centering
    \caption{Feature Inference MSE on CIFAR-10 in Vanilla SL with or without Access to the Architecture of $f$ (Full Results Available in Table~\ref{tab:diff_sim_full} in the Appendix)}
    \label{tab:diff_sim}
    \setlength{\tabcolsep}{6pt}
    \begin{tabular}{ccccc}
        \toprule
        \multirow{2}{*}{lv.} & \multicolumn{2}{c}{Same architecture} & \multicolumn{2}{c}{Different architecture}\\
        \cmidrule(lr){2-3}
        \cmidrule(lr){4-5}
        & SDAR & PCAT \cite{gao2023pcat} & SDAR & PCAT \cite{gao2023pcat} \\ \midrule
        4 & \bfseries0.0084 (0.0002) & 0.0237 (0.0021) & \bfseries 0.0103 (0.0017) & 0.0257 (0.0034)\\
        5 & \bfseries0.0118 (0.0008) & 0.0337 (0.0048) & \bfseries0.0143 (0.0005) & 0.0289 (0.0038)\\
        6 & \bfseries0.0140 (0.0009) & 0.0382 (0.0062) & \bfseries0.0165 (0.0010) & 0.0433 (0.0114)\\
        7 & \bfseries0.0212 (0.0019) & 0.0568 (0.0062) & \bfseries0.0258 (0.0031) & 0.0545 (0.0039)\\\bottomrule
    \end{tabular}
\end{table}

Intuitively, with $D'$ of smaller size or with many classes missing, one would expect the simulator/decoder trained on $D'$ to be more prone to overfitting and thus to learn less useful information about client's private data. However, the core idea of SDAR is to reduce such overfitting by adversarial regularization. The positive results presented in this section demonstrate the effectiveness of adversarial regularization and show that SDAR can still work effectively even with a limited size of additional data, or with a substantial distributional discrepancy between $D'$ and $D$.
At last, it is worth noting that SDAR requires the auxiliary dataset's classes to be a subset of the target dataset's classes for simulator training, therefore we cannot evaluate the attack performance when the auxiliary dataset contains classes not present in the target dataset, for example, using CIFAR-100 to attack CIFAR-10. The same limitation also applies to PCAT \cite{gao2023pcat}. However, as discussed previously, it is reasonable to assume that the server has knowledge of the client's data domain and can curate an auxiliary dataset from various sources, such as public datasets or even synthetic data.

\paragraph{Effects of simulator architecture}
We have assumed that the server knows the architecture of the client's partial model $f$ and can initialize a simulator with identical architecture. Although this is a realistic assumption in practice, it is interesting to investigate whether SDAR still has superior attack performance when the server does not know the architecture of $f$. To this end, we conduct additional experiments with SDAR and PCAT \cite{gao2023pcat} in vanilla SL, where the server only knows the input and output dimensions of $f$. In particular, we use the standard ResNet-20 model for $f$ and $g$, but we use plain convolutional layers for the server's simulator $\tilde f$ as the server does not know the architecture of the client's model but only its input and output dimensions.
 We report the attack performance of SDAR and PCAT in Table~\ref{tab:diff_sim}. We observe no significant loss in both attacks' performance compared to the scenario where the server's simulator shares an identical architecture as the client, and SDAR is still effective in reconstructing high-quality training examples and outperforms PCAT by a large margin. This demonstrates that although the additional knowledge of the client's model architecture can sometimes benefit the adversary of our proposed attack, such knowledge has limited effects on attack performance and SDAR does not require the server to know the architecture of $f$ to be effective.

\begin{table}[t]
    \centering
    \caption{Computational Complexity of Split Learning with Various Inference Attacks}
    \label{tab:complexity}
    \begin{tabular}{ll}
        \toprule
        Method & FLOPs per iteration\\\midrule
        SL w/o attack & $\mathcal{O}(b(F_f+F_g))$\\
        SL w/ PCAT & $\mathcal{O}(b(F_f+F_g+F_{\tilde{f}}+F_{\tilde f^{-1}}))$\\
        SL w/ SDAR & $\mathcal{O}(b(F_f+F_g+F_{\tilde{f}}+F_{\tilde f^{-1}}+F_{d_1}+F_{d_2}))$\\
        SL w/ UnSplit & $\mathcal{O}(b(F_f+F_g+mF_{\tilde f}))$\\\bottomrule
    \end{tabular}

    \normalsize
    \vskip 1.5\baselineskip
    \scriptsize

    \centering
    \caption{Average Runtime per Iteration (s) of Split Learning with Various Inference Attacks on CIFAR-10 with ResNet-20 at Different Split Levels in Vanilla SL}
    \label{tab:runtime}
    \setlength{\tabcolsep}{5pt}
    \begin{tabular}{lcccc}
        \toprule
        Split level & 4 & 5 & 6 & 7\\\midrule
        SL w/ PCAT & 0.113 (0.000) & 0.120 (0.000) & 0.124 (0.009) & 0.132 (0.008)\\
        SL w/ SDAR & 0.164 (0.001) & 0.164 (0.004) & 0.169 (0.001) & 0.174 (0.002)\\
        SL w/ UnSplit & 4212.754 & 5019.434 & 5718.145 & 6929.952 \\\bottomrule
    \end{tabular}
\end{table}

\paragraph{Computational complexity} As analyzed in Section~\ref{sec:complexity}, SL training with SDAR is asymptotically as computational efficient as the standard SL training process, when fixing model complexity. We present the FLOPs required by SL with PCAT, SDAR, and UnSplit attacks in Table~\ref{tab:complexity}. The computational complexity of SDAR and PCAT follows the analysis in Section~\ref{sec:complexity}, while in UnSplit, the server needs to perform an inner loop of iterative optimization process to attack each batch of private examples for every iteration. Let the number of optimization steps be $m$, then the UnSplit requires a total of $\mathcal{O}(b(F_f+F_g+mF_{\tilde f}))$ FLOPs per iteration for SL training and private feature reconstruction, significantly more computationally expensive than SDAR and PCAT.
We also present the runtime of SL with different inference attacks in a controlled environment with one half of an A100 80GB GPU (as a virtual MIG device) in Table~\ref{tab:runtime}. We observe that SDAR is only slightly slower than PCAT, and both are significantly faster than UnSplit. The difference between SDAR and PCAT, albeit limited, indicates the additional costs of adversarial regularization. This shows that, compared to PCAT, SDAR achieves high-quality attack performance with a reasonable trade-off in computational cost.

\paragraph{Ablation studies} To conclude this section, we conduct ablation studies in Table~\ref{tab:ablation} to evaluate the effectiveness of each component in SDAR. We observe that all components are crucial, as removing any of them results in worse attack performance. In particular, attack performance degenerates the most when the server does not use adversarial regularization (case 4), showing its effectiveness to reduce overfitting and improve the generalization of the simulator/decoder. Between the two adversarial regularizers for $\tilde f$ and $\tilde f^{-1}$, SDAR is more sensitive to the former (case 2 vs 3). This is because the successful generalization of the decoder heavily relies on the quality of the simulator. At last, we notice a marginal performance gain by utilizing labels of the private examples in the style of CGANs \cite{mirza2014conditional} (case 1 vs 5), when the discriminators and decoders are fed with additional information.%

\subsection{Experiments on U-shaped SL attacks}

\begin{table}[t]
    \caption{Ablation Studies on CIFAR-10 with ResNet-20 at Split Level 7 in the Vanilla Split Learning Setting}
    \label{tab:ablation}
    \centering
    \begin{tabular}{ccccc}
        \toprule
        & $d_1$ (for $\tilde f$) & $d_2$ (for $\tilde f^{-1}$) & Conditional & Attack MSE\\\midrule
        1 & \ding{51} & \ding{51} & \ding{51} & \bfseries 0.0212 (0.0019)\\
        2 & \ding{55} & \ding{51} & \ding{51} & 0.0579 (0.0053)\\
        3 & \ding{51} & \ding{55} & \ding{51} & 0.0234 (0.0015)\\
        4 & \ding{55} & \ding{55} & \ding{51} & 0.0607 (0.0093)\\
        5 & \ding{51} & \ding{51} & \ding{55} & 0.0229 (0.0021)\\\bottomrule
    \end{tabular}
\end{table}

\begin{figure}[t]
    \centering
    \includegraphics{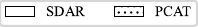}

    \vskip .5\baselineskip

    \includegraphics{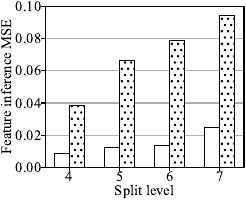}
    \hfil
    \includegraphics{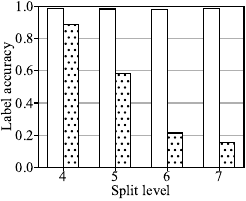}

    \caption{Mean feature inference MSE (left) and label inference accuracy (right) with ResNet-20 at the split levels of 4--7 on CIFAR-10 in U-shaped SL. Refer to Table~\ref{tab:u_mse_full} and Table~\ref{tab:u_acc_full} in the appendix for full results.}
    \label{fig:u_cifar10}

    \vskip 1.5\baselineskip

    \small
    \renewcommand{\arraystretch}{0}
    \begin{tabular}{@{}m{4mm}@{}m{4.2cm}@{\hspace{2pt}}m{4.2cm}@{}}
         & \includegraphics[width=0.99\linewidth]{img/examples/original/cifar10/cifar10_cropped_6.png} & \includegraphics[width=0.99\linewidth]{img/examples/original/cifar10/cifar10_cropped_6.png}\\\midrule
        4 & \includegraphics[width=0.99\linewidth]{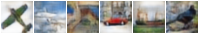} & \includegraphics[width=0.99\linewidth]{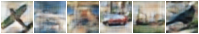}\\
        5 & \includegraphics[width=0.99\linewidth]{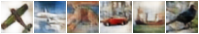} & \includegraphics[width=0.99\linewidth]{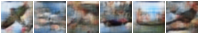}\\
        6 & \includegraphics[width=0.99\linewidth]{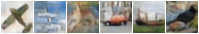} & \includegraphics[width=0.99\linewidth]{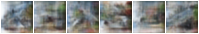}\\
        7 & \includegraphics[width=0.99\linewidth]{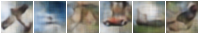} & \includegraphics[width=0.99\linewidth]{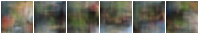}\\
        & \centering \vspace{6pt} \footnotesize SDAR  & \centering \vspace{6pt} \footnotesize PCAT \cite{gao2023pcat}
    \end{tabular}
    \normalsize
    \captionof{figure}{Examples of feature inference attack results with ResNet-20 on CIFAR-10 in U-shaped SL. See Fig.~\ref{fig:vanilla_examples} for detailed descriptions. Refer to Fig.~\ref{fig:u_examples_full} in the appendix for full results.}
    \label{fig:u_examples}

    \vskip 1.5\baselineskip

    \centering
    \includegraphics{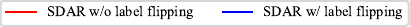}
    \vskip .5\baselineskip

    \includegraphics{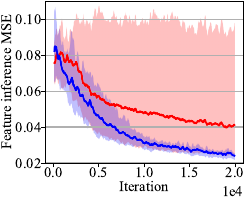}
    \hfil
    \includegraphics{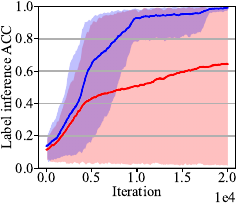}
    \caption{Effects of random label flipping on SDAR performed on CIFAR-10 with ResNet-20 at split level 7 in U-shaped SL.}
    \label{fig:label_flip}
\end{figure}

\paragraph{Attack effectiveness} Under U-shaped split learning, we report the feature reconstruction MSE and the label inference accuracy of SDAR in Fig.~\ref{fig:u_cifar10}, and present examples of private feature reconstructions in Fig.~\ref{fig:u_examples}. Note that we no longer present results of UnSplit \cite{erdougan2022unsplit} here due to limited space and its inability to produce distinguishable results even in the vanilla SL setting. 
Similar to feature inference attacks in vanilla SL, SDAR can reconstruct private features of high quality across various split levels, and there is no significant performance drop compared to vanilla SL. 
The reason is that although the server no longer has access to the last part of the model, i.e., $h$, it manages to train a simulator $\tilde h$ that learns the same representations, as supported by the high label inference accuracy which consistently reaches around 98\% on CIFAR-10 across all split levels. 
As the server trains $\tilde h\circ g\circ\tilde f$ solely on $D'$ (while $g$ is trained on $D$), its high test accuracy on unseen examples in $D$ demonstrates that information about $D$ is indeed leaked to the server via the trained parameters of $g$. In contrast, PCAT \cite{gao2023pcat} achieves visibly worse feature reconstruction performance than in vanilla SL, since PCAT is unable to train an effective simulator to $h$, as supported by its poor label inference accuracy. With deeper split levels and fewer parameters in the server's $g$, less information about $D$ is leaked via $g$, and hence the label inference accuracy of PCAT dramatically degenerates to the level of random guessing when the split level increases to 7.

\paragraph{Effects of random label flipping} Many of the properties of SDAR in vanilla SL discussed in Section~\ref{sec:exp_vanilla}, e.g. the effects of target model width, auxiliary data, simulator architecture and the components of adversarial regularizers, directly apply to the U-shaped SL, and here we focus on the effects of random label flipping on the training of $\tilde h$. Fig.~\ref{fig:label_flip} shows that after removing random label flipping, the label inference accuracy of SDAR %
drops significantly, and in the worst case, the label inference accuracy quickly vanishes to near-zero, even worse than random guessing. Thus, without label flipping, server's model $\tilde h$ will quickly overfit to its own dataset $D'$ without simulating the behavior of $h$ on $D$. As a result, the simulator $\tilde f$ trained together with $\tilde h$ will not learn the same representations as $f$ and hence the decoder can no longer generalize to decode the client's $f$, leading to poor feature inference performance as well. With random label flipping, the server's model $\tilde h$ is forced to learn the general representations over time instead of overfitting to $D'$.%

\subsection{Non-i.i.d. multi-client scenarios}\label{sec:multi_client}
\begin{figure}[t]
    \scriptsize
    \centering
    \captionof{table}{Effects of Client Heterogeneity on SDAR Performed on Tiny ImageNet with ResNet-20 at Split Level 7 in Vanilla SL}
    \label{tab:hetero}
    \begin{tabular}{ccccc}
    \toprule
            No. of hetero. clients & 1 & 2 & 5 & 10 \\\midrule
            \multirow{2}{*}{Attack MSE} & 0.0300 & 0.0383 & 0.0395 & 0.0419\\
            & (0.0016) & (0.0012) & (0.0023) & (0.0025)\\\bottomrule
    \end{tabular}
    \normalsize

    \vskip 1.5\baselineskip

    \small
    \renewcommand{\arraystretch}{0}
    \begin{tabular}{@{}m{2.95cm}@{}m{5.95cm}@{}}
        Original & \includegraphics[width=0.99\linewidth]{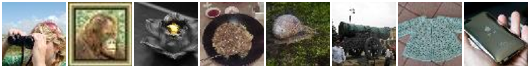} \\\midrule
        I.i.d. data & \includegraphics[width=0.99\linewidth]{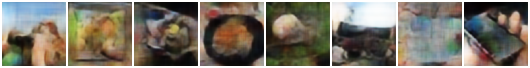}\\
        5 non-i.i.d. clients & \includegraphics[width=0.99\linewidth]{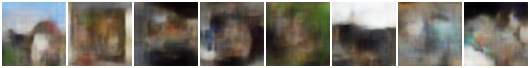}\\
        10 non-i.i.d. clients & \includegraphics[width=0.99\linewidth]{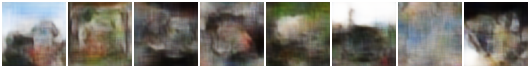}\\
    \end{tabular}

    \captionof{figure}{Examples of feature inference results of SDAR performed in single-client (equivalently i.i.d. multi-client) vs heterogeneous multi-client settings on Tiny ImageNet with ResNet-20 at split level 7 in the vanilla SL setting.}
    \label{fig:multi}
\end{figure}

So far, our experiments focus on the setting where all batches of private training data are independently and identically distributed. This is naturally true for a single-client scenario and remains so for multiple homogeneous clients.
In real-world practice, the server often serves multiple clients, who possess heterogeneous data distributions.
To investigate the effects of data heterogeneity on SDAR, we conduct additional evaluations under a heterogeneous $k$-client setting, where $k\in\{2,5,10\}$ clients possess different data distributions. In this setting, each client holds $200/k$ classes of Tiny ImageNet, and they take turns to interact with the server.
We report SDAR's feature inference performance in Table~\ref{tab:hetero} and provide examples of reconstruction results in Fig.~\ref{fig:multi}. We observe that the attack performance slightly degrades as the number of heterogeneous clients increases, but the visual quality of the reconstructed features remains acceptable. This suggests that SDAR can still be effective in the multi-client setting, even when clients possess heterogeneous data distributions.
The performance degradation could be attributed to the fact that in the non-i.i.d. setting, the updates from different clients are not homogeneous, which may result in issues with the convergence of partial models, thus a worse simulator on the server side. 
Nevertheless, although the incoming batches of representations are not homogeneous, the server knows which client the batch is coming from and can curate a specific auxiliary dataset for each client to ensure $\tilde f$ to faithfully simulate the client's model. Therefore, we observe limited performance drop in the heterogeneous multi-client setting.

\section{Potential Countermeasures}\label{sec:defense}

\begin{figure}[t]
\centering
\includegraphics{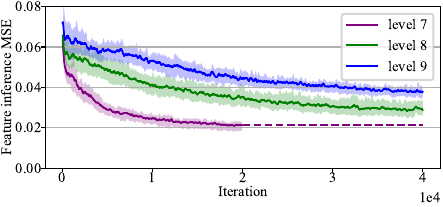}
\caption{Attack MSE of SDAR on CIFAR-10 with ResNet-20 at deeper split levels of 7--9 in the vanilla SL configuration.}
\label{fig:deeper_cifar10}

\vskip 1.5\baselineskip

\scriptsize
\captionof{table}{SDAR Attack MSE and SL Training Accuracy on CIFAR-100 with ResNet-20 at Split Level 7 in Vanilla SL at Various Dropout Rates}
\label{tab:dropout_cifar100}
\centering
\begin{tabular}{cccccc}
\toprule
Dropout rate & 0.0 & 0.1 & 0.2 & 0.4 & 0.8\\\midrule
\multirow{2}{*}{Attack MSE} & 0.0220 & 0.0212 & 0.0226 & 0.0237 & 0.0250 \\
& (0.0013) & (0.0001) & (0.0001) & (0.0019) & (0.0026) \\\midrule
\multirow{2}{*}{Train ACC (\%)} & 96.32 & 93.03 & 91.43 & 86.07 & 40.19 \\
& (0.49) & (0.40) & (0.20) & (0.34) & (0.61) \\\bottomrule
\end{tabular}
\normalsize

\vskip 1.5\baselineskip

\scriptsize
\captionof{table}{SDAR Attack MSE on CIFAR-10 with ResNet-20 at Split Level 7 in Vanilla SL with $\ell_1/\ell_2$ Regularization}
\label{tab:l1l2}
\centering
\begin{tabular}{ccccc}
\toprule
Regularization factor  & 0.0 & 0.001 & 0.01 & 0.1\\\midrule
\multirow{2}{*}{\shortstack{Attack MSE on SL\\with $\ell_1$ regularization}} & 0.0212  & 0.0.0208  & 0.0228  & 0.0221 \\
& (0.0019) & (0.0012) & (0.0017) & (0.0017)\\\midrule
\multirow{2}{*}{\shortstack{Attack MSE on SL\\with $\ell_2$ regularization}} & 0.0212  & 0.0231  & 0.0209  & 0.0220 \\
& (0.0019) & (0.0020) & (0.0008) & (0.0019)\\\bottomrule
\end{tabular}
\normalsize
\end{figure}

\paragraph{Change of model architectures} We have demonstrated the effectiveness of SDAR on ResNet-20 and PlainNet-20 (in Fig.~\ref{fig:vsl_bars_cifar} and Fig.~\ref{fig:u_cifar10}) and their narrower variants (in Fig.~\ref{fig:width_cifar}) at the deep split level of seven. Since narrower models at deeper split levels are relatively more robust to inference attacks, a natural defense against SDAR is to assign more layers to the client's model.
We evaluate SDAR at split levels up to nine and present its attack performance in Fig.~\ref{fig:deeper_cifar10}.
Note that at split level 9, the client hosts all convolutional layers, and the server has merely 650 parameters. This is an extreme case that is not practical in real-world deployments, as this invalidates SL's purpose of relieving the client's computational burden with a capable server. 
From Fig.~\ref{fig:deeper_cifar10}, we see that even at level 9, SDAR's attack MSE continues to decrease over time without saturation. Therefore, we conclude that SDAR is robust against different split levels, and changing the model architecture is not a viable defense against our attacks.

\paragraph{Regularization} Overfitting is a key concern in inference attacks on ML models \cite{shokri2017membership,nasr2019comprehensive}, as ML models may overfit to and unintentionally memorize private training data. Dropout \cite{srivastava2014dropout} and $\ell_1/\ell_2$ regularization \cite{ng2004feature} are straightforward yet effective techniques to prevent overfitting, and recent studies \cite{shokri2017membership,luo2021feature,salem2019ml,jiang2024on} have shown that they can also mitigate inference attacks on ML models.
Therefore, we consider these two methods as a potential defense and evaluate SDAR against them.
As shown in Table~\ref{tab:dropout_cifar100} and Table~\ref{tab:l1l2}, the introduction of dropout or $\ell_1/\ell_2$ regularization with a higher rate or factor slightly degrades the attack performance of SDAR, but the attack MSE remains at a low level, indicating the attack effectiveness against dropout. 
Table~\ref{tab:dropout_cifar100} also shows that large dropout rates can lead to significant degradation in the training accuracy of SL. 
The reason that our attacks are robust against regularization defense is due to the fact that the same regularization is also applied in the training of the simulator, so the $\tilde f$ can simulate the regularized behavior of the client's $f$. Therefore, dropout and $\ell_1/\ell_2$ regularization are not viable defenses against SDAR.

\paragraph{Decorrelation} A recent method for enhancing the privacy of SL is to decorrelate the intermediate representations $Z_s$ with the input private features $X$ \cite{vepakomma2020nopeek,vepakomma2021nopeek}. This is done by replacing $g\circ f$'s loss function $\ell(g(f(X)),Y)$ with $$
    (1-\alpha)\ell(g(f(X)),Y) + \alpha\cdot\text{dCol}(f(X), X)$$ while training, 
    where $\alpha\in[0,1]$ is the decorrelation parameter and $\text{dCol}(\cdot,\cdot)$ is the distance correlation \cite{szekely2007measuring}. The decorrelation term encourages the intermediate representations $Z_s$ to be decorrelated with the input private features $X$, thus making it harder for the server to infer private training data from $Z_s$. Larger $\alpha$ values result in stronger decorrelation and hence better privacy preservation, but also greater loss in the model utility of the original task.

\begin{figure}[t]
    \scriptsize
    \centering
    \captionof{table}{SDAR Attack MSE on CIFAR-100 with ResNet-20 at Level 7 in Vanilla SL with the Decorrelation Defense}
    \label{tab:decorrelation}
    \begin{tabular}{cccccc}
        \toprule
        $\alpha$ & 0.0 & 0.1 & 0.2 & 0.4 & 0.8\\\midrule
        \multirow{2}{*}{Attack MSE} & 0.0220 & 0.0257 & 0.0314 & 0.0414 & 0.0433 \\
        & (0.0013) & (0.0012) &(0.0023) & (0.0017) & (0.0024)\\\midrule
        \multirow{2}{*}{Train ACC (\%)} & 96.32 & 96.31 & 94.83 & 94.31 & 74.57 \\
        & (0.49) & (0.30) & (0.21) & (0.39) & (2.98) \\\bottomrule
    \end{tabular}
    \normalsize

    \vskip 1.5\baselineskip

    \centering
    \small
    \renewcommand{\arraystretch}{0}
    \begin{tabular}{@{}m{1.3cm}@{}m{7.6cm}@{}}
        Original & \includegraphics[width=0.99\linewidth]{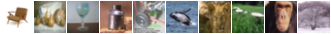} \\\midrule
        $\alpha=0.0$ & \includegraphics[width=0.99\linewidth]{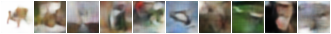}\\
        $\alpha=0.8$ & \includegraphics[width=0.99\linewidth]{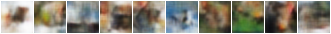}\\
    \end{tabular}
    \captionof{figure}{Examples of reconstruction results on CIFAR-100 with ResNet-20 at split level 7 in vanilla SL with decorrelation defense with $\alpha\in\{0,0.8\}$.}
    \label{fig:decorrelation_examples}
\end{figure}

To evaluate the effectiveness of SDAR against this decorrelation defense, we introduce a decorrelation term in Eq. \eqref{eq:e_loss} to decorrelate simulator's output $\tilde f(X')$ with auxiliary data $X'$ in SDAR, such that the simulator can learn the same decorrelation behaviors of client's $f$. 
We perform SDAR on SL with decorrelation with $\alpha\in\{0.1,0.2,0.4,0.8\}$. As shown in Table~\ref{tab:decorrelation} and Fig.~\ref{fig:decorrelation_examples}, although decorrelation manages to degrade the attack quality when $\alpha$ increases, SDAR still  produces visibly effective reconstructions even at $\alpha=0.8$. Also, Table~\ref{tab:decorrelation} shows that the decorrelation defense with larger $\alpha$ results in a notable loss in accuracy of the original task, rendering this defense impractical in real-world applications due to the trade-off between privacy and utility. Therefore, decorrelation is not a viable defense against SDAR.

\paragraph{Homomorphic encryption}
The most straightforward idea to mitigate privacy issues in SL is to encrypt the intermediate representations $Z_s$ and send only the ciphertext to the server s.t. the server cannot use it to infer private features. This usually requires homomorphic encryption (HE) \cite{takabi2016privacy,wood2020homomorphic} where the server can only perform computation on the ciphertext without decrypting it and return the ciphertext of the result to the client. In the context of SL, this means that the server must execute forward pass, loss evaluation, and backward propagation all in ciphertext, and at last sends the gradients $\nabla\theta_{s+1}$ in ciphertext back to the client. However, basic computations in HE schemes are limited to additions and multiplications~\cite{keller2020spdz}, and it is computationally prohibitive to execute the forward pass and backward propagation in ciphertext, which typically involves non-linear operations and necessitates the use of approximation techniques~\cite{takabi2016privacy}. 
Therefore, HE can hardly be applicable to SL, especially for deep NN models.

\paragraph{Secure multi-party computation}
Another widely-used cryptographic technique to protect privacy is secure multi-party computation (MPC) \cite{wu2020privacy,yao1982mpc,bonawitz2017practical,keller2020spdz}, where multiple parties jointly compute a function on their private inputs without revealing their inputs to each other. To the best of our knowledge, there has been no prior work improving the privacy of SL using MPC. Even if it is possible to apply MPC such that the client and the server can train their models without revealing private data of the client or model parameters of both parties, such protocol will introduce significant computational overhead and communication cost, completely defeating the purpose and promises of split learning of communication and computation efficiency.

\paragraph{Differential privacy} Currently, the state-of-the-art technique to quantify and reduce information disclosure about individuals is differential privacy (DP) \cite{xiao2010differential,abadi2016deep,dwork2014algorithmic}, which mathematically bounds the influence of private data on the released information. In the context of SL, this would mean that the intermediate representations $Z_s$ are differentially private with regard to the client's private features $X$. If we are to achieve DP in split learning, random noise will be injected by the client, such that even if the client arbitrarily modifies the input data $X$, the intermediate representations $Z_s$ will remain almost unchanged. That is, client's model $f$ will output similar representations for any input data, which practically renders the model trivial. Therefore, DP is not applicable to our problem.

\begin{table*}[t]
    \caption{Properties and Threat Models of Server-Side Training-Time Inference Attacks on SL$^*$}
    \label{tab:threat}
    \centering
    \begin{tabular}{lccccccc}
        \toprule
        \multirow{2}{*}{Attack} & \multirow{2}{*}{Infer feature} & \multirow{2}{*}{Infer label} & \multirow{2}{*}{Passive} & \multicolumn{2}{c}{Data assumption} & \multicolumn{2}{c}{Model assumption}\\
        \cmidrule(lr){5-6} \cmidrule(lr){7-8}
        & & & & Auxiliary feature & Auxiliary label & Model architecture & Model weights \\ \midrule
        FSHA (Pasquini et al.~\cite{pasquini2021unleashing}) &  \ding{51} & \ding{55} & \ding{55} & \CIRCLE & \Circle & \LEFTcircle & \Circle \\
        UnSplit (Erdo{\u{g}}an et al.~\cite{erdougan2022unsplit}) & \ding{51} & \ding{51} & \ding{51} & \Circle & \Circle  & \CIRCLE & \Circle \\
        EXACT (Qiu et al.~\cite{qiu2023exact}) & \ding{51} & \ding{51} & \ding{51} & \Circle & \Circle & \CIRCLE & \CIRCLE\\
        PCAT (Gao and Zhang~\cite{gao2023pcat}) & \ding{51} & \ding{51} & \ding{51} & \CIRCLE & \CIRCLE & \LEFTcircle & \Circle \\
        \midrule
        \textbf{SDAR (Ours)} & \ding{51} & \ding{51} & \ding{51} & \CIRCLE & \CIRCLE & \LEFTcircle & \Circle \\
        \bottomrule\\
    \end{tabular}

    \par\footnotesize $^*$Symbol \protect{\CIRCLE} indicates that the attack requires the corresponding assumption to be satisfied, while \protect{\Circle} indicates that the attack does not exploit the corresponding assumption. Symbol \protect{\LEFTcircle} indicates that the attack does not require the corresponding assumption but can benefit from it.
\end{table*}

\section{Related work}\label{sec:related}
 
\paragraph{Split learning} Split learning (SL) \cite{poirot2019split,gupta2018distributed,vepakomma2018split,vepakomma2022split} is a privacy preserving protocol for distributed training and inference of deep neural networks, and has gain increasing attention due to its simplicity, efficiency and scalability \cite{kairouz2021advances,thapa2021advancements,yin2021comprehensive,singh2019detailed}. Since its proposal, many works have extended SL to various configurations, including U-shaped SL that protects the client's labels \cite{gupta2018distributed}, SL with multiple clients with horizontal \cite{vepakomma2018split} or vertical \cite{ceballos2020split} partition. We refer the audience to Section~\ref{sec:pre} and available surveys \cite{kairouz2021advances,thapa2021advancements,yin2021comprehensive,singh2019detailed} for detailed review.

\paragraph{Inference attacks against split learning} The security of SL has been a focus of the research community since the proposal of SL. Various privacy attacks have been proposed to infer clients' labels or features under different threat models. We characterize these threat models via three dimensions: passiveness, the data assumption and the model assumption. \textit{Passiveness} refers to whether the adversary fully complies with the SL protocol without actively manipulating the training process (by hijacking gradients, for example).
A malicious server that actively manipulates the training process is considered an active or dishonest adversary, while a server that only passively observes the training process is considered a passive adversary, also known as a semi-honest or, interchangeably, an honest-but-curious adversary.
\textit{Data assumption} refers to whether the adversary has access to an auxiliary dataset of a similar distribution as the target private training data. \textit{Model assumption} refers to whether the adversary has access to the architecture or weights of the victim client's model. Existing attacks all adopt at least one of the two assumptions. We summarize them in Table~\ref{tab:threat} and discuss the feasibility of inference attacks with neither assumptions in Section~\ref{sec:conclusion}.

Pasquini et al.~\cite{pasquini2021unleashing} propose FSHA, an active feature inference attack where the server hijacks the gradients sent back to the client to control its updates such that the client's model behaves similarly to the decodable encoder trained by the server as part of an autoencoder. FSHA requires the adversary to have access to an (unlabelled) auxiliary dataset similar to the target data to train the autoencoder. Although FSHA does not require the server to know the architecture of the client's model, it benefits from such knowledge when the encoder shares the same architecture as the client's model.
Erdo{\u{g}}an et al.~\cite{erdougan2022unsplit} design UnSplit, a passive feature and label inference attack via alternating optimization of a surrogate model and inferred private inputs. UnSplit achieves passiveness and requires no knowledge of the client's data, but requires the client's model architecture to be known to the server.
 Similarly, Qiu et al.~\cite{qiu2023exact} also consider passive feature and label inference attacks without the data assumption, but with the extremely strong model assumption that the server knows the model parameters of the victim clients, thus can utilize the excessive information on client's model parameters and gradients.
 Concurrently with our work, Gao and Zhang~\cite{gao2023pcat} propose a passive feature and label inference attack against SL, namely PCAT, which adopts a similar framework as our na\"ive SDA discussed in the beginning of Section~\ref{sec:vanilla}, with additional improvements such as label alignment and delayed training. SDAR and PCAT share the same threat model, where an honest-but-curious adversary has access to a labelled auxiliary dataset similar to the target data, but has no access to the weights of the client's model. Both attacks can function without the knowledge of the client's model architecture, but may benefit from such knowledge. We demonstrate with extensive experiments in Section~\ref{sec:exp} that SDAR consistently outperforms PCAT across various, and in particular, challenging settings.

On a separate line of research, several works \cite{li2022label,liu2022clustering} focus on label inference attacks. This is not our focus as we aim to develop effective attacks that can infer both the clients' labels and features. At last, several attacks \cite{he2019model,yin2023ginver} have been proposed against split inference, also known as collaborative inference. However, split inference is more vulnerable as the intermediate representations are products of a fixed client's model, while in training-time SL attacks, the server has to invert the output of a constantly updated model. Therefore, attacks against split inference are not our focus in this work.

\paragraph{Privacy-preserving improvements of SL} In response to the privacy vulnerabilities of split learning, various privacy-preserving improvements have been proposed. U-shaped SL \cite{gupta2018distributed} enhances privacy by preventing label exposure to the server. However, our work demonstrates that SDAR can still infer private labels with high accuracy. Targeting active hijacking attacks such as FSHA \cite{pasquini2021unleashing}, Erdo{\u{g}}an et al.~\cite{erdogan2022splitguard}, Fu et al.~\cite{fu2023focusing} detect and defend against the active manipulation of back-propagated gradients. Vepakomma et al.~\cite{vepakomma2020nopeek,vepakomma2021nopeek} propose to improve the privacy of SL by minimizing the correlation between the private inputs and the shared representations via injecting a correlation regularization term to the loss function of the original SL task, which we investigate in Section~\ref{sec:defense}.

\section{Conclusion}
\label{sec:conclusion}

In this paper, we investigate the privacy risks of split learning under the notion of an honest-but-curious server. We identify that existing server-side passive attacks often target impractical and vulnerable settings to gain extra advantage for the adversary. We present new passive attacks on split learning, namely, SDAR, by utilizing GAN-inspired adversarial regularization to learn a decodable simulator of client's private model that produces representations indistinguishable from the client's. Empirically, we show that SDAR is effective in inferring private features of the client in vanilla split learning, and both private features and labels in the U-shaped split learning, under challenging settings where existing passive attacks fail to produce non-trivial results. At last, we propose and evaluate potential defenses against our attacks and highlight the need for improving SL to further protect the client's private data.

Split learning is a promising protocol that enables distributed training and inference of neural networks on devices with limited computational resources. There are many open challenges and potential directions regarding its security.
One limitation of SDAR (and PCAT) is that they require the adversary to have access to a \textit{labelled} auxiliary dataset in the same domain as the target data. While state-of-the-art active attacks like FSHA do not require the labels of the auxiliary dataset, it is interesting to investigate whether passive attacks are feasible only with an unlabelled auxiliary dataset. Also, it is worth noting that all attacks discussed in Section~\ref{sec:related} and listed in Table~\ref{tab:threat} assume at least one of the model and the data assumptions, and it remains an open question whether it is possible to attack SL with neither assumptions. Intuitively, if the server has no knowledge on either the structure/weights of the client's model $f$ or the domain/distribution of its data $X$, it becomes very challenging, if not impossible, to recover $X$ merely by observing $Z=f(X)$. It's interesting to explore more capable attacks that invert an inaccessible and unknown function operating on an unknown domain.
One may also study the privacy guarantee of SL from a theoretical lens and rigorously measure the information leakage of the protocol. At last, it will be an interesting direction to design privacy-preserving mechanisms to mitigate our proposed attacks and improve the privacy of SL.

\section*{Acknowledgements}
This research is supported by the National Research Foundation, Singapore under its AI Singapore Programme (AISG Award No: AISG3-RP-2022-029). Any opinions, findings and conclusions or recommendations expressed in this material are those of the author(s) and do not reflect the views of National Research Foundation, Singapore. Yuncheng Wu's work is supported by the Fundamental Research Funds for the Central Universities, and the Research Funds of Renmin University of China (24XNKJ08).

\bibliographystyle{IEEEtranS}
\bibliography{ref}
\appendices
\section{Pseudocode}\label{app:pseudocode}

We provide the pseudocode of SDAR against vanilla SL in Fig.~\ref{alg:sdar}, and against U-shaped SL in Fig.~\ref{alg:sdar_u}.

\begin{figure}[t]
  \small
  \begin{algorithmic}[1]
    \Procedure{Client:}{}
    \State Initializes model $f$ with parameters $\theta_f$
    \For{$i\in\{1,2,\ldots\}$}
      \State Samples a batch of examples $(X,Y)\in D$
      \State $Z_s\gets f(X)$
      \State Sends $(Z_s,Y)$ to the server \Comment{Timestamp $t_{i1}$}
      \State Receives $\nabla\theta_{s+1}$ from the server \Comment{Timestamp $t_{i2}$}
      \State Calculates $\nabla\theta_f$ using $\nabla\theta_{s+1}$
      \State $\theta_f\gets \theta_f-\eta\nabla\theta_f$\Comment{Original SL task}
    \EndFor
    \EndProcedure
  
    \State 
    \Procedure{Server:}{}
      \State Initializes model $g$ with parameters $\theta_g$
      \State Initializes models $\tilde f,\tilde f^{-1},d_1,d_2$ w/ $\theta_{\tilde f},\theta_{\tilde f^{-1}},\theta_{d_1},\theta_{d_2}$
      \For{$i\in\{1,2,\ldots\}$}
        \State Receives $(Z_s,Y)$ from the client \Comment{Timestamp $t_{i1}$}
        \State $\hat Y\gets g(Z_s)$
        \State $\mathcal L\gets\ell(\hat Y,Y)$
        \State $\nabla\theta_g\gets\partial \mathcal L/\partial\theta_g$
        \State Sends $\nabla\theta_{s+1}$ to the client \Comment{Timestamp $t_{i2}$}
        \State $\theta_g\gets\theta_g-\eta\nabla\theta_g$\Comment{Original SL task}
        \State
        \State Samples a batch of examples $(X',Y')\in D'$
        \State $Z_s'\gets \tilde f(X')$; $\hat Y'\gets g(Z_s')$
        \State $\hat X'\gets\tilde f^{-1}(Z_s',Y')$; $\hat X\gets\tilde f^{-1}(Z_s,Y)$
        \State $\mathcal L_{\tilde f}\gets\ell(\hat Y',Y') + \lambda_1\ell_{\text{BCE}}(d_1(Z_s',Y'),1)$
        \State $\mathcal L_{d_1}\gets\ell_{\text{BCE}}(d_1(Z_s',Y'),0)+\ell_{\text{BCE}}(d_1(Z_s,Y),1)$
        \State $\mathcal L_{\tilde f^{-1}}\gets\ell_{\text{MSE}}(X',\hat X')+\lambda_2\cdot\ell_{\text{BCE}}(d_2(\hat X,Y),1)$
        \State $\mathcal L_{d_2}\gets\ell_{\text{BCE}}(d_2(\hat X,0)+\ell_{\text{BCE}}(d_2(X',Y'),1)$
        \State 
        \State $\nabla\theta_{\tilde f}\gets\partial\mathcal L_{\tilde f}/\partial\theta_{\tilde f}$; $\theta_{\tilde f}\gets \theta_{\tilde f}-\eta_{\tilde f}\nabla\theta_{\tilde f}$
        \State $\nabla\theta_{d_1}\gets\partial\mathcal L_{d_1}/\partial\theta_{d_1}$; $\theta_{d_1}\gets \theta_{d_1}-\eta_{d_1}\nabla\theta_{d_1}$
        \State $\nabla\theta_{\tilde f^{-1}}\gets\partial\mathcal L_{\tilde f^{-1}}/\partial\theta_{\tilde f^{-1}}$; $\theta_{\tilde f^{-1}}\gets \theta_{\tilde f^{-1}}-\eta_{\tilde f^{-1}}\nabla\theta_{\tilde f^{-1}}$
        \State $\nabla\theta_{d_2}\gets\partial\mathcal L_{d_2}/\partial\theta_{d_2}$; $\theta_{d_2}\gets \theta_{d_2}-\eta_{d_2}\nabla\theta_{d_2}$
        \State
        \State $\hat X\gets\tilde f^{-1}(Z_s,Y)$\Comment{Reconstructed private features}
      \EndFor
    \EndProcedure
    \end{algorithmic}
    \caption{SDAR against vanilla split learning. $\ell(\cdot,\cdot)$ is the loss function of the original SL task; $s$ is the split level; $\eta$ is the learning rate of the original SL task; $\eta_{\tilde f},\eta_{\tilde f^{-1}},\eta_{d_1},\eta_{d_2}$ are learning rates for the server's models $\tilde f,\tilde f^{-1},d_1,d_2$.}\label{alg:sdar}
\end{figure}

\begin{figure}[t]
  \small
  \begin{algorithmic}[1]
  \Procedure{Client:}{}
  \State Initializes models $f,h$ with parameters $\theta_f,\theta_h$
  \For{$i\in\{1,2,\ldots\}$}
    \State Samples a batch of examples $(X,Y)\in D$
    \State $Z_s\gets f(X)$
    \State Sends $Z_s$ to the server \Comment{Timestamp $t_{i1}$}
    \State Receives $Z_t$ from the server \Comment{Timestamp $t_{i2}$}
    \State $\hat Y\gets h(Z_t)$; $\mathcal L\gets\ell(\hat Y, Y)$
    \State $\nabla\theta_h\gets\partial\mathcal L/\partial\theta_h$; $\theta_h\gets\theta_h-\eta\nabla\theta_h$ \Comment{Original SL task}
    \State Sends $\nabla\theta_{t+1}$ to the server \Comment{Timestamp $t_{i3}$}
    \State Receives $\nabla\theta_{s+1}$ from the server \Comment{Timestamp $t_{i4}$}
    \State Calculates $\nabla\theta_f$ using $\nabla\theta_{s+1}$
    \State $\theta_f\gets \theta_f-\eta\nabla\theta_f$\Comment{Original SL task}
  \EndFor
  \EndProcedure

  \State 
  \Procedure{Server:}{}
    \State Initializes $g$ with parameters $\theta_g$
    \State Initializes $\tilde h,\tilde f,\tilde f^{-1},d_1,d_2$ with param. $\theta_{\tilde h},\theta_{\tilde f},\theta_{\tilde f^{-1}},\theta_{d_1},\theta_{d_2}$
    \For{$i\in\{1,2,\ldots\}$}
      \State Receives $(Z_s,Y)$ from the client \Comment{Timestamp $t_{i1}$}
      \State $Z_t=g(Z_s)$
      \State Sends $Z_t$ to the client \Comment{Timestamp $t_{i2}$}
      \State Receives $\nabla\theta_{t+1}$ from the client \Comment{Timestamp $t_{i3}$}
      \State Calculates $\nabla\theta_g$ using $\nabla\theta_{t+1}$
      \State $\theta_g\gets\theta_g-\eta\nabla\theta_g$ \Comment{Original SL task}
      \State Sends $\nabla\theta_{s+1}$ to the client \Comment{Timestamp $t_{i4}$}
      \State
      \State Samples a batch of examples $(X',Y')\in D'$
      \State $Z_s'\gets\tilde{f}(X')$; $Z_t'\gets g(Z_s')$; $\hat Y'\gets\tilde h(Z_t')$
      \State $\hat X'\gets\tilde f^{-1}(Z_s')$; $\hat X\gets\tilde f^{-1}(Z_s)$
      \State $\tilde Y_i'\gets \begin{cases}
        Y_i' & \text{w.p. } 1-p\\
        \text{Uniform}(\mathcal Y) & \text{w.p. } p
      \end{cases}$\Comment{Random flipping}
      \State $\mathcal L_{\tilde h}\gets \ell(\hat Y',\tilde Y')$
      \State $\mathcal L_{\tilde f}\gets\ell(\hat Y',\tilde Y') + \lambda_1 \ell_{\text{BCE}}(d_1(Z_s'),1)$
      \State $\mathcal L_{d_1}\gets\ell_{\text{BCE}}(d_1(Z_s'),0) + \ell_{\text{BCE}}(d_1(Z_s),1)$
      \State $\mathcal L_{\tilde f^{-1}}\gets\ell_{\text{MSE}}(X',\hat X')+\lambda_2\ell_{\text{BCE}}(d_2(\hat X),1)$
      \State $\mathcal L_{d_2}\gets\ell_{\text{BCE}}(d_2(\hat X),0)+\ell_{\text{BCE}}(d_2(X'),1)$
      \State 
      \State $\nabla\theta_{\tilde h}\gets\partial\mathcal L_{\tilde h}/\partial\theta_{\tilde h}$; $\theta_{\tilde h}\gets \theta_{\tilde h}-\eta_{\tilde h}\nabla\theta_{\tilde h}$
      \State $\nabla\theta_{\tilde f}\gets\partial\mathcal L_{\tilde f}/\partial\theta_{\tilde f}$; $\theta_{\tilde f}\gets \theta_{\tilde f}-\eta_{\tilde f}\nabla\theta_{\tilde f}$
      \State $\nabla\theta_{d_1}\gets\partial\mathcal L_{d_1}/\partial\theta_{d_1}$; $\theta_{d_1}\gets \theta_{d_1}-\eta_{d_1}\nabla\theta_{d_1}$
      \State $\nabla\theta_{\tilde f^{-1}}\gets\partial\mathcal L_{\tilde f^{-1}}/\partial\theta_{\tilde f^{-1}}$; $\theta_{\tilde f^{-1}}\gets \theta_{\tilde f^{-1}}-\eta_{\tilde f^{-1}}\nabla\theta_{\tilde f^{-1}}$
      \State $\nabla\theta_{d_2}\gets\partial\mathcal L_{d_2}/\partial\theta_{d_2}$; $\theta_{d_2}\gets \theta_{d_2}-\eta_{d_2}\nabla\theta_{d_2}$
      \State
      \State $\hat X\gets\tilde f^{-1}(Z_s)$\Comment{Reconstructed private features}
      \State $\hat Y\gets\tilde h(Z_t)$\Comment{Inferred private labels}
    \EndFor
  \EndProcedure

  \end{algorithmic}
  \caption{SDAR against U-shaped split learning. $\ell(\cdot,\cdot)$ is the loss function of the original SL task; $s,t$ are the split levels; $\eta$ is the learning rate of the original SL task; $\eta_{\tilde h},\eta_{\tilde f},\eta_{\tilde f^{-1}},\eta_{d_1},\eta_{d_2}$ are learning rates for the server's models $\tilde h,\tilde f,\tilde f^{-1},d_1,d_2$ respectively. $p$ is the label flipping probability. $\mathcal Y$ is the set of all possible labels.}\label{alg:sdar_u}
\end{figure}

\section{Implementation details}\label{app:impl_detail}

\subsection{Experiment setup}\label{app:exp_setup}
The experiments are conducted on machines running Ubuntu 20.04 LTS, equipped with two Intel\textsuperscript{\tiny\textregistered} Xeon\textsuperscript{\tiny\textregistered} Gold 6326 CPUs, 256GB of RAM and an NVIDIA\textsuperscript{\tiny\textregistered} A100 80GB GPU. We implement our attack and other baselines in Python and TensorFlow. Our code base is available at \url{https://github.com/zhxchd/SDAR_SplitNN/} for reproducibility.

\subsection{Split learning setup}\label{app:exp_sl}
\begin{figure}[t]%
  \subfloat[ResNet-20]{%
  \includegraphics[height=66mm]{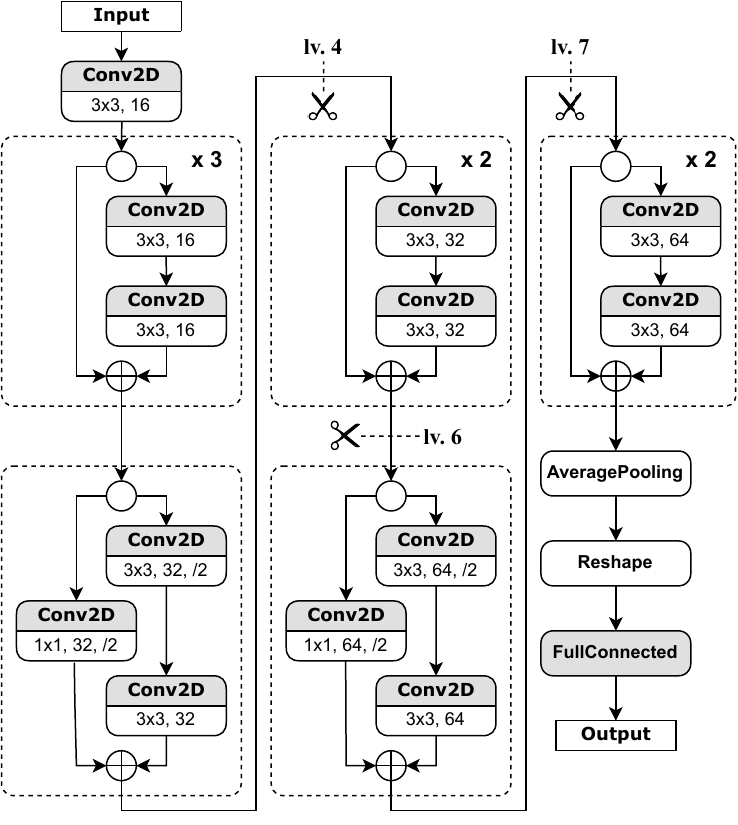}
  }
  \hfil
  \subfloat[PlainNet-20]{%
  \includegraphics[height=66mm]{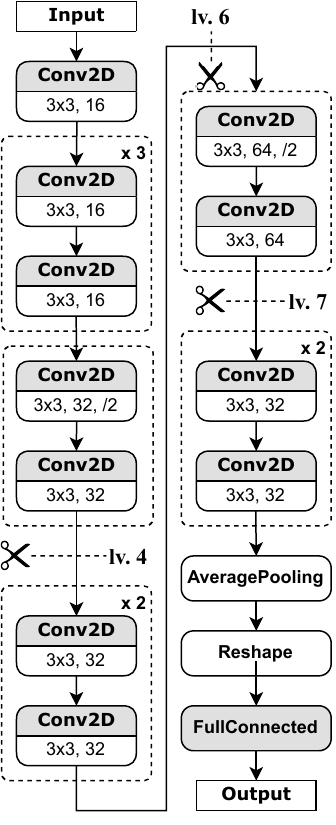}
  }
  \caption{Model architectures and split configurations used in the experiments. Each box marked as "Conv2D" represents a convolutional layer with the specified kernel size, number of filters, and stride (1 if absent), followed by batch normalization and ReLU.}
  \label{fig:model}
\end{figure}

We target the same SL setup across all experiments, where the complete model $H$ is either ResNet-20 or PlainNet-20, and we experiment with various split configurations of each model. We show the model architectures of the target models and their split configurations in Fig.~\ref{fig:model}. For all experiments, we use Adam optimizer to optimize $H$ (that is, $g\circ f$ in vanilla SL and $h\circ g\circ f$ in U-shaped SL) with default initial learning rate $\eta=0.001$. We use 128 examples per batch as per the original ResNet paper suggests \cite{he2016deep} except STL-10, where we use a batch size of 32 as STL-10 only contains 13,000 images in total. We train the model $H$ with split learning protocol for 20000 iterations (i.e., batches). For all attacks, on all datasets, with both models, at all considered split levels, and under both vanilla and U-shaped settings, we run five trials for statistical significance.

\subsection{Implementation details of SDAR}\label{app:sdar_impl}

\paragraph{Models} Under our firstly considered threat model where the server knows the model architecture of $H=g\circ f$, the server uses the same model architecture for its simulator $\tilde f$ as the client's model $f$, but with random initialization. When experimenting with the relaxed assumption that the server does not know the architecture of the client model (e.g. in Table~\ref{tab:diff_sim}), we use ResNet-20 for $H$ while making the server to use a plain convolutional network as a simulator. This model structure is obtained by removing the skip connections in the client's residual network. For all experiments, the server's decoder $\tilde f^{-1}$ is a deep convolutional network, where each building block of $\tilde f$ corresponds to a deconvolution block (deconvolutional layer, followed by batch normalization and ReLU) in reverse order. Note that we replace deconvolutional layers with strides 2 with a combination of an upsampling layer and a convolutional layer, to reduce the checkerboard effects \cite{odena2016deconvolution}. For the discriminator $d_1$ which distinguishes intermediate representations, we use a deep convolutional network with up to 256 filters, where each convolutional layer is followed by LeakyReLU and batch normalization except the last layer. The discriminator $d_2$ that distinguishes real and fake images follows a rather standard architecture, consists of convolutional layers, batch normalization and LeakyReLU. In the vanilla setting where labels are used as part of the input to the decoder and discriminators, we first transform the labels into embeddings of 50 units, and then further transform the embeddings by a learnable fully connected layer, before concatenating with the input to the decoder and discriminators. Note that we use the same model structures for the decoder and discriminators for both cases of the target model (i.e. ResNet-20 and PlainNet-20). We list the detailed model structures for the attacker's models $\tilde f^{-1},d_1,d_2$ under vanilla SL in Table~\ref{tab:model_structure}.

\begin{table}[t]
  \centering
  \caption{Hyperparameters used in SDAR}
  \label{tab:hyper}
  \begin{tabular}{cccccc}
  \toprule
  Setting & Dataset & Model & $\lambda_1$ & $\lambda_2$ & $p$\\\midrule
  Vanilla & All datasets & Both models & 0.02 & 0.00001 & NA\\
  U Shaped & CIFAR-10 & ResNet-20 & 0.02 & 0.00001 & 0.2\\
  U Shaped & CIFAR-100 & ResNet-20 & 0.04 & 0.00001 & 0.2\\
  U Shaped & Tiny ImageNet & ResNet-20 & 0.04 & 0.00001 & 0.2\\
  U Shaped & STL-10 & ResNet-20 & 0.04 & 0.00001 & 0.2\\
  U Shaped & CIFAR-10 & PlainNet-20 & 0.04 & 0.00001 & 0.1\\
  U Shaped & CIFAR-100 & PlainNet-20 & 0.04 & 0.00001 & 0.1\\
  U Shaped & Tiny ImageNet & PlainNet-20 & 0.04 & 0.00001 & 0.4\\
  U Shaped & STL-10 & PlainNet-20 & 0.04 & 0.00001 & 0.4\\\bottomrule
  \end{tabular}
\end{table}

\paragraph{Hyperparameters} As simulators to $f$ and $h$ respectively, we choose the learning rate of $\tilde f$ and $\tilde h$ the same as $\eta$, and we always keep the learning rate of the decoder half of that of the simulator, i.e., $2\eta_{\tilde f^{-1}}=\eta_{\tilde f}=\eta_{\tilde h}=\eta$. We choose the regularization factors $\lambda_1,\lambda_2$ in a way that the regularization term will not dominate the total loss, but still non-negligible.
For the adversarial regularization of $\tilde f$, this means that $\lambda_1$ should be set s.t. the regularization term $\lambda_1\ell_{\text{BCE}}(d_1(Z_s'),1)$ should not surpass the classification loss $\ell(\hat Y',\tilde Y')$.
We expect that, when the training stabilizes, the discriminator loss $\ell_{\text{BCE}}(d_1(Z_s'),1)$ should be around the binary cross entropy loss of random guess, i.e., $\log(2)\approx0.6931$, to ensure that the simulator and the discriminator can be continually improved via adversarial training. Therefore, to ensure that the regularization term $\lambda_1\ell_{\text{BCE}}(d_1(Z_s'),1)$ does not dominate the total loss, $\lambda_1$ is chosen to at the order of magnitude of $10^{-2}$, such that the regularization term will have the order of magnitude of $10^{-3}$, which does not dominate the training classification loss while being non-negligible.
For the adversarial regularization of $\tilde f^{-1}$, we choose $\lambda_2$ to be much smaller, at $10^{-5}$, as the reconstruction loss $\ell_{\text{MSE}}(X',\hat X')$ is expected to be much smaller than the classification loss, and the discriminator loss $\ell_{\text{BCE}}(d_2(\hat X),1)$ should be around $\log(2)$ as well.
After determining the regularization factors $\lambda_1,\lambda_2$, the learning rates for the discriminators are chosen as $\eta_{d_1}=\lambda_1\eta_{\tilde f}$ and $\eta_{d_2}=\lambda_2\eta_{\tilde f}$, to ensure that the step size to minimize generator loss (as a regularization term) is on par with that of the discriminator loss.

Guided by the above considerations, we set $\lambda_1=0.02$ and $\lambda_2=10^{-5}$ for all experiments in vanilla SL. For U-shaped SL, the attacks are more sensitive to the choices of $\lambda_1$ and the flip probability $p$, and we test with different configurations for different datasets and models and choose the best-performing combination. In particular, we test $\lambda_1\in\{0.2,0.4\}$ and $p\in\{0.1,0.2,0.4\}$. Note that such hyperparameter search is feasible in practice, as the server can always run the SDAR attack multiple times on the observed representations with different configurations and collect all reconstructed images. The hyperparameters used in the experiments are shown in Table~\ref{tab:hyper}.

\paragraph{Training instability} Due to the inherent instability of GANs, we observe, though very rarely, cases where SDAR fails to converge, resulting in a high reconstruction loss. In such cases, we reinitialize the attacker's models and restart the training process. We only observe such cases in the U-shaped setting, and the reinitialization process is only needed in very few (fewer than 5) trials among our extensive experiments. Note that such process is feasible in practice, as the server can always run the SDAR attack multiple times on the observed representations and collect all reconstructed images.
\subsection{Implementation details of baseline methods}\label{app:baseline_impl}

\paragraph{PCAT} We implement PCAT \cite{gao2023pcat} with the exact same model architectures for simulator and decoder as SDAR for a fair comparison. We also give the server in PCAT the same auxiliary dataset as SDAR. Compared to na\"ive SDA, PCAT uses two unique techniques to improve its performance. First, in vanilla SL where the server knows the labels of the private examples, PCAT aligns these labels by sampling auxiliary examples with the same labels as the received private examples. Second, in both vanilla and U-shaped SL, PCAT introduces a delay in attacking, where the server does not train its simulator and decoder until a certain number of iterations have been executed, to avoid the noisy early-stage training process disturbing the attacks. We implement both techniques, and choose a delay period of 100 iterations, as used in the original paper \cite{gao2023pcat}. Note that by the time our research was conducted, the code of PCAT was not publicly available, so we implement PCAT from scratch based on the description in the original paper \cite{gao2023pcat}.

\paragraph{UnSplit} UnSplit \cite{erdougan2022unsplit} infers the private features $X$ by minimizing $\ell_\text{MSE}(\tilde f(\hat X),Z_s)$ via alternating optimization of $\tilde f$ and $\hat X$, where $\tilde f$ is a surrogate model of the same architecture as $f$ but randomly initialized, and $\hat X$ is initialized as tensors filled with 0.5. We follow the implementation of the original paper \cite{erdougan2022unsplit} and use the same configuration for the alternating optimization algorithm. We use Adam optimizer with learning rate of 0.001 for both $\tilde f$ and $\hat X$. The alternating optimization process consists of 1000 rounds, each having 100 $\tilde f$ optimization steps and 100 $\hat X$ optimization steps. Following \cite{erdougan2022unsplit}, total variation is used to regularize the optimization of $\hat X$. In each training iteration of SL, UnSplit infers the private training examples $X$ by running the iterative alternating optimization procedure. However, due to the extremely high computational cost of the iterative optimization process, we are unable to run the optimization algorithm for every training iteration, but only attack the last batch of training examples.

\paragraph{FSHA} We implement FSHA \cite{pasquini2021unleashing} with the same model architectures for the encoder, decoder and simulator discriminator as SDAR for a fair comparison. As FSHA utilizes Wassertein GAN loss with gradiant penalty (WGAN-GP) \cite{gulrajani2017improved} for training, which is not compatible with batch normalization layers in the discriminator, we remove such layers in the discriminator model. In FSHA, the client's model $f$ is hijacked by the server to be updated to simulate the behaviors of the server's own encoder $\tilde f$. Following the original paper, we use learning rate 0.00001 for the training of $f$ as a generator and use learning rate 0.0001 for the discriminators with gradient penalty coefficient of 500. The encoder and decoder are trained in an autoencoder fashion \cite{hinton2006reducing} with learning rate 0.00001.

\begin{table}[t]
    \caption{Extended Tables and Figures and Their Corresponding Versions in the Main Text}
    \label{tab:ext}
    \centering
    \begin{tabular}{cc}
      \toprule
      Figure/table in main text & Extended figure/table \\\midrule
      Fig.~\ref{fig:vsl_bars_cifar} & Table~\ref{tab:vanilla_full}\\
      Fig.~\ref{fig:vanilla_examples} & Fig.~\ref{fig:v_examples_full}\\
      Fig.~\ref{fig:sdar_cifar10_curve} & Fig.~\ref{fig:v_sdar}\\
      Fig.~\ref{fig:fsha_cifar10} & Fig.~\ref{fig:fsha_full}\\
      Fig.~\ref{fig:width_cifar} & Fig.~\ref{fig:width_full}\\
      Table~\ref{tab:aux_data_size} & Table~\ref{tab:aux_data_size_full}\\
      Table~\ref{tab:diff_sim} & Table~\ref{tab:diff_sim_full}\\
      Fig.~\ref{fig:u_cifar10} & Table~\ref{tab:u_mse_full} and \ref{tab:u_acc_full}\\
      Fig.~\ref{fig:u_examples} & Fig.~\ref{fig:u_examples_full}\\\bottomrule
    \end{tabular}
\end{table}

\begin{table*}[t]
    \centering
    \caption{Attacker's Model Structures for $\tilde f^{-1}, d_1, d_2$ in Vanilla SL$^*$}
    \label{tab:model_structure}
    \begin{tabular}{@{}lllll@{}}
    \toprule
     & Split Level 4 & Split Level 5 & Split Level 6 & Split Level 7 \\ \midrule
    $\tilde f^{-1}$ & \begin{tabular}[t]{@{}l@{}}y = Embedding(num\_classes, 50)(y)\\ y = Dense(x.shape{[}0{]} * x.shape{[}1{]})(y)\\ y = Reshape((x.shape{[}0{]}, x.shape{[}1{]})(y)\\ x = Concatenate()({[}x,y{]})\\ x = UpSampling2D((2,2))(x)\\ x = Conv(32, 3, (1,1))(x)\\ x = ReLU()(BatchNorm()(x))\\ x = ConvTranspose(16, 3, (1,1))(x)\\ x = ReLU()(BatchNorm()(x))\\ x = ConvTranspose(16, 3, (1,1))(x)\\ x = ReLU()(BatchNorm()(x))\\ x = ConvTranspose(16, 3, (1,1))(x)\\ x = ReLU()(BatchNorm()(x))\\ x = Conv(3, 3, (1, 1))(x)\\ x = Sigmoid()(x)\end{tabular} & \begin{tabular}[t]{@{}l@{}}y = Embedding(num\_classes, 50)(y)\\ y = Dense(x.shape{[}0{]} * x.shape{[}1{]})(y)\\ y = Reshape((x.shape{[}0{]}, x.shape{[}1{]})(y)\\ x = Concatenate()({[}x,y{]})\\ x = ConvTranspose(32, 3, (1,1))(x)\\ x = ReLU()(BatchNorm()(x))\\ x = UpSampling2D((2,2))(x)\\ x = Conv(32, 3, (1,1))(x)\\ x = ReLU()(BatchNorm()(x))\\ x = ConvTranspose(16, 3, (1,1))(x)\\ x = ReLU()(BatchNorm()(x))\\ x = ConvTranspose(16, 3, (1,1))(x)\\ x = ReLU()(BatchNorm()(x))\\ x = ConvTranspose(16, 3, (1,1))(x)\\ x = ReLU()(BatchNorm()(x))\\ x = Conv(3, 3, (1, 1))(x)\\ x = Sigmoid()(x)\end{tabular} & \begin{tabular}[t]{@{}l@{}}y = Embedding(num\_classes, 50)(y)\\ y = Dense(x.shape{[}0{]} * x.shape{[}1{]})(y)\\ y = Reshape((x.shape{[}0{]}, x.shape{[}1{]})(y)\\ x = Concatenate()({[}x,y{]})\\ x = ConvTranspose(32, 3, (1,1))(x)\\ x = ReLU()(BatchNorm()(x))\\ x = ConvTranspose(32, 3, (1,1))(x)\\ x = ReLU()(BatchNorm()(x))\\ x = UpSampling2D((2,2))(x)\\ x = Conv(32, 3, (1,1))(x)\\ x = ReLU()(BatchNorm()(x))\\ x = ConvTranspose(16, 3, (1,1))(x)\\ x = ReLU()(BatchNorm()(x))\\ x = ConvTranspose(16, 3, (1,1))(x)\\ x = ReLU()(BatchNorm()(x))\\ x = ConvTranspose(16, 3, (1,1))(x)\\ x = ReLU()(BatchNorm()(x))\\ x = Conv(3, 3, (1, 1))(x)\\ x = Sigmoid()(x)\end{tabular} & \begin{tabular}[t]{@{}l@{}}y = Embedding(num\_classes, 50)(y)\\ y = Dense(x.shape{[}0{]} * x.shape{[}1{]})(y)\\ y = Reshape((x.shape{[}0{]}, x.shape{[}1{]})(y)\\ x = Concatenate()({[}x,y{]})\\ x = UpSampling2D((2,2))(x)\\ x = Conv(64, 3, (1,1))(x)\\ x = ReLU()(BatchNorm()(x))\\ x = ConvTranspose(32, 3, (1,1))(x)\\ x = ReLU()(BatchNorm()(x))\\ x = ConvTranspose(32, 3, (1,1))(x)\\ x = ReLU()(BatchNorm()(x))\\ x = UpSampling2D((2,2))(x)\\ x = Conv(32, 3, (1,1))(x)\\ x = ReLU()(BatchNorm()(x))\\ x = ConvTranspose(16, 3, (1,1))(x)\\ x = ReLU()(BatchNorm()(x))\\ x = ConvTranspose(16, 3, (1,1))(x)\\ x = ReLU()(BatchNorm()(x))\\ x = ConvTranspose(16, 3, (1,1))(x)\\ x = ReLU()(BatchNorm()(x))\\ x = Conv(3, 3, (1, 1))(x)\\ x = Sigmoid()(x)\end{tabular} \\ \midrule
    $d_1$ & \begin{tabular}[t]{@{}l@{}}y = Embedding(num\_classes, 50)(y)\\ y = Dense(x.shape{[}0{]} * x.shape{[}1{]})(y)\\ y = Reshape((x.shape{[}0{]}, x.shape{[}1{]})(y)\\ x = Concatenate()({[}x,y{]})\\ x = Conv(64, 3, (1,1))(x)\\ x = LeakyReLU()(x)\\ x = Conv(128, 3, (2,2))(x)\\ x = LeakyReLU()(BatchNorm()(x))\\ x = Conv(256, 3, (1,1))(x)\\ x = LeakyReLU()(BatchNorm()(x))\\ x = Conv(256, 3, (1,1))(x)\\ x = LeakyReLU()(BatchNorm()(x))\\ x = Conv(256, 3, (1,1))(x)\\ x = LeakyReLU()(BatchNorm()(x))\\ x = Conv(256, 3, (2,2))(x)\\ x = Flatten()(x)\\ x = Dropout(0.4)(x)\\ x = Dense(1)(x)\end{tabular} & \begin{tabular}[t]{@{}l@{}}y = Embedding(num\_classes, 50)(y)\\ y = Dense(x.shape{[}0{]} * x.shape{[}1{]})(y)\\ y = Reshape((x.shape{[}0{]}, x.shape{[}1{]})(y)\\ x = Concatenate()({[}x,y{]})\\ x = Conv(64, 3, (1,1))(x)\\ x = LeakyReLU()(x)\\ x = Conv(128, 3, (2,2))(x)\\ x = LeakyReLU()(BatchNorm()(x))\\ x = Conv(256, 3, (1,1))(x)\\ x = LeakyReLU()(BatchNorm()(x))\\ x = Conv(256, 3, (1,1))(x)\\ x = LeakyReLU()(BatchNorm()(x))\\ x = Conv(256, 3, (1,1))(x)\\ x = LeakyReLU()(BatchNorm()(x))\\ x = Conv(256, 3, (2,2))(x)\\ x = Flatten()(x)\\ x = Dropout(0.4)(x)\\ x = Dense(1)(x)\end{tabular} & \begin{tabular}[t]{@{}l@{}}y = Embedding(num\_classes, 50)(y)\\ y = Dense(x.shape{[}0{]} * x.shape{[}1{]})(y)\\ y = Reshape((x.shape{[}0{]}, x.shape{[}1{]})(y)\\ x = Concatenate()({[}x,y{]})\\ x = Conv(64, 3, (1,1))(x)\\ x = LeakyReLU()(x)\\ x = Conv(128, 3, (2,2))(x)\\ x = LeakyReLU()(BatchNorm()(x))\\ x = Conv(256, 3, (1,1))(x)\\ x = LeakyReLU()(BatchNorm()(x))\\ x = Conv(256, 3, (1,1))(x)\\ x = LeakyReLU()(BatchNorm()(x))\\ x = Conv(256, 3, (1,1))(x)\\ x = LeakyReLU()(BatchNorm()(x))\\ x = Conv(256, 3, (2,2))(x)\\ x = Flatten()(x)\\ x = Dropout(0.4)(x)\\ x = Dense(1)(x)\end{tabular} & \begin{tabular}[t]{@{}l@{}}y = Embedding(num\_classes, 50)(y)\\ y = Dense(x.shape{[}0{]} * x.shape{[}1{]})(y)\\ y = Reshape((x.shape{[}0{]}, x.shape{[}1{]})(y)\\ x = Concatenate()({[}x,y{]})\\ x = Conv(128, 3, (1,1))(x)\\ x = LeakyReLU()(x)\\ x = Conv(256, 3, (1,1))(x)\\ x = LeakyReLU()(BatchNorm()(x))\\ x = Conv(256, 3, (1,1))(x)\\ x = LeakyReLU()(BatchNorm()(x))\\ x = Conv(256, 3, (1,1))(x)\\ x = LeakyReLU()(BatchNorm()(x))\\ x = Conv(256, 3, (2,2))(x)\\ x = Flatten()(x)\\ x = Dropout(0.4)(x)\\ x = Dense(1)(x)\end{tabular} \\ \midrule
    $d_2$ & \begin{tabular}[t]{@{}l@{}}y = Embedding(num\_classes, 50)(y)\\ y = Dense(x.shape{[}0{]} * x.shape{[}1{]})(y)\\ y = Reshape((x.shape{[}0{]}, x.shape{[}1{]})(y)\\ x = Concatenate()({[}x,y{]})\\ x = Conv(64, 3, (1,1))(x)\\ x = LeakyReLU()(x)\\ x = Conv(128, 3, (2,2))(x)\\ x = LeakyReLU()(BatchNorm()(x))\\ x = Conv(128, 3, (2,2))(x)\\ x = LeakyReLU()(BatchNorm()(x))\\ x = Conv(256, 3, (2,2))(x)\\ x = LeakyReLU()(x)\\ x = Flatten()(x)\\ x = Dropout(0.4)(x)\\ x = Dense(1)(x)\end{tabular} & \begin{tabular}[t]{@{}l@{}}y = Embedding(num\_classes, 50)(y)\\ y = Dense(x.shape{[}0{]} * x.shape{[}1{]})(y)\\ y = Reshape((x.shape{[}0{]}, x.shape{[}1{]})(y)\\ x = Concatenate()({[}x,y{]})\\ x = Conv(64, 3, (1,1))(x)\\ x = LeakyReLU()(x)\\ x = Conv(128, 3, (2,2))(x)\\ x = LeakyReLU()(BatchNorm()(x))\\ x = Conv(128, 3, (2,2))(x)\\ x = LeakyReLU()(BatchNorm()(x))\\ x = Conv(256, 3, (2,2))(x)\\ x = LeakyReLU()(x)\\ x = Flatten()(x)\\ x = Dropout(0.4)(x)\\ x = Dense(1)(x)\end{tabular} & \begin{tabular}[t]{@{}l@{}}y = Embedding(num\_classes, 50)(y)\\ y = Dense(x.shape{[}0{]} * x.shape{[}1{]})(y)\\ y = Reshape((x.shape{[}0{]}, x.shape{[}1{]})(y)\\ x = Concatenate()({[}x,y{]})\\ x = Conv(64, 3, (1,1))(x)\\ x = LeakyReLU()(x)\\ x = Conv(128, 3, (2,2))(x)\\ x = LeakyReLU()(BatchNorm()(x))\\ x = Conv(128, 3, (2,2))(x)\\ x = LeakyReLU()(BatchNorm()(x))\\ x = Conv(256, 3, (2,2))(x)\\ x = LeakyReLU()(x)\\ x = Flatten()(x)\\ x = Dropout(0.4)(x)\\ x = Dense(1)(x)\end{tabular} & \begin{tabular}[t]{@{}l@{}}y = Embedding(num\_classes, 50)(y)\\ y = Dense(x.shape{[}0{]} * x.shape{[}1{]})(y)\\ y = Reshape((x.shape{[}0{]}, x.shape{[}1{]})(y)\\ x = Concatenate()({[}x,y{]})\\ x = Conv(64, 3, (1,1))(x)\\ x = LeakyReLU()(x)\\ x = Conv(128, 3, (2,2))(x)\\ x = LeakyReLU()(BatchNorm()(x))\\ x = Conv(128, 3, (2,2))(x)\\ x = LeakyReLU()(BatchNorm()(x))\\ x = Conv(256, 3, (2,2))(x)\\ x = LeakyReLU()(x)\\ x = Flatten()(x)\\ x = Dropout(0.4)(x)\\ x = Dense(1)(x)\end{tabular} \\ \bottomrule\\
    \end{tabular}
    \par\footnotesize $^*$Under U-shaped SL where the server no longer knows the labels of the private data, the model structures follow this table after the removal of label embeddings and their concatenation with the input.
  \end{table*}

\section{Full Experimental Results}\label{app:exp_res}
\begin{figure}[p]
    \centering
    \includegraphics{img/curves/legend.pdf}

    \subfloat[ResNet-20 on CIFAR-10]{%
        \begin{minipage}{0.49\linewidth}%
            \includegraphics{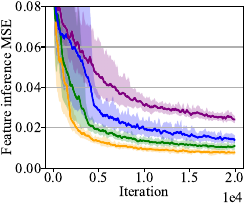}
            \hfil
            \includegraphics{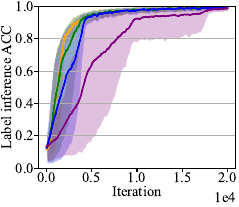}
        \end{minipage}
    }
    \hfil
    \subfloat[ResNet-20 on CIFAR-100]{%
        \begin{minipage}{0.49\linewidth}
            \includegraphics{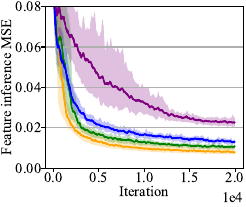}
            \hfil
            \includegraphics{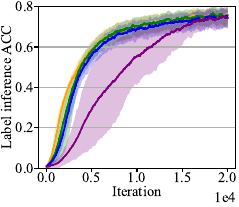}
        \end{minipage}
    }
    \hfil
    \subfloat[ResNet-20 on Tiny ImageNet]{%
        \begin{minipage}{0.49\linewidth}%
            \includegraphics{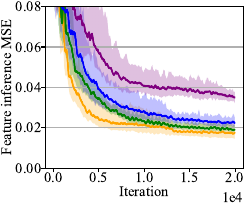}
            \hfil
            \includegraphics{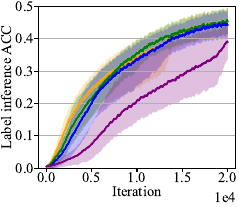}
        \end{minipage}
    }
    \hfil
    \subfloat[ResNet-20 on STL-10]{%
        \begin{minipage}{0.49\linewidth}
            \includegraphics{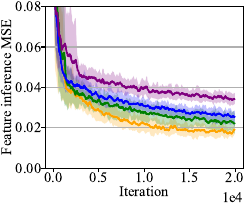}
            \hfil
            \includegraphics{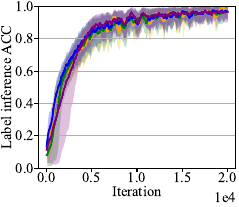}
        \end{minipage}
    }
    \hfil
    \subfloat[PlainNet-20 on CIFAR-10]{%
        \begin{minipage}{0.49\linewidth}
            \includegraphics{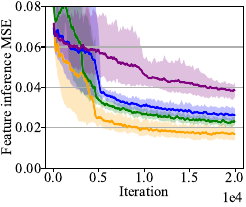}
            \hfil
            \includegraphics{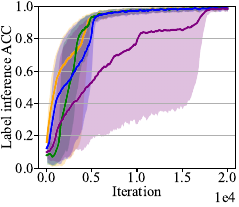}
        \end{minipage}
    }
    \hfil
    \subfloat[PlainNet-20 on CIFAR-100]{%
        \begin{minipage}{0.49\linewidth}
            \includegraphics{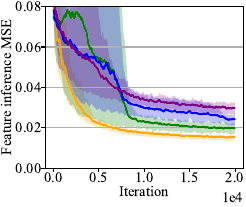}
            \hfil
            \includegraphics{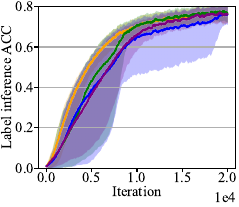}
        \end{minipage}
    }
    \hfil
    \subfloat[PlainNet-20 on Tiny ImageNet]{%
        \begin{minipage}{0.49\linewidth}
            \includegraphics{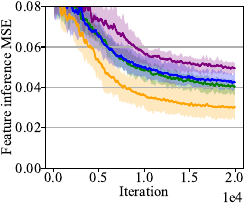}
            \hfil
            \includegraphics{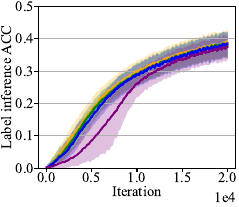}
        \end{minipage}
    }
    \hfil
    \subfloat[PlainNet-20 on STL-10]{%
        \begin{minipage}{0.49\linewidth}
            \includegraphics{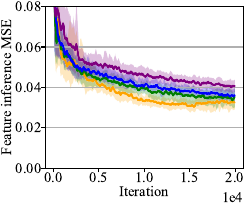}
            \hfil
            \includegraphics{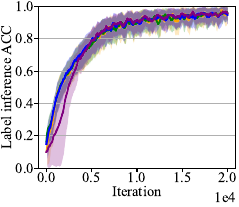}
        \end{minipage}
    }
    \caption{Feature inference MSE (left) and label inference accuracy (right) of SDAR over training iterations on all datasets with both models at split levels 4--7 on U-shaped SL.}
    \label{fig:u_curves_full}
\end{figure}

  In Section~\ref{sec:exp}, we only present experimental results on a limited number of datasets due to limited space. We hereby provide the full results of our experiments on all four datasets with both models. We list all extended tables and figures and their corresponding versions in the main text in Table~\ref{tab:ext}. In addition, we also report the attack MSE and label inference accuracy for U-shaped SL of SDAR over training iterations across all configurations in Fig.~\ref{fig:u_curves_full}.

\newpage
\onecolumn

\begin{table}[p]
    \centering
    \caption{Mean Values with Standard Deviations of the Feature Inference MSE over 5 trials on All Datasets and with Both Models in the Vanilla SL Setting}
    \label{tab:vanilla_full}
    
\begin{tabular}{lllccccc}
\toprule
\multirow{2}{*}{Dataset} & \multirow{2}{*}{Model} & \multirow{2}{*}{Method} & \multicolumn{4}{c}{Split Level} \\
\cmidrule(lr){4-7}
& & & 4 & 5 & 6 & 7 \\
\midrule
\multirow{6}{*}{CIFAR-10}  & \multirow{3}{*}{ResNet-20}  & SDAR & \bfseries 0.0084 (0.0003) & \bfseries 0.0118 (0.0008) & \bfseries 0.0140 (0.0009) & \bfseries 0.0212 (0.0019) \\
    & & PCAT & 0.0237 (0.0021) & 0.0337 (0.0048) & 0.0382 (0.0062) & 0.0568 (0.0062) \\
    & & UnSplit & 0.0667 (0.0003) & 0.0673 (0.0001) & 0.0679 (0.0001) & 0.0663 (0.0000) \\
\cmidrule(lr){2-7}  & \multirow{3}{*}{PlainNet-20}  &SDAR & \bfseries 0.0144 (0.0010) & \bfseries 0.0217 (0.0021) & \bfseries 0.0260 (0.0025) & \bfseries 0.0350 (0.0014) \\
    & &PCAT &0.0341 (0.0049) &0.0434 (0.0038) &0.0461 (0.0030) &0.0591 (0.0018) \\
    & &UnSplit &0.0666 (0.0001) &0.0668 (0.0000) &0.0668 (0.0001) &0.0664 (0.0001) \\
\midrule
\multirow{6}{*}{CIFAR-100}  & \multirow{3}{*}{ResNet-20}  & SDAR & \bfseries 0.0096 (0.0005) & \bfseries 0.0128 (0.0005) & \bfseries 0.0143 (0.0007) & \bfseries 0.0220 (0.0014) \\
    & & PCAT & 0.0185 (0.0032) & 0.0244 (0.0028) & 0.0300 (0.0014) & 0.0430 (0.0014) \\
    & & UnSplit & 0.0732 (0.0006) & 0.0732 (0.0002) & 0.0733 (0.0002) & 0.0720 (0.0001) \\
\cmidrule(lr){2-7}  & \multirow{3}{*}{PlainNet-20}  &SDAR &\bfseries 0.0185 (0.0016) & \bfseries 0.0227 (0.0011) & \bfseries 0.0247 (0.0009) & \bfseries 0.0301 (0.0014) \\
    & &PCAT &0.0341 (0.0032) &0.0386 (0.0034) & 0.0389 (0.0036) &0.0578 (0.0031) \\
    & & UnSplit &0.0726 (0.0003) &0.0726 (0.0001) &0.0726 (0.0000) &0.0721 (0.0000) \\
\midrule
\multirow{6}{*}{Tiny ImageNet}  & \multirow{3}{*}{ResNet-20}  & SDAR & \bfseries 0.0174 (0.0030) & \bfseries 0.0185 (0.0011) & \bfseries 0.0214 (0.0014) & \bfseries 0.0300 (0.0016) \\
    & & PCAT & 0.0390 (0.0083) & 0.0400 (0.0024) & 0.0439 (0.0038) & 0.0484 (0.0026) \\
    & & UnSplit & 0.0781 (0.0002) & 0.0797 (0.0002) & 0.0801 (0.0001) & 0.0786 (0.0000) \\
\cmidrule(lr){2-7}  & \multirow{3}{*}{PlainNet-20}  &SDAR & \bfseries 0.0299 (0.0029) & \bfseries 0.0369 (0.0016) &\bfseries 0.0393 (0.0020) &\bfseries 0.0419 (0.0030) \\
    & &PCAT & 0.0634 (0.0073) &0.0567 (0.0065) &0.0579 (0.0036) &0.0610 (0.0015) \\
    & &UnSplit &0.0787 (0.0003) &0.0791 (0.0001) &0.0788 (0.0001) &0.0787 (0.0000) \\
\midrule
\multirow{6}{*}{STL-10}  & \multirow{3}{*}{ResNet-20}  & SDAR & \bfseries 0.0234 (0.0036) & \bfseries 0.0254 (0.0039) & \bfseries 0.0279 (0.0019) & \bfseries 0.0340 (0.0026) \\
    & & PCAT & 0.0692 (0.0210) & 0.0710 (0.0098) & 0.0821 (0.0095) & 0.0747 (0.0039) \\
    & & UnSplit & 0.0711 (0.0006) & 0.0706 (0.0001) & 0.0710 (0.0003) & 0.0698 (0.0001) \\
\cmidrule(lr){2-7}  & \multirow{3}{*}{PlainNet-20}  &SDAR &\bfseries 0.0327 (0.0057) &\bfseries 0.0383 (0.0046) &\bfseries 0.0383 (0.0015) & \bfseries 0.0410 (0.0024) \\
    & &PCAT &0.0710 (0.0159) &0.0693 (0.0101) & 0.0850 (0.0094) & 0.0812 (0.0084) \\
    & & UnSplit & 0.0702 (0.0002) & 0.0701 (0.0001) & 0.0700 (0.0001) & 0.0699 (0.0000) \\
\bottomrule
\end{tabular}

  \end{table}

\begin{figure}[p]
    \small
    \renewcommand{\arraystretch}{0}
    \subfloat[ResNet-20 on CIFAR-10]{%
        \small%
        \renewcommand{\arraystretch}{0}%
        \begin{tabular}{@{}m{3mm}@{}m{4.3cm}@{\hspace{2pt}}m{4.3cm}@{}}
            & \includegraphics[width=0.99\linewidth]{img/examples/original/cifar10/cifar10_cropped_6.png} & \includegraphics[width=0.99\linewidth]{img/examples/original/cifar10/cifar10_cropped_6.png}\\\midrule
            4 & \includegraphics[width=0.99\linewidth]{img/examples/sdar_results/vsl/cifar10/resnet_l4_cropped_6.png} & \includegraphics[width=0.99\linewidth]{img/examples/pcat_results/vsl/cifar10/resnet_l4_cropped_6.png}\\
            5 & \includegraphics[width=0.99\linewidth]{img/examples/sdar_results/vsl/cifar10/resnet_l5_cropped_6.png} & \includegraphics[width=0.99\linewidth]{img/examples/pcat_results/vsl/cifar10/resnet_l5_cropped_6.png}\\
            6 & \includegraphics[width=0.99\linewidth]{img/examples/sdar_results/vsl/cifar10/resnet_l6_cropped_6.png} & \includegraphics[width=0.99\linewidth]{img/examples/pcat_results/vsl/cifar10/resnet_l6_cropped_6.png}\\
            7 & \includegraphics[width=0.99\linewidth]{img/examples/sdar_results/vsl/cifar10/resnet_l7_cropped_6.png} & \includegraphics[width=0.99\linewidth]{img/examples/pcat_results/vsl/cifar10/resnet_l7_cropped_6.png}\\\\[6pt]
            &\centering \footnotesize SDAR & \centering \footnotesize PCAT \cite{gao2023pcat}
        \end{tabular}
    }
    \hfill
    \subfloat[ResNet-20 on CIFAR-100]{%
    \renewcommand{\arraystretch}{0}%
        \small%
        \begin{tabular}{@{}m{3mm}@{}m{4.3cm}@{\hspace{2pt}}m{4.3cm}@{}}
            & \includegraphics[width=0.99\linewidth]{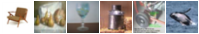} & \includegraphics[width=0.99\linewidth]{img/examples/original/cifar100/cifar100_cropped_6.png}\\\midrule
           4 & \includegraphics[width=0.99\linewidth]{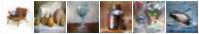} & \includegraphics[width=0.99\linewidth]{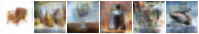}\\
           5 & \includegraphics[width=0.99\linewidth]{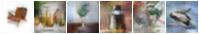} & \includegraphics[width=0.99\linewidth]{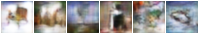}\\
           6 & \includegraphics[width=0.99\linewidth]{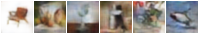} & \includegraphics[width=0.99\linewidth]{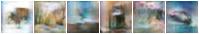}\\
           7 & \includegraphics[width=0.99\linewidth]{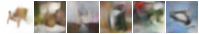} & \includegraphics[width=0.99\linewidth]{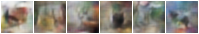}\\\\[6pt]
           &\centering \footnotesize SDAR & \centering \footnotesize PCAT \cite{gao2023pcat}
       \end{tabular}
    }
    \hfill
    \subfloat[ResNet-20 on Tiny ImageNet]{%
    \renewcommand{\arraystretch}{0}%
        \small%
        \begin{tabular}{@{}m{3mm}@{}m{4.3cm}@{\hspace{2pt}}m{4.3cm}@{}}
            & \includegraphics[width=0.99\linewidth]{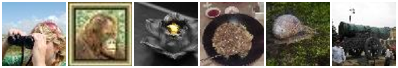} & \includegraphics[width=0.99\linewidth]{img/examples/original/tiny64/tiny64_cropped_6.png}\\\midrule
           4 & \includegraphics[width=0.99\linewidth]{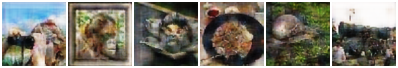} & \includegraphics[width=0.99\linewidth]{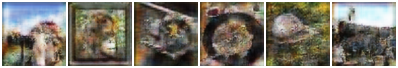}\\
           5 & \includegraphics[width=0.99\linewidth]{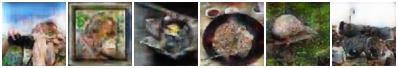} & \includegraphics[width=0.99\linewidth]{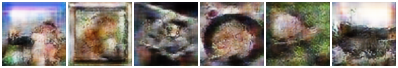}\\
           6 & \includegraphics[width=0.99\linewidth]{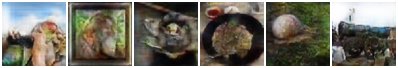} & \includegraphics[width=0.99\linewidth]{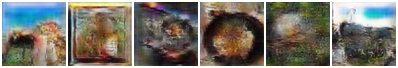}\\
           7 & \includegraphics[width=0.99\linewidth]{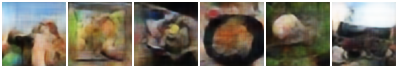} & \includegraphics[width=0.99\linewidth]{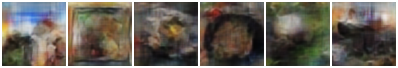}\\\\[6pt]
           &\centering \footnotesize SDAR & \centering \footnotesize PCAT \cite{gao2023pcat}
       \end{tabular}
    }
    \hfill
    \subfloat[ResNet-20 on STL-10]{%
    \renewcommand{\arraystretch}{0}%
        \small%
        \begin{tabular}{@{}m{3mm}@{}m{4.3cm}@{\hspace{2pt}}m{4.3cm}@{}}
            & \includegraphics[width=0.99\linewidth]{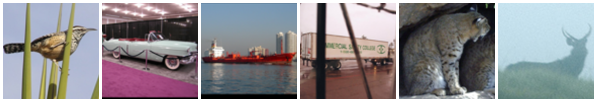} & \includegraphics[width=0.99\linewidth]{img/examples/original/stl10/stl10_cropped_6.png}\\\midrule
           4 & \includegraphics[width=0.99\linewidth]{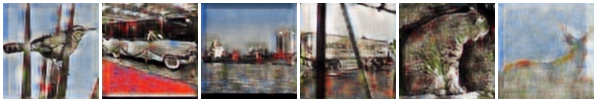} & \includegraphics[width=0.99\linewidth]{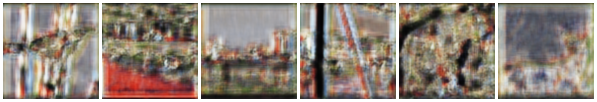}\\
           5 & \includegraphics[width=0.99\linewidth]{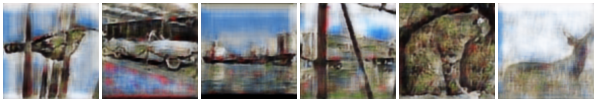} & \includegraphics[width=0.99\linewidth]{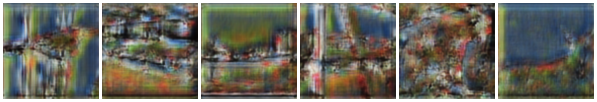}\\
           6 & \includegraphics[width=0.99\linewidth]{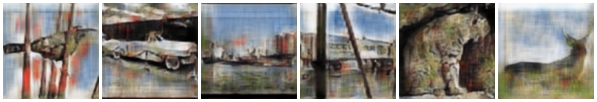} & \includegraphics[width=0.99\linewidth]{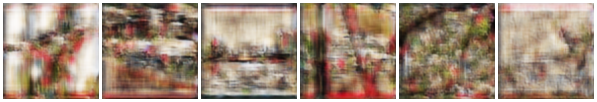}\\
           7 & \includegraphics[width=0.99\linewidth]{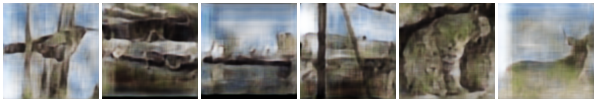} & \includegraphics[width=0.99\linewidth]{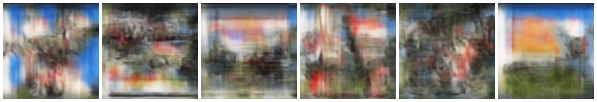}\\\\[6pt]
           &\centering \footnotesize SDAR & \centering \footnotesize PCAT \cite{gao2023pcat}
       \end{tabular}
    }
    \hfill
    \subfloat[PlainNet-20 on CIFAR-10]{%
        \small%
        \renewcommand{\arraystretch}{0}%
        \begin{tabular}{@{}m{3mm}@{}m{4.3cm}@{\hspace{2pt}}m{4.3cm}@{}}
            & \includegraphics[width=0.99\linewidth]{img/examples/original/cifar10/cifar10_cropped_6.png} & \includegraphics[width=0.99\linewidth]{img/examples/original/cifar10/cifar10_cropped_6.png}\\\midrule
            4 & \includegraphics[width=0.99\linewidth]{img/examples/sdar_results/vsl/cifar10/plainnet_l4_cropped_6.png} & \includegraphics[width=0.99\linewidth]{img/examples/pcat_results/vsl/cifar10/plainnet_l4_cropped_6.png}\\
            5 & \includegraphics[width=0.99\linewidth]{img/examples/sdar_results/vsl/cifar10/plainnet_l5_cropped_6.png} & \includegraphics[width=0.99\linewidth]{img/examples/pcat_results/vsl/cifar10/plainnet_l5_cropped_6.png}\\
            6 & \includegraphics[width=0.99\linewidth]{img/examples/sdar_results/vsl/cifar10/plainnet_l6_cropped_6.png} & \includegraphics[width=0.99\linewidth]{img/examples/pcat_results/vsl/cifar10/plainnet_l6_cropped_6.png}\\
            7 & \includegraphics[width=0.99\linewidth]{img/examples/sdar_results/vsl/cifar10/plainnet_l7_cropped_6.png} & \includegraphics[width=0.99\linewidth]{img/examples/pcat_results/vsl/cifar10/plainnet_l7_cropped_6.png}\\\\[6pt]
            &\centering \footnotesize SDAR & \centering \footnotesize PCAT \cite{gao2023pcat}
        \end{tabular}
    }
    \hfill
    \subfloat[PlainNet-20 on CIFAR-100]{%
    \renewcommand{\arraystretch}{0}%
        \small%
        \begin{tabular}{@{}m{3mm}@{}m{4.3cm}@{\hspace{2pt}}m{4.3cm}@{}}
            & \includegraphics[width=0.99\linewidth]{img/examples/original/cifar100/cifar100_cropped_6.png} & \includegraphics[width=0.99\linewidth]{img/examples/original/cifar100/cifar100_cropped_6.png}\\\midrule
           4 & \includegraphics[width=0.99\linewidth]{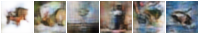} & \includegraphics[width=0.99\linewidth]{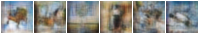}\\
           5 & \includegraphics[width=0.99\linewidth]{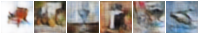} & \includegraphics[width=0.99\linewidth]{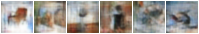}\\
           6 & \includegraphics[width=0.99\linewidth]{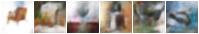} & \includegraphics[width=0.99\linewidth]{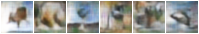}\\
           7 & \includegraphics[width=0.99\linewidth]{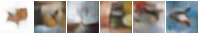} & \includegraphics[width=0.99\linewidth]{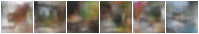}\\\\[6pt]
           &\centering \footnotesize SDAR & \centering \footnotesize PCAT \cite{gao2023pcat}
       \end{tabular}
    }
    \hfill
    \subfloat[PlainNet-20 on Tiny ImageNet]{%
    \renewcommand{\arraystretch}{0}%
        \small%
        \begin{tabular}{@{}m{3mm}@{}m{4.3cm}@{\hspace{2pt}}m{4.3cm}@{}}
            & \includegraphics[width=0.99\linewidth]{img/examples/original/tiny64/tiny64_cropped_6.png} & \includegraphics[width=0.99\linewidth]{img/examples/original/tiny64/tiny64_cropped_6.png}\\\midrule
           4 & \includegraphics[width=0.99\linewidth]{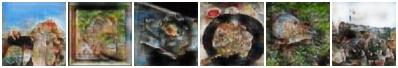} & \includegraphics[width=0.99\linewidth]{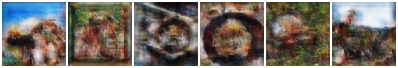}\\
           5 & \includegraphics[width=0.99\linewidth]{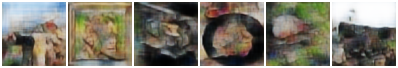} & \includegraphics[width=0.99\linewidth]{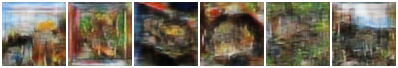}\\
           6 & \includegraphics[width=0.99\linewidth]{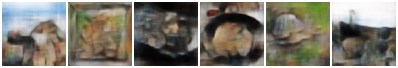} & \includegraphics[width=0.99\linewidth]{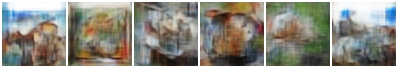}\\
           7 & \includegraphics[width=0.99\linewidth]{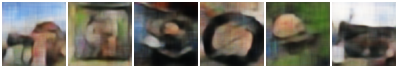} & \includegraphics[width=0.99\linewidth]{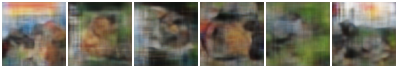}\\\\[6pt]
           &\centering \footnotesize SDAR & \centering \footnotesize PCAT \cite{gao2023pcat}
       \end{tabular}
    }
    \hfill
    \subfloat[PlainNet-20 on STL-10]{%
    \renewcommand{\arraystretch}{0}%
        \small%
        \begin{tabular}{@{}m{3mm}@{}m{4.3cm}@{\hspace{2pt}}m{4.3cm}@{}}
            & \includegraphics[width=0.99\linewidth]{img/examples/original/stl10/stl10_cropped_6.png} & \includegraphics[width=0.99\linewidth]{img/examples/original/stl10/stl10_cropped_6.png}\\\midrule
           4 & \includegraphics[width=0.99\linewidth]{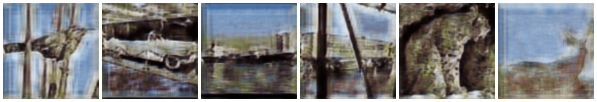} & \includegraphics[width=0.99\linewidth]{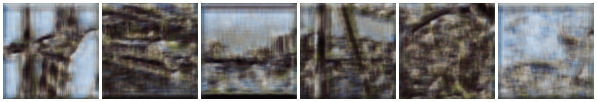}\\
           5 & \includegraphics[width=0.99\linewidth]{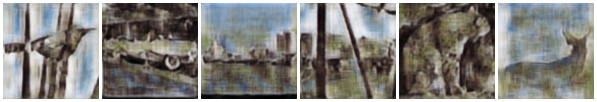} & \includegraphics[width=0.99\linewidth]{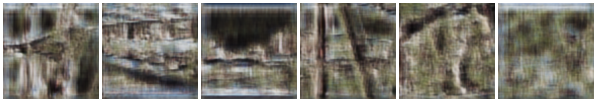}\\
           6 & \includegraphics[width=0.99\linewidth]{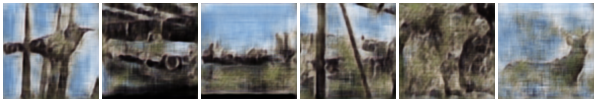} & \includegraphics[width=0.99\linewidth]{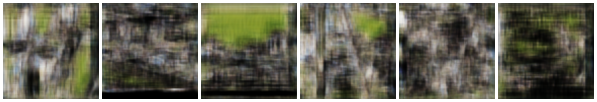}\\
           7 & \includegraphics[width=0.99\linewidth]{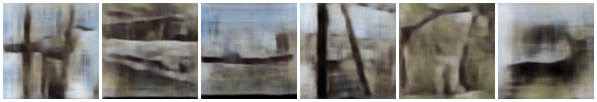} & \includegraphics[width=0.99\linewidth]{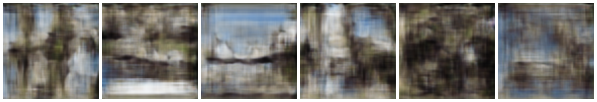}\\\\[6pt]
           &\centering \footnotesize SDAR & \centering \footnotesize PCAT \cite{gao2023pcat}
       \end{tabular}
    }
    \caption{Examples of feature inference attack results of SDAR on all datasets in vanilla SL.}
    \label{fig:v_examples_full}
\end{figure}

\begin{figure}[p]
    \centering
    \includegraphics{img/curves/legend.pdf}

    \subfloat[ResNet-20 on CIFAR-10]{\includegraphics{img/curves/vsl_resnet_cifar10_mse.pdf}}
    \hfil
    \subfloat[ResNet-20 on CIFAR-100]{\includegraphics{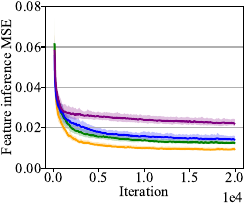}}
    \hfil
    \subfloat[ResNet-20 on Tiny ImageNet]{\includegraphics{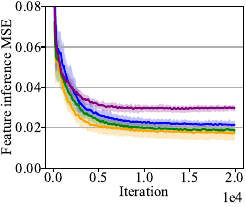}}
    \hfil
    \subfloat[ResNet-20 on STL-10]{\includegraphics{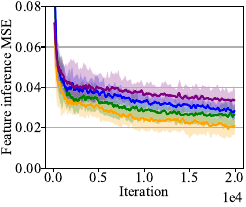}}
    \hfil
    \subfloat[PlainNet-20 on CIFAR-10]{\includegraphics{img/curves/vsl_plainnet_cifar10_mse.pdf}}
    \hfil
    \subfloat[PlainNet-20 on CIFAR-100]{\includegraphics{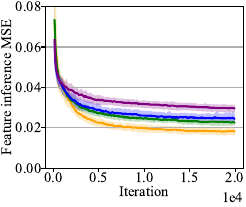}}
    \hfil
    \subfloat[PlainNet-20 on Tiny ImageNet]{\includegraphics{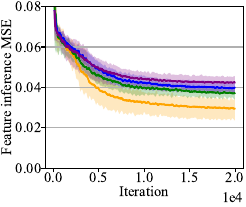}}
    \hfil
    \subfloat[PlainNet-20 on STL-10]{\includegraphics{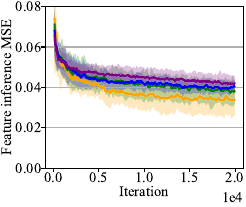}}
    \caption{Feature inference MSE of SDAR over training iterations on all datasets with both models at split levels 4--7 in vanilla SL.}
    \label{fig:v_sdar}
\end{figure}

\begin{figure}[p]

    \centering
    \includegraphics{img/fsha/legend.pdf}

    \subfloat[ResNet-20 on CIFAR-10]{%
        \begin{minipage}{0.49\linewidth}%
            \includegraphics{img/fsha/resnet_cifar10_attack_loss.pdf}
            \hfil
            \includegraphics{img/fsha/resnet_cifar10_training_loss.pdf}
        \end{minipage}
    }
    \hfil
    \subfloat[ResNet-20 on CIFAR-100]{%
        \begin{minipage}{0.49\linewidth}
            \includegraphics{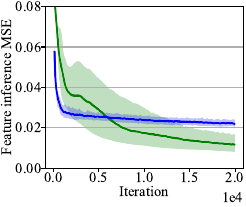}
            \hfil
            \includegraphics{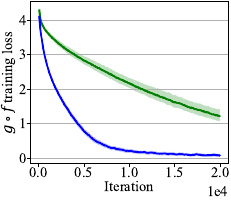}
        \end{minipage}
    }
    \hfil
    \subfloat[ResNet-20 on Tiny ImageNet]{%
        \begin{minipage}{0.49\linewidth}%
            \includegraphics{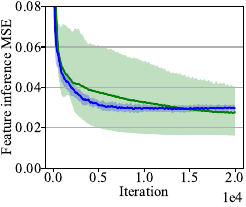}
            \hfil
            \includegraphics{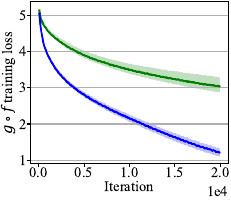}
        \end{minipage}
    }
    \hfil
    \subfloat[ResNet-20 on STL-10]{%
        \begin{minipage}{0.49\linewidth}
            \includegraphics{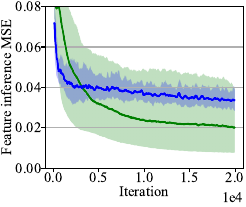}
            \hfil
            \includegraphics{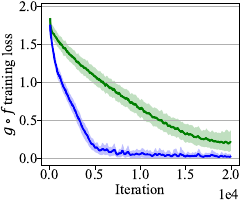}
        \end{minipage}
    }
    \hfil
    \subfloat[PlainNet-20 on CIFAR-10]{%
        \begin{minipage}{0.49\linewidth}
            \includegraphics{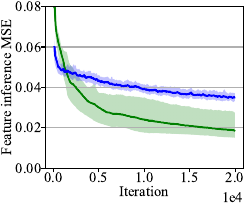}
            \hfil
            \includegraphics{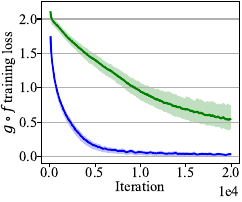}
        \end{minipage}
    }
    \hfil
    \subfloat[PlainNet-20 on CIFAR-100]{%
        \begin{minipage}{0.49\linewidth}
            \includegraphics{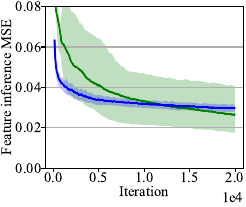}
            \hfil
            \includegraphics{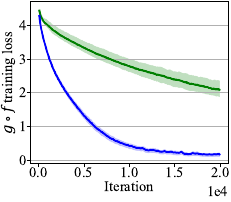}
        \end{minipage}
    }
    \hfil
    \subfloat[PlainNet-20 on Tiny ImageNet]{%
        \begin{minipage}{0.49\linewidth}%
            \includegraphics{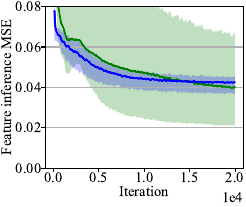}
            \hfil
            \includegraphics{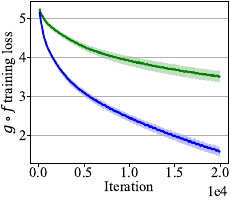}
        \end{minipage}
    }
    \hfil
    \subfloat[PlainNet-20 on STL-10]{%
        \begin{minipage}{0.49\linewidth}
            \includegraphics{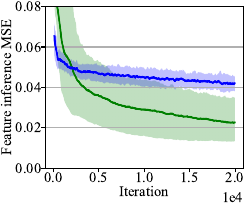}
            \hfil
            \includegraphics{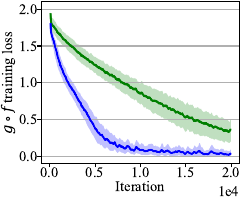}
        \end{minipage}
    }
    \caption{Attack MSE (left) and training losses on the original task (right) of SDAR vs FSHA on all datasets with both models at split level 7 in vanilla SL.}
    \label{fig:fsha_full}
\end{figure}

\begin{figure}[p]
    \centering
    \includegraphics{img/width/legend.pdf}

    \subfloat[ResNet-20 on CIFAR-10]{\includegraphics{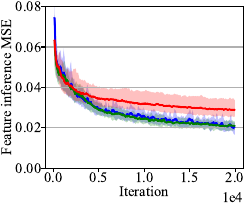}}
    \hfil
    \subfloat[ResNet-20 on CIFAR-100]{\includegraphics{img/width/resnet_cifar100.pdf}}
    \hfil
    \subfloat[ResNet-20 on Tiny ImageNet]{\includegraphics{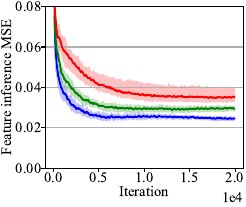}}
    \hfil
    \subfloat[ResNet-20 on STL-10]{\includegraphics{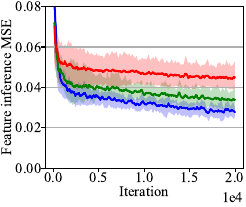}}
    \hfil
    \subfloat[PlainNet-20 on CIFAR-10]{\includegraphics{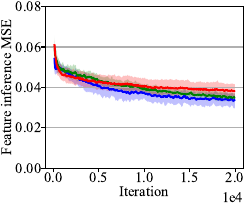}}
    \hfil
    \subfloat[PlainNet-20 on CIFAR-100]{\includegraphics{img/width/plainnet_cifar100.pdf}}
    \hfil
    \subfloat[PlainNet-20 on Tiny ImageNet]{\includegraphics{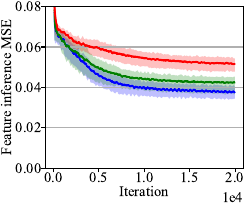}}
    \hfil
    \subfloat[PlainNet-20 on STL-10]{\includegraphics{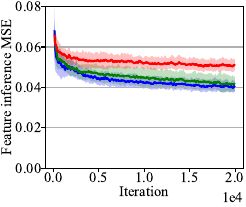}}
    \caption{Effects of target model widths on SDAR performed on all datasets with both models at split level 7 in vanilla SL. Note that the same trend is observed for all configurations with one exception in the case of ResNet-20 on CIFAR-10, where the improvement in attack performance is less evident with wider models. This is likely due to the fact that the standard model is already wide enough to capture the input data distribution as the setting of ResNet-20 on CIFAR-10 is relatively easier to attack compared to other configurations.}
    \label{fig:width_full}
\end{figure}

\begin{table}[p]
    \centering
    \caption{Effects of Different Sizes of the Auxiliary Data $D'$ on SDAR Performed on All Datasets with Both Models at Split Level 7 in the Vanilla SL Setting}
    \label{tab:aux_data_size_full}
    \begin{tabular}{llccccc}
\toprule
\multirow{2}{*}{Dataset} & \multirow{2}{*}{Model} & \multicolumn{5}{c}{$|D'|/|D|$}\\
\cmidrule(lr){3-7} & & 1.0 & 0.5 & 0.2 & 0.1 & 0.05 \\
\midrule
\multirow{2}{*}{CIFAR-10}  & ResNet-20 & 0.0212 (0.0019) & 0.0225 (0.0014) & 0.0232 (0.0027) & 0.0248 (0.0024) & 0.0297 (0.0011) \\
    & PlainNet-20 & 0.0350 (0.0014) & 0.0353 (0.0025) & 0.0365 (0.0011) & 0.0363 (0.0014) & 0.0435 (0.0025) \\
\midrule
\multirow{2}{*}{CIFAR-100}  & ResNet-20 & 0.0220 (0.0014) & 0.0222 (0.0010) & 0.0226 (0.0010) & 0.0257 (0.0012) & 0.0294 (0.0005) \\
    & PlainNet-20 & 0.0301 (0.0014) & 0.0311 (0.0016) & 0.0362 (0.0036) & 0.0403 (0.0010) & 0.0436 (0.0008) \\
\midrule
\multirow{2}{*}{Tiny ImageNet}  & ResNet-20 & 0.0300 (0.0016) & 0.0310 (0.0016) & 0.0342 (0.0015) & 0.0365 (0.0016) & 0.0409 (0.0020) \\
    & PlainNet-20 & 0.0419 (0.0030) & 0.0458 (0.0043) & 0.0557 (0.0017) & 0.0605 (0.0019) & 0.0650 (0.0056) \\
\midrule
\multirow{2}{*}{ STL-10}  &  ResNet-20 &  0.0340 (0.0026) &  0.0348 (0.0026) &  0.0366 (0.0023) &  0.0393 (0.0012) &  0.0457 (0.0013) \\
    & PlainNet-20 & 0.0410 (0.0024) & 0.0427 (0.0025) & 0.0464 (0.0020) & 0.0496 (0.0006) & 0.0574 (0.0027) \\
\bottomrule
\end{tabular}

\end{table}

\begin{table}[p]
    \centering
    \caption{Mean Values with Standard Deviations of the Feature Inference MSE of SDAR over 5 Trials Performed on All Datasets with in the Vanilla SL Setting with or without Access to the Architecture of $f$}
    \label{tab:diff_sim_full}
    \begin{tabular}{llcccc}
\toprule
\multirow{2}{*}{Dataset} & \multirow{2}{*}{Simulator architecture} & \multicolumn{4}{c}{Split Level}\\
\cmidrule(lr){3-6} & & 4 & 5 & 6 & 7 \\
\midrule
\multirow{2}{*}{CIFAR-10}  & Same architecture & 0.0084 (0.0003) & 0.0118 (0.0008) & 0.0140 (0.0009) & 0.0212 (0.0019) \\
    & Different architecture & 0.0103 (0.0017) & 0.0143 (0.0005) & 0.0165 (0.0010) & 0.0258 (0.0031) \\
\midrule
\multirow{2}{*}{CIFAR-100}  & Same architecture & 0.0096 (0.0005) & 0.0128 (0.0005) & 0.0143 (0.0007) & 0.0220 (0.0014) \\
    & Different architecture & 0.0101 (0.0015) & 0.0132 (0.0014) & 0.0163 (0.0018) & 0.0232 (0.0019) \\
\midrule
\multirow{2}{*}{Tiny ImageNet}  & Same architecture & 0.0174 (0.0030) & 0.0185 (0.0011) & 0.0214 (0.0014) & 0.0300 (0.0016) \\
    & Different architecture & 0.0177 (0.0024) & 0.0221 (0.0012) & 0.0252 (0.0029) & 0.0307 (0.0010) \\
\midrule
\multirow{2}{*}{STL-10} &  Same architecture &  0.0234 (0.0036) & 0.0254 (0.0039) & 0.0279 (0.0019) & 0.0340 (0.0026) \\
    & Different architecture & 0.0260 (0.0049) & 0.0299 (0.0039) & 0.0311 (0.0028) & 0.0370 (0.0030) \\
\bottomrule
\end{tabular}

\end{table}

\begin{table}[p]
    \centering
    \caption{Mean Values with Standard Deviations of the Feature Inference MSE over 5 Trials on All Datasets with Both Models in U-shaped SL}
    \label{tab:u_mse_full}
    
\begin{tabular}{lllccccc}
\toprule
\multirow{2}{*}{Dataset} & \multirow{2}{*}{Model} & \multirow{2}{*}{Method} & \multicolumn{4}{c}{Split Level} \\
\cmidrule(lr){4-7}
& & & 4 & 5 & 6 & 7 \\
\midrule
\multirow{4}{*}{CIFAR-10}  & \multirow{2}{*}{ResNet-20}  &SDAR & \bfseries 0.0085 (0.0013) & \bfseries 0.0120 (0.0017) & \bfseries 0.0138 (0.0013) & \bfseries 0.0247 (0.0009) \\
    & &PCAT &  0.0387 (0.0262) &  0.0665 (0.0262) &  0.0791 (0.0122) &  0.0947 (0.0184) \\
\cmidrule(lr){2-7}  & \multirow{2}{*}{PlainNet-20}  &SDAR & \bfseries 0.0176 (0.0026) & \bfseries 0.0222 (0.0008) & \bfseries 0.0266 (0.0024) & \bfseries 0.0392 (0.0025) \\
    & &PCAT &  0.0599 (0.0189) &  0.0740 (0.0162) &  0.0652 (0.0172) &  0.0756 (0.0021) \\
\midrule
\multirow{4}{*}{CIFAR-100}  & \multirow{2}{*}{ResNet-20}  &SDAR & \bfseries 0.0084 (0.0015) & \bfseries 0.0115 (0.0010) & \bfseries 0.0132 (0.0006) & \bfseries 0.0227 (0.0007) \\
    & &PCAT &  0.0547 (0.0418) &  0.0708 (0.0200) &  0.0998 (0.0091) &  0.0938 (0.0062) \\
\cmidrule(lr){2-7}  & \multirow{2}{*}{PlainNet-20}  &SDAR & \bfseries 0.0157 (0.0014) & \bfseries 0.0198 (0.0020) & \bfseries 0.0239 (0.0015) & \bfseries 0.0294 (0.0014) \\
    & &PCAT &  0.0361 (0.0059) &  0.0630 (0.0196) &  0.0742 (0.0193) &  0.0807 (0.0084) \\
\midrule
\multirow{4}{*}{Tiny ImageNet}  & \multirow{2}{*}{ResNet-20}  &SDAR & \bfseries 0.0184 (0.0031) & \bfseries 0.0192 (0.0015) & \bfseries 0.0223 (0.0025) & \bfseries 0.0348 (0.0020) \\
    & &PCAT &  0.0551 (0.0386) &  0.0764 (0.0259) &  0.0913 (0.0270) &  0.1085 (0.0102) \\
\cmidrule(lr){2-7}  & \multirow{2}{*}{PlainNet-20}  &SDAR & \bfseries 0.0296 (0.0075) & \bfseries 0.0403 (0.0015) & \bfseries 0.0425 (0.0037) & \bfseries 0.0487 (0.0018) \\
    & &PCAT &  0.0757 (0.0160) &  0.0785 (0.0165) &  0.0823 (0.0169) &  0.0843 (0.0034) \\
\midrule
\multirow{4}{*}{STL-10}  & \multirow{2}{*}{ResNet-20}  &SDAR & \bfseries 0.0201 (0.0036) & \bfseries 0.0225 (0.0029) & \bfseries 0.0257 (0.0024) & \bfseries 0.0323 (0.0024) \\
    & &PCAT &  0.1137 (0.0207) &  0.0831 (0.0124) &  0.0869 (0.0093) &  0.0925 (0.0053) \\
\cmidrule(lr){2-7}  & \multirow{2}{*}{PlainNet-20}  &SDAR & \bfseries 0.0323 (0.0033) & \bfseries 0.0354 (0.0022) & \bfseries 0.0356 (0.0023) & \bfseries 0.0413 (0.0028) \\
    & &PCAT &  0.0937 (0.0152) &  0.0854 (0.0030) &  0.0919 (0.0139) &  0.1043 (0.0163) \\
\bottomrule
\end{tabular}

\end{table}

\begin{table}[p]
    \centering
    \caption{Mean Values with Standard Deviations of the Label Inference accuracy (\%) over 5 Trials on All Datasets with Both Models in U-shaped SL}
    \label{tab:u_acc_full}
    
\begin{tabular}{lllccccc}
\toprule
\multirow{2}{*}{Dataset} & \multirow{2}{*}{Model} & \multirow{2}{*}{Method} & \multicolumn{4}{c}{Split Level} \\
\cmidrule(lr){4-7}
& & & 4 & 5 & 6 & 7 \\
\midrule
\multirow{4}{*}{CIFAR-10}  & \multirow{2}{*}{ResNet-20}  & SDAR & \bfseries 98.61 (0.26) & \bfseries 98.49 (0.25) & \bfseries 98.15 (0.51) & \bfseries 98.71 (0.77) \\
    & & PCAT & 88.66 (18.88) & 58.11 (34.33) & 21.34 (25.29) & 15.45 (9.99) \\
\cmidrule(lr){2-7}  & \multirow{2}{*}{PlainNet-20}  & SDAR & \bfseries 98.10 (0.53) & \bfseries 98.18 (0.29) & \bfseries 98.15 (0.74) & \bfseries 97.93 (0.72) \\
    & & PCAT & 42.08 (45.74) & 30.44 (36.04) & 39.54 (40.51) & 15.68 (9.54) \\
\midrule
\multirow{4}{*}{CIFAR-100}  & \multirow{2}{*}{ResNet-20}  & SDAR & \bfseries 75.16 (1.46) & \bfseries 74.84 (1.63) & \bfseries 73.72 (2.36) & \bfseries 73.83 (1.20) \\
    & & PCAT & 45.33 (33.78) & 16.91 (18.71) & 3.44 (2.51) & 2.32 (1.75) \\
\cmidrule(lr){2-7}  & \multirow{2}{*}{PlainNet-20}  & SDAR & 74.37 (0.60) & \bfseries 72.80 (0.86) & \bfseries 72.86 (2.10) & \bfseries 73.91 (0.78) \\
    & & PCAT &\bfseries 74.60 (1.68) & 33.03 (22.73) & 20.25 (28.64) & 9.92 (16.77) \\
\midrule
\multirow{4}{*}{Tiny ImageNet}  & \multirow{2}{*}{ResNet-20}  & SDAR & \bfseries 41.22 (1.16) & \bfseries 42.27 (0.58) & \bfseries 41.50 (0.44) & \bfseries 37.19 (2.36) \\
    & & PCAT & 33.32 (15.90) & 15.89 (8.53) & 13.83 (14.63) & 0.98 (0.36) \\
\cmidrule(lr){2-7}  & \multirow{2}{*}{PlainNet-20}  & SDAR & \bfseries 34.94 (1.05) & \bfseries 35.47 (1.10) & \bfseries 35.63 (0.48) & \bfseries 34.91 (0.46) \\
    & & PCAT & 29.75 (15.41) & 33.43 (6.40) & 11.05 (10.76) & 2.78 (1.87) \\
\midrule
\multirow{4}{*}{STL-10}  & \multirow{2}{*}{ResNet-20}  & SDAR & \bfseries 94.86 (4.83) & \bfseries 97.66 (1.06) & \bfseries 91.85 (9.39) & \bfseries 96.21 (2.71) \\
    & & PCAT & 24.31 (23.38) & 44.64 (36.91) & 5.09 (4.21) & 18.11 (14.48) \\
\cmidrule(lr){2-7}  & \multirow{2}{*}{PlainNet-20}  & SDAR & \bfseries 93.79 (2.90) & \bfseries 90.45 (5.59) & \bfseries 93.90 (2.85) & \bfseries 93.44 (2.49) \\
    & & PCAT & 47.79 (30.14) & 16.39 (12.64) & 56.13 (38.54) & 13.12 (12.27) \\
\bottomrule
\end{tabular}

\end{table}

\begin{figure}[p]
    \small
    \renewcommand{\arraystretch}{0}
    \subfloat[ResNet-20 on CIFAR-10]{%
        \small%
        \renewcommand{\arraystretch}{0}%
        \begin{tabular}{@{}m{3mm}@{}m{4.3cm}@{\hspace{2pt}}m{4.3cm}@{}}
            & \includegraphics[width=0.99\linewidth]{img/examples/original/cifar10/cifar10_cropped_6.png} & \includegraphics[width=0.99\linewidth]{img/examples/original/cifar10/cifar10_cropped_6.png}\\\midrule
            4 & \includegraphics[width=0.99\linewidth]{img/examples/sdar_results/usl/cifar10/resnet_l4_cropped_6.png} & \includegraphics[width=0.99\linewidth]{img/examples/pcat_results/usl/cifar10/resnet_l4_cropped_6.png}\\
            5 & \includegraphics[width=0.99\linewidth]{img/examples/sdar_results/usl/cifar10/resnet_l5_cropped_6.png} & \includegraphics[width=0.99\linewidth]{img/examples/pcat_results/usl/cifar10/resnet_l5_cropped_6.png}\\
            6 & \includegraphics[width=0.99\linewidth]{img/examples/sdar_results/usl/cifar10/resnet_l6_cropped_6.png} & \includegraphics[width=0.99\linewidth]{img/examples/pcat_results/usl/cifar10/resnet_l6_cropped_6.png}\\
            7 & \includegraphics[width=0.99\linewidth]{img/examples/sdar_results/usl/cifar10/resnet_l7_cropped_6.png} & \includegraphics[width=0.99\linewidth]{img/examples/pcat_results/usl/cifar10/resnet_l7_cropped_6.png}\\\\[6pt]
            &\centering \footnotesize SDAR & \centering \footnotesize PCAT \cite{gao2023pcat}
        \end{tabular}
    }
    \hfill
    \subfloat[ResNet-20 on CIFAR-100]{%
    \renewcommand{\arraystretch}{0}%
        \small%
        \begin{tabular}{@{}m{3mm}@{}m{4.3cm}@{\hspace{2pt}}m{4.3cm}@{}}
            & \includegraphics[width=0.99\linewidth]{img/examples/original/cifar100/cifar100_cropped_6.png} & \includegraphics[width=0.99\linewidth]{img/examples/original/cifar100/cifar100_cropped_6.png}\\\midrule
           4 & \includegraphics[width=0.99\linewidth]{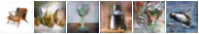} & \includegraphics[width=0.99\linewidth]{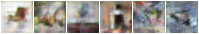}\\
           5 & \includegraphics[width=0.99\linewidth]{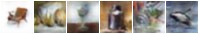} & \includegraphics[width=0.99\linewidth]{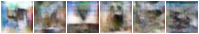}\\
           6 & \includegraphics[width=0.99\linewidth]{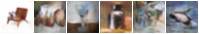} & \includegraphics[width=0.99\linewidth]{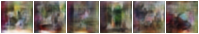}\\
           7 & \includegraphics[width=0.99\linewidth]{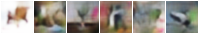} & \includegraphics[width=0.99\linewidth]{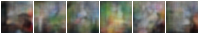}\\\\[6pt]
           &\centering \footnotesize SDAR & \centering \footnotesize PCAT \cite{gao2023pcat}
       \end{tabular}
    }
    \hfill
    \subfloat[ResNet-20 on Tiny ImageNet]{%
    \renewcommand{\arraystretch}{0}%
        \small%
        \begin{tabular}{@{}m{3mm}@{}m{4.3cm}@{\hspace{2pt}}m{4.3cm}@{}}
            & \includegraphics[width=0.99\linewidth]{img/examples/original/tiny64/tiny64_cropped_6.png} & \includegraphics[width=0.99\linewidth]{img/examples/original/tiny64/tiny64_cropped_6.png}\\\midrule
           4 & \includegraphics[width=0.99\linewidth]{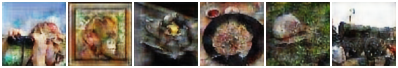} & \includegraphics[width=0.99\linewidth]{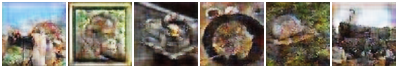}\\
           5 & \includegraphics[width=0.99\linewidth]{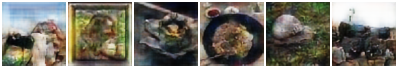} & \includegraphics[width=0.99\linewidth]{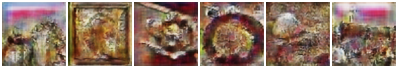}\\
           6 & \includegraphics[width=0.99\linewidth]{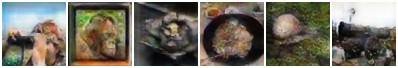} & \includegraphics[width=0.99\linewidth]{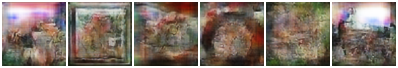}\\
           7 & \includegraphics[width=0.99\linewidth]{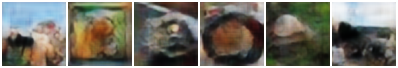} & \includegraphics[width=0.99\linewidth]{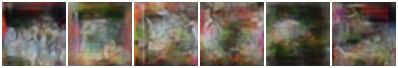}\\\\[6pt]
           &\centering \footnotesize SDAR & \centering \footnotesize PCAT \cite{gao2023pcat}
       \end{tabular}
    }
    \hfill
    \subfloat[ResNet-20 on STL-10]{%
    \renewcommand{\arraystretch}{0}%
        \small%
        \begin{tabular}{@{}m{3mm}@{}m{4.3cm}@{\hspace{2pt}}m{4.3cm}@{}}
            & \includegraphics[width=0.99\linewidth]{img/examples/original/stl10/stl10_cropped_6.png} & \includegraphics[width=0.99\linewidth]{img/examples/original/stl10/stl10_cropped_6.png}\\\midrule
           4 & \includegraphics[width=0.99\linewidth]{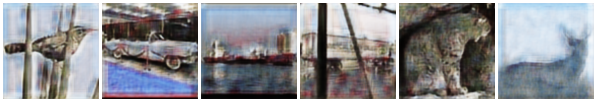} & \includegraphics[width=0.99\linewidth]{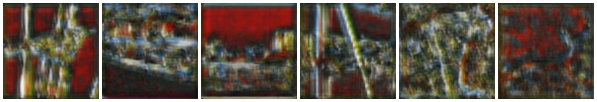}\\
           5 & \includegraphics[width=0.99\linewidth]{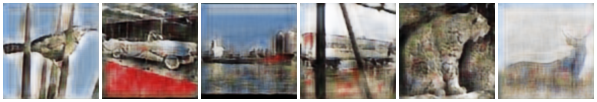} & \includegraphics[width=0.99\linewidth]{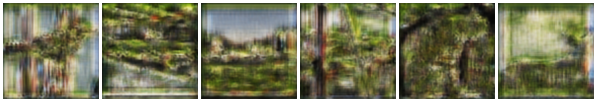}\\
           6 & \includegraphics[width=0.99\linewidth]{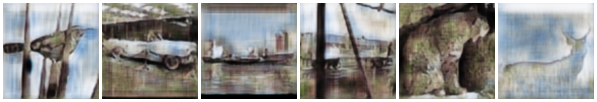} & \includegraphics[width=0.99\linewidth]{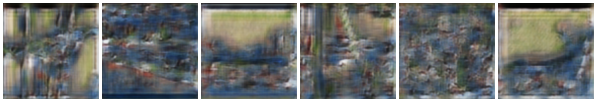}\\
           7 & \includegraphics[width=0.99\linewidth]{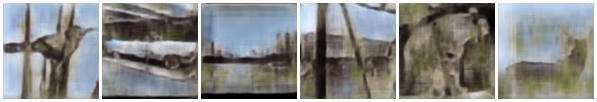} & \includegraphics[width=0.99\linewidth]{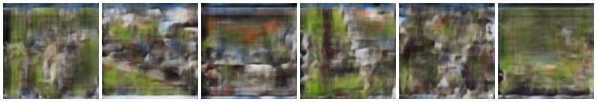}\\\\[6pt]
           &\centering \footnotesize SDAR & \centering \footnotesize PCAT \cite{gao2023pcat}
       \end{tabular}
    }
    \hfill
    \subfloat[PlainNet-20 on CIFAR-10]{%
        \small%
        \renewcommand{\arraystretch}{0}%
        \begin{tabular}{@{}m{3mm}@{}m{4.3cm}@{\hspace{2pt}}m{4.3cm}@{}}
            & \includegraphics[width=0.99\linewidth]{img/examples/original/cifar10/cifar10_cropped_6.png} & \includegraphics[width=0.99\linewidth]{img/examples/original/cifar10/cifar10_cropped_6.png}\\\midrule
            4 & \includegraphics[width=0.99\linewidth]{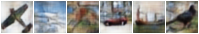} & \includegraphics[width=0.99\linewidth]{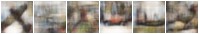}\\
            5 & \includegraphics[width=0.99\linewidth]{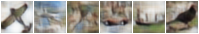} & \includegraphics[width=0.99\linewidth]{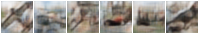}\\
            6 & \includegraphics[width=0.99\linewidth]{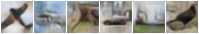} & \includegraphics[width=0.99\linewidth]{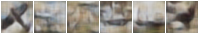}\\
            7 & \includegraphics[width=0.99\linewidth]{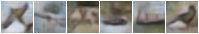} & \includegraphics[width=0.99\linewidth]{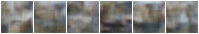}\\\\[6pt]
            &\centering \footnotesize SDAR & \centering \footnotesize PCAT \cite{gao2023pcat}
        \end{tabular}
    }
    \hfill
    \subfloat[PlainNet-20 on CIFAR-100]{%
    \renewcommand{\arraystretch}{0}%
        \small%
        \begin{tabular}{@{}m{3mm}@{}m{4.3cm}@{\hspace{2pt}}m{4.3cm}@{}}
            & \includegraphics[width=0.99\linewidth]{img/examples/original/cifar100/cifar100_cropped_6.png} & \includegraphics[width=0.99\linewidth]{img/examples/original/cifar100/cifar100_cropped_6.png}\\\midrule
           4 & \includegraphics[width=0.99\linewidth]{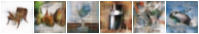} & \includegraphics[width=0.99\linewidth]{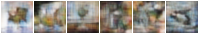}\\
           5 & \includegraphics[width=0.99\linewidth]{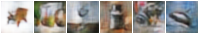} & \includegraphics[width=0.99\linewidth]{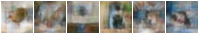}\\
           6 & \includegraphics[width=0.99\linewidth]{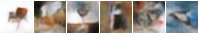} & \includegraphics[width=0.99\linewidth]{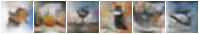}\\
           7 & \includegraphics[width=0.99\linewidth]{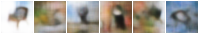} & \includegraphics[width=0.99\linewidth]{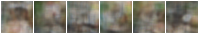}\\\\[6pt]
           &\centering \footnotesize SDAR & \centering \footnotesize PCAT \cite{gao2023pcat}
       \end{tabular}
    }
    \hfill
    \subfloat[PlainNet-20 on Tiny ImageNet]{%
    \renewcommand{\arraystretch}{0}%
        \small%
        \begin{tabular}{@{}m{3mm}@{}m{4.3cm}@{\hspace{2pt}}m{4.3cm}@{}}
            & \includegraphics[width=0.99\linewidth]{img/examples/original/tiny64/tiny64_cropped_6.png} & \includegraphics[width=0.99\linewidth]{img/examples/original/tiny64/tiny64_cropped_6.png}\\\midrule
           4 & \includegraphics[width=0.99\linewidth]{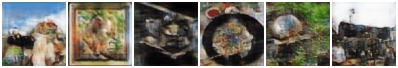} & \includegraphics[width=0.99\linewidth]{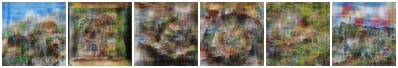}\\
           5 & \includegraphics[width=0.99\linewidth]{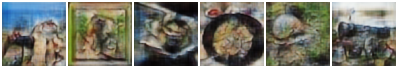} & \includegraphics[width=0.99\linewidth]{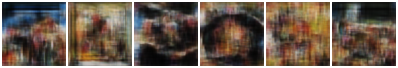}\\
           6 & \includegraphics[width=0.99\linewidth]{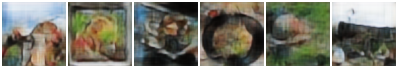} & \includegraphics[width=0.99\linewidth]{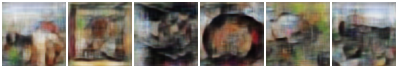}\\
           7 & \includegraphics[width=0.99\linewidth]{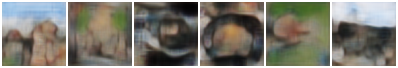} & \includegraphics[width=0.99\linewidth]{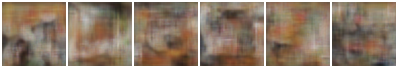}\\\\[6pt]
           &\centering \footnotesize SDAR & \centering \footnotesize PCAT \cite{gao2023pcat}
       \end{tabular}
    }
    \hfill
    \subfloat[PlainNet-20 on STL-10]{%
    \renewcommand{\arraystretch}{0}%
        \small%
        \begin{tabular}{@{}m{3mm}@{}m{4.3cm}@{\hspace{2pt}}m{4.3cm}@{}}
            & \includegraphics[width=0.99\linewidth]{img/examples/original/stl10/stl10_cropped_6.png} & \includegraphics[width=0.99\linewidth]{img/examples/original/stl10/stl10_cropped_6.png}\\\midrule
           4 & \includegraphics[width=0.99\linewidth]{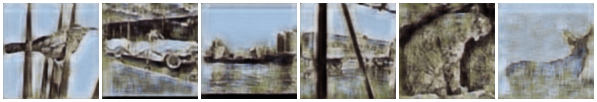} & \includegraphics[width=0.99\linewidth]{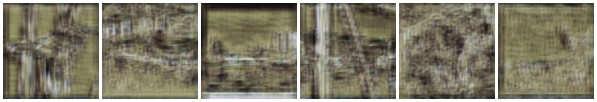}\\
           5 & \includegraphics[width=0.99\linewidth]{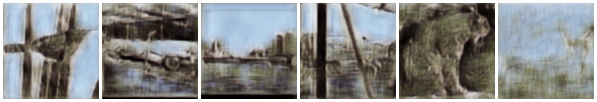} & \includegraphics[width=0.99\linewidth]{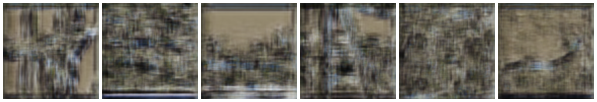}\\
           6 & \includegraphics[width=0.99\linewidth]{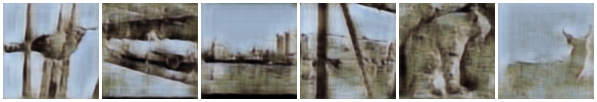} & \includegraphics[width=0.99\linewidth]{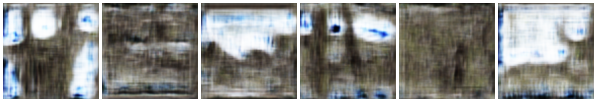}\\
           7 & \includegraphics[width=0.99\linewidth]{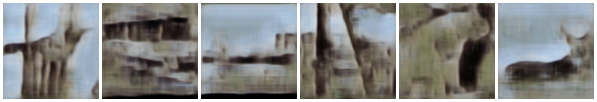} & \includegraphics[width=0.99\linewidth]{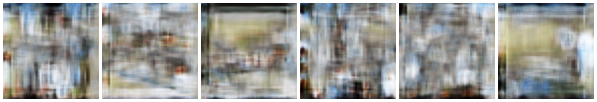}\\\\[6pt]
           &\centering \footnotesize SDAR & \centering \footnotesize PCAT \cite{gao2023pcat}
       \end{tabular}
    }
    \caption{Examples of feature inference attack results of SDAR on all datasets with both models in U-shaped SL.}
    \label{fig:u_examples_full}
\end{figure}

\end{document}